\documentclass[12pt,a4paper]{article}
\usepackage{amsmath,amssymb,cite,comment,mathtools}
\usepackage[dvipdfmx]{graphicx}
\usepackage[usenames]{color}
\usepackage[bookmarksopen,colorlinks=true,linkcolor=dark_green,citecolor=dark_red,urlcolor=dark_red,linktocpage=false]{hyperref}

\definecolor{dark_blue}{rgb}{0,0,0.6}
\definecolor{dark_green}{rgb}{0,0.4,0}
\definecolor{dark_red}{rgb}{0.6,0,0}

\usepackage[height=22.5cm,width=16.5cm,centering]{geometry}

\def\thefootnote{\fnsymbol{footnote}}

\renewcommand{\thefootnote}{\fnsymbol{footnote}}
\begin{document}
%%%%%%%%%%%%%%%%%%%%%%%%%%%%%%%%%%%%%%%%%%%%%%%%%%

%%%%%%%%%%%%%%%%%%%%%%%%%%%%%%%%%%%%%%%%%%%%%%%%%%
\begin{titlepage}

\begin{center}

\hfill CTPU-PTC-18-26 \\
\hfill KEK-TH-2073 \\
\hfill OU-HET-977 \\
\hfill UT-HET-128

\vskip .25in

{\fontsize{17pt}{0pt} \bf
Fingerprinting models of first-order phase transitions 
}
\\ \vspace{5mm} 
{\fontsize{17pt}{0pt} \bf
by the synergy between
}
\\ \vspace{5mm} 
{\fontsize{17pt}{0pt} \bf
collider and gravitational-wave experiments
}

\vskip .5in

{\large
Katsuya Hashino$^{a,b}$, 
Ryusuke Jinno$^{c,d}$, 
Mitsuru Kakizaki$^{b}$,
\\ \vspace{3mm} 
Shinya Kanemura$^{a}$, 
Tomo Takahashi$^{e}$ 
and 
Masahiro Takimoto$^{d,f}$
}

\vskip 0.25in

\begin{tabular}{ll}
$^{a}$ &\!\! 
{\em Department of Physics, Osaka University, Toyonaka, Osaka 560-0043, Japan} \\[.3em]
$^{b}$ &\!\! 
{\em Department of Physics, University of Toyama, 3190 Gofuku, Toyama 930-8555, Japan} \\[.3em]
$^{c}$ &\!\! 
{\em Center for Theoretical Physics of the Universe, Institute for Basic Science (IBS),} \\
&{\em Daejeon 34051, Korea} \\[.3em]
$^{d}$ &\!\! {\em Theory Center, High Energy Accelerator Research Organization (KEK),} \\
&{\em Oho, Tsukuba, Ibaraki 305-0801, Japan} \\[.3em]
$^{e}$ &\!\! 
{\em Department of Physics, Saga University, Saga 840-8502, Japan} \\[.3em]
$^{f}$ &\!\! 
{\em Department of Particle Physics and Astrophysics, Weizmann Institute of Science,} \\
&{\em Rehovot 7610001, Israel} \\[.3em]
\end{tabular}

\end{center}
\vskip .10in

\begin{abstract}
We investigate the sensitivity of future space-based interferometers such as LISA and DECIGO
to the parameters of new particle physics models which drive a first-order phase transition
in the early Universe.
We first perform a Fisher matrix analysis on 
the quantities characterizing the gravitational wave spectrum resulting from the phase transition,
such as the peak frequency and amplitude.
We next perform a Fisher analysis for the quantities which determine the properties of the phase transition,
such as the latent heat and the time dependence of the bubble nucleation rate.
Since these quantities are determined by the model parameters of the new physics,
we can estimate the expected sensitivities to such parameters.
We illustrate this point by taking three new physics models for example:
(1) models with additional isospin singlet scalars
(2) a model with an extra real Higgs singlet,
and 
(3) a classically conformal $B-L$ model.
We find that future gravitational wave observations play complementary roles to future collider experiments 
in pinning down the parameters of new physics models driving a first-order phase transition.
\end{abstract}

\end{titlepage}

\tableofcontents
\thispagestyle{empty}

\renewcommand{\thepage}{\arabic{page}}
\setcounter{page}{1}
\renewcommand{\thefootnote}{$\diamondsuit$\arabic{footnote}}
\setcounter{footnote}{0}
%%%%%%%%%%%%%%%%%%%%%%%%%%%%%%%%%%%%%%%%%%%%%%%%%%

\newpage
\setcounter{page}{1}

%%%%%%%%%%%%%%%%%%%%%%%%%%%%%%%%%%%%%%%%%%%%%%%%%%
\section{Introduction}
\label{sec:Intro}
\setcounter{equation}{0}
%%%%%%%%%%%%%%%%%%%%%%%%%%%%%%%%%%%%%%%%%%%%%%%%%%

The discovery of the Higgs boson $h$ at the CERN Large Hadron Collider (LHC) is
one of the most prominent scientific developments in the past 
decades~\cite{Aad:2012tfa,Chatrchyan:2012xdj},
establishing the spontaneous symmetry breaking and mass generation mechanism experimentally.
Nevertheless, the whole picture of the Higgs sector remains unclear.
Namely, the type and the number of Higgs multiplets, the shape of the Higgs potential and the dynamics of the electroweak phase transition are all unknown.
Understanding the nature of the Higgs sector is important not only for establishing the mechanism for the origin of mass but also for 
unraveling its connection to physics beyond the standard model (SM),  
such as neutrino oscillations, the existence of dark matter, baryon asymmetry of the Universe
and cosmic inflation. 
For example, electroweak baryogenesis in the early Universe~\cite{Kuzmin:1985mm}
is an excellent physics case in which the Higgs sector leads us to new physics.

The conventional way to explore new physics models is to discover new particles 
and/or measure deviations from the SM predictions at collider experiments.
So far, no new particle other than the Higgs boson has been found at the LHC.
As for the deviations in various Higgs boson couplings, 
the expected accuracy is of a few percent level at the High-Luminosity LHC, 
and is improved to a permille level at future electron-positron colliders  
such as the International Linear Collider (ILC)~\cite{BrauJames:2007aa,Djouadi:2007ik,Phinney:2007gp,Behnke:2007gj,Behnke:2013lya}, the Compact LInear Collider  (CLIC)~\cite{Battaglia:2004mw,Linssen:2012hp}, 
the Future Circular Collider of electrons and positrons (FCC-ee)~\cite{Gomez-Ceballos:2013zzn} 
and the Circular Electron Positron Collider (CEPC)~\cite{CEPC-SPPCStudyGroup:2015csa,CEPC-SPPCStudyGroup:2015esa}. 
With such a precision, we may be able to detect deviations in various coupling constants of the Higgs boson 
with a distinct pattern, by which we can fingerprint new physics models indirectly. 

The shape of the Higgs potential can be directly reconstructed by measuring
the triple Higgs boson coupling (the $hhh$ coupling), which is expected to be determined with order of one deviation at the HL-LHC.
If the ILC with the center-of-mass energy of 1~TeV is realized, the error
for the $hhh$ coupling can be reduced to $10\%$~\cite{Asner:2013psa,Moortgat-Picka:2015yla,Fujii:2015jha}, 
which is sufficient to test the scenario of electroweak baryogenesis. 
However, it has recently been discussed that the collision energy of the ILC is reduced to 250~GeV 
with the integrated luminosity to be 2~ab$^{-1}$~\cite{Fujii:2017vwa} to make it a Higgs factory, 
where the Higgs boson decays can be measured very precisely while the measurement of the $hhh$ coupling 
and the top Yukawa coupling are left for the far future.      
If this is the case, there may be little hope for the precise determination of the Higgs potential for a long time.

Fortunately, observation of gravitational waves (GWs) provides us with 
an exciting possibility of probing the early Universe well before the Big Bang Nucleosynthesis.
The detection of GWs from black hole binaries~\cite{Abbott:2016nmj,Abbott:2016blz,Abbott:2017vtc}
and from neutron star mergers~\cite{TheLIGOScientific:2017qsa}
has already signaled a new era of GW astronomy, 
and in the future, space interferometers such as Laser Interferometer Space Antenna (LISA)~\cite{Audley:2017drz} 
and DECi-hertz Interferometer Gravitational wave Observatory (DECIGO)~\cite{Seto:2001qf}
will open up an era of GW cosmology.~\footnote{
For other proposals for space interferometry, see e.g. Refs.~\cite{Luo:2015ght,Dimopoulos:2008sv}.
}
Especially, the LISA project has already been approved and will start its operation in 2034,
making it possible to test various extensions of the SM that predict stochastic GWs.
First-order phase transition is one of the best-motivated GW sources
not only because it is a crucial element for successful electroweak baryogenesis,
but also because the resulting GW spectrum is typically peaked around the interferometer frequency band: milli- to deci-Hertz.
Particle physics models which generate detectable GWs have been vigorously discussed by many authors~\cite{
Apreda:2001tj,Apreda:2001us,Grojean:2006bp,Huber:2007vva,Espinosa:2008kw,
Ashoorioon:2009nf,Kang:2009rd,Jarvinen:2009mh,Konstandin:2010cd,
No:2011fi,Wainwright:2011qy,Barger:2011vm,Leitao:2012tx,
Dorsch:2014qpa,Kozaczuk:2014kva,Schwaller:2015tja,Kakizaki:2015wua,Jinno:2015doa,Huber:2015znp,Leitao:2015fmj,
Huang:2016odd,Garcia-Pepin:2016hvs,Jaeckel:2016jlh,Dev:2016feu,Hashino:2016rvx,
Jinno:2016knw,Barenboim:2016mjm,Kobakhidze:2016mch,Hashino:2016xoj,Artymowski:2016tme,
Kubo:2016kpb,Balazs:2016tbi,Vaskonen:2016yiu,Dorsch:2016nrg,
Huang:2017laj,Baldes:2017rcu,Chao:2017vrq,Beniwal:2017eik,Addazi:2017gpt,Kobakhidze:2017mru,
Tsumura:2017knk,Marzola:2017jzl,Bian:2017wfv,Huang:2017rzf,Iso:2017uuu,Addazi:2017oge,Kang:2017mkl,
Cai:2017tmh,Chao:2017ilw,Aoki:2017aws,Huang:2017kzu,Demidov:2017lzf,Chen:2017cyc,
Chala:2018ari,Hashino:2018zsi,Vieu:2018zze,Croon:2018new,Bruggisser:2018mus,
Wan:2018udw,Huang:2018aja,Bruggisser:2018mrt,Axen:2018zvb,Megias:2018sxv,Alves:2018oct,Baldes:2018emh},
and the resulting GW spectrum has been studied in great detail both from analytic and numerical viewpoints~\cite{
Kosowsky:1991ua,Kosowsky:1992rz,Kosowsky:1992vn,Kamionkowski:1993fg,
Caprini:2006jb,Caprini:2007xq,Huber:2008hg,Kahniashvili:2008pf,Kahniashvili:2008pe,
Caprini:2009fx,Caprini:2009yp,Kahniashvili:2009mf,
Child:2012qg,Hindmarsh:2013xza,Giblin:2014qia,Hindmarsh:2015qta,Kisslinger:2015hua,Caprini:2015zlo,
Hindmarsh:2017gnf,Jinno:2017fby,Jinno:2016vai,Jinno:2017ixd,Konstandin:2017sat,
Cutting:2018tjt,Niksa:2018ofa}.
Therefore, by using this accumulated knowledge,
we may be able to explore the Higgs potential through the observation of GWs at future space-based interferometers.
Importantly, around the time of the LISA project,
precision measurements of the Higgs boson couplings can be made 
at future collider experiments such as the ILC250~\cite{Fujii:2017vwa},
and hence we expect a great synergy between GW observations and collider experiments.

Although many papers have investigated the possibility of detecting GWs from phase transition at future experiments, 
most of them perform a relatively simple analysis in which 
it is discussed whether the predicted GW spectrum comes above or below the sensitivity curves.
This type of analysis gives a rough estimate on 
what kind of models or which parameter space generate a detectable amount of GWs.
However, it cannot quantify to what extent the model parameters can be measured once GWs are detected, 
or what constraints can be derived when future experiments actually give us the data. 
In view of the recent growing interest in GWs,
it is of great importance to study the attainable precision of the future GW experiments in exploring the Higgs sector 
and their complementarity to collider experiments.

In light of these considerations, in this paper, we adopt the method of Fisher matrix analysis 
and study expected constraints in future GW experiments such as LISA and DECIGO.
We also consider a experiment like Big-Bang Observer~\cite{BBO}.
We investigate possible future constraints on parameters characterizing the spectral shape
and those characterizing the properties of the transition.
Since these quantities are determined by fundamental parameters in the underlying particle physics model, 
we can also estimate the expected sensitivities to such parameters.
Then we compare/add them with possible future constraints from collider experiments
to investigate the synergy between GW and collider experiments.  

The organization of the paper is as follows.
In Sec.~\ref{sec:Setup} we summarize our setup for the Fisher matrix analysis,
and explain how we constrain model parameters 
by assuming the specifications of future GW experiments such as LISA, DECIGO and BBO.
In Sec.~\ref{sec:General} we perform a Fisher analysis on a general peaky spectrum, 
taking the peak frequency, its amplitude and spectral slopes as free parameters.
In Sec.~\ref{sec:Transition} we perform a Fisher analysis on transition parameters, 
i.e. $\alpha$, $\beta/H_*$, $T_*$ and so on (which we define later), 
using the GW spectral shapes in the literature.
In Sec.~\ref{sec:Model} we adopt specific particle physics models 
to illustrate that their model parameters can indeed be constrained by future GW experiments,
and discuss their complementarity to collider experiments.
We finally conclude in Sec.~\ref{sec:Conc}.
Some results based on different model setups are also presented in Appendix.

%%%%%%%%%%%%%%%%%%%%%%%%%%%%%%%%%%%%%%%%%%%%%%%%%%
\section{Setup}
\label{sec:Setup}
\setcounter{equation}{0}
%%%%%%%%%%%%%%%%%%%%%%%%%%%%%%%%%%%%%%%%%%%%%%%%%%

In this section, we summarize the formalism adopted in our analysis. 
The GW spectrum from first-order phase transitions is also briefly discussed.

%%%%%%%%%%%%%%%%%%%%%%%%%%%%%%%%%%%%%%%%%%%%%%%%%%
\subsection{Gravitational wave spectrum}
\label{subsec:Power}
%%%%%%%%%%%%%%%%%%%%%%%%%%%%%%%%%%%%%%%%%%%%%%%%%%

Gravitational waves $h_{ij}$ are given as the transverse-traceless part of the metric:
\begin{align}
ds^2
&= 
-dt^2 + a^2(t)(\delta_{ij} + h_{ij}(t,\vec{x}))dx^idx^j.
\end{align}
In the following we consider quantities such as the GW spectrum at the present time 
$t = t_0$ and take $a(t_0) = 1$.
We expand $h_{ij}$ as
\begin{align}
h_{ij} (t, \vec{x})
&= 
\sum_{\lambda = +,\times} \int_{-\infty}^\infty df \int d^2\hat{n} 
~h_\lambda(f,\hat{n}) \epsilon_{ij}^\lambda(\hat{n}) e^{2\pi i f(t - \hat{n} \cdot \vec{x})},
\end{align}
with $\lambda$ being the label for GW polarization,
and we impose the normalization condition 
$\epsilon_{ij}^\lambda(\hat{n})\epsilon_{ij}^{\lambda'}(\hat{n}) = 2\delta_{\lambda \lambda'}$
and the reality condition $\epsilon_{ij}^{\lambda*}(\hat{n}) = \epsilon_{ij}^\lambda(\hat{n})$ 
on the polarization tensor. 
Then GWs $h_\lambda$ satisfy $h_\lambda^*(f,\hat{n}) = h_\lambda(-f,\hat{n})$ 
from the reality of $h_{ij}$.
Now we define the power spectrum $S_h$ by
\begin{align}
\left< 
h_\lambda (f,\hat{n})
h_{\lambda'}^* (f',\hat{n}')
\right>
&= 
\frac{1}{16\pi} 
\delta(f - f')
\delta(\hat{n} - \hat{n}')
\delta_{\lambda \lambda'}
S_h(f).
\label{eq:ShDef}
\end{align}
Here $\left< \cdots \right>$ denotes the ensemble average,
and we assume that the two polarizations of GWs are uncorrelated and have the same amplitude.
This power spectrum satisfies $S_h(f) = S_h(-f)$.

The intensity of GWs is also expressed by the ratio of their energy density to the critical energy density of the Universe.
The former is given by (see e.g. Ref.~\cite{Maggiore:1900zz})
\begin{align}
\rho_{\rm GW} (t)
&= 
\frac{M_P^2}{4} \left< \dot{h}_{ij}^2(t, \vec{x}) \right>_{\rm osc},
\end{align}
where $M_P = (8 \pi G)^{-1/2}$ is the reduced Planck mass 
and $\left< \cdots \right>_{\rm osc}$ means taking both ensemble average and oscillation average.
Note that the L.H.S. does not depend on $\vec{x}$.
Also, $t$ is implicitly taken to be around the present cosmic age and omitted in the following.
We decompose the total energy density into the contributions from each frequency as
\begin{align}
\rho_{\rm GW}
&= 
\int_0^\infty df 
~\frac{d\rho_{\rm GW}}{d\ln f} (f).
\end{align}
Then the GW energy density per logarithmic frequency is written as 
\begin{align}
\frac{d\rho_{\rm GW}}{d\ln f} (f)
&= 
2\pi^2M_P^2 f^3 S_h(f).
\end{align}
We define $\Omega_{\rm GW}$ to be the ratio of the GW energy density to 
the critical energy density $\rho_c$ of the present Universe
\begin{align}
\Omega_{\rm GW}(f)
&\equiv
\frac{1}{\rho_c}
\frac{d\rho_{\rm GW}}{d\ln f}(f),
\end{align}
which is related to the spectral density $S_h$ as
\begin{align}
S_h(f)
&= 
\frac{3H_0^2}{2\pi^2}
\frac{1}{f^3}
\Omega_{\rm GW}(f).
\end{align}
%%

%%%%%%%%%%%%%%%%%%%%%%%%%%%%%%%%%%%%%%%%%%%%%%%%%%
\subsection{Statistical analysis}
\label{subsec:Stat}
%%%%%%%%%%%%%%%%%%%%%%%%%%%%%%%%%%%%%%%%%%%%%%%%%%

In this subsection we summarize the formalism we use for the statistical analysis for GW experiments. 
We use the Fisher matrix analysis, which is essentially a Gaussian approximation of the likelihood function.
As we see below, the Fisher information matrix ${\mathcal F}_{ab}$ is given by 
the curvature of the logarithm of this Gaussian-approximated likelihood around the fiducial parameter point.
The inverse of this Fisher matrix gives the covariance matrix, which characterizes the uncertainties in the parameters.

In Secs.~\ref{sec:General}--\ref{sec:Model} we assume LISA, DECIGO 
and BBO-like (which we denote simply as BBO in the following) experiments.
For cross-correlated detectors such as DECIGO and BBO (here we assume cross-correlated DECIGO detector),
the signal-to-noise ratio and $\delta \chi^2$, the latter of which is given by the logarithm of the likelihood function ${\cal L}$,
are calculated as~\cite{Allen:1997ad,Seto:2005qy}
\begin{align}
\left( \frac{S}{N} \right)^2
&=
2T_{\rm obs}
\sum_{(I,I')}
\int_0^\infty df
~
\frac{\Gamma_{II'}^2(f) S_h^2(f,\left\{ \hat{p} \right\})}
{\sigma_{II'}^{{\rm (null)}2}(f)},
\label{eq:SN}
\end{align}
and
\begin{align}
\delta \chi^2 (\left\{ p \right \},\left\{ \hat{p} \right \})
&=
-2\ln {\mathcal L}(\left\{ p \right\}, \left\{ \hat{p} \right\})
=
2T_{\rm obs}
\sum_{(I,I')}
\int_0^\infty df
~
\frac{\Gamma_{II'}^2(f)
\left[ S_h(f,\left\{ p \right\}) - S_h(f,\left\{ \hat{p} \right\}) \right]^2}
{\sigma_{II'}^2(f)}.
\label{eq:deltachi2}
\end{align}
Here $T_{\rm obs}$ is the observation period and 
$S_h(f,\left\{ p \right\})$ denotes the GW spectrum realized with 
a set of fundamental parameters $\left\{ p \right\}$.
In Secs.~\ref{sec:General}--\ref{sec:Model} we take different parameter sets for $\left\{ p \right\}$.
Throughout this paper $\left\{ \hat{p} \right\}$ denotes fiducial values for $\left\{ p \right\}$.
Also, $I$ and $I'$ run over different interferometer channels.
In addition, $\Gamma_{II'}$ is the overlap reduction function, 
which accounts for the insensitivity to the GW signal due to the geometry of detectors $I$ and $I'$.\footnote{
For the calculation of the overlap reduction function, see e.g. Refs.~\cite{Cornish:2001qi,Cornish:2001bb,Corbin:2005ny}.
For the Fisher analysis including the overlap reduction function with two units of triangular configuration,
see e.g. Refs.~\cite{Seto:2005qy,Kuroyanagi:2009br,Kuroyanagi:2011fy,Kuroyanagi:2012wm,Jinno:2014qka}.
}
In Eqs.~(\ref{eq:SN}) and (\ref{eq:deltachi2}), $\sigma_{II'}$ in the denominator is given by
\begin{align}
\sigma_{II'}^2(f)
&=
\left[ S_I(f) + \Gamma_{II}(f)S_h(f,\left\{ \hat{p} \right\}) \right]
\left[ S_{I'}(f) + \Gamma_{I'I'}(f)S_h(f,\left\{ \hat{p} \right\}) \right]
+
\Gamma_{II'}^2(f)
S_h^2(f,\left\{ \hat{p} \right\}).
\label{eq:sigma}
\end{align}
In this expression we included the effect beyond weak-signal limit~\cite{Kudoh:2005as}.
Also, $\sigma_{II'}^{{\rm (null)}}$ is given by taking $S_h \to 0$ limit in Eq.~(\ref{eq:sigma}).
The Fisher information matrix ${\mathcal F}_{ab}$,
or the inverse of the covariance matrix $\left< \Delta p_a \Delta p_b \right>$,
can be obtained from the expression (\ref{eq:deltachi2}) as
(see e.g. Ref.~\cite{Seto:2005qy})
\begin{align}
{\mathcal F}_{ab}
&=
\left< \Delta p_a \Delta p_b \right>^{-1}
=
2T_{\rm obs}
\sum_{(I,I')}
\int_0^\infty df
~
\frac{\Gamma_{II'}^2(f)
\partial_{p_a} S_h(f,\left\{ \hat{p} \right\})
\partial_{p_b} S_h(f,\left\{ \hat{p} \right\})}
{\sigma_{II'}^2(f)}.
\label{eq:Fab}
\end{align}
Here $\partial_{p_a}$ denotes the derivative with respect to parameter $p_a$.
As a result, $\delta \chi^2$ is approximated as
\begin{align}
\delta \chi^2 (\left\{ p \right \},\left\{ \hat{p} \right \})
&\simeq 
{\mathcal F}_{ab} (p_a - \hat{p}_a)(p_b - \hat{p}_b).
\label{eq:deltachi2Approx}
\end{align}
In the analysis in Secs.~\ref{sec:General}--\ref{sec:Model},
we adopt the effective sensitivity\footnote{
This common definition does not take into account a relatively large factor $T_{\rm obs} \int df \sim T_{\rm obs} \times f_{\rm typ}$ 
(with $f_{\rm typ}$ being the typical peak frequency of the GW spectrum) which appears in Eq.~(\ref{eq:deltachi2Seff}).
To take this into account, one may instead use power-law sensitivity curve: see Ref.~\cite{Thrane:2013oya}.
}
\begin{align}
S_{\rm eff}(f)
&=
\left[
\sum_{(I,I')}
\frac{\Gamma_{II'}^2(f)}{\sigma_{II'}^{{\rm (null)}2}(f)}
\right]^{-1/2},
\label{eq:Seff}
\end{align}
and approximate the expressions for $\delta \chi^2$ and ${\cal F}_{ab}$ as
\begin{align}
\delta \chi^2 (\left\{ p \right \},\left\{ \hat{p} \right \})
&=
2T_{\rm obs}
\int_0^\infty df
~
\frac{\left[ S_h(f,\left\{ p \right\}) - S_h(f,\left\{ \hat{p} \right\}) \right]^2}
{\left[ S_{\rm eff}(f) + S_h(f,\left\{ \hat{p} \right\}) \right]^2},
\label{eq:deltachi2Seff}
\end{align}
and
\begin{align}
{\mathcal F}_{ab}
&=
2T_{\rm obs}
\int_0^\infty df
~
\frac{\partial_{p_a} S_h(f,\left\{ \hat{p} \right\}) \partial_{p_b} S_h(f,\left\{ \hat{p} \right\})}
{\left[ S_{\rm eff}(f) + S_h(f,\left\{ \hat{p} \right\}) \right]^2}.
\label{eq:FabSeff}
\end{align}
This approximation is justified as long as $\Gamma_{II}$, $\Gamma_{I'I'}$ and $\Gamma_{II'}$ are of the same order.

Now we discuss the case of LISA.
LISA is a single-detector and therefore the above expression for cross-correlated detectors may not be applied directly.
As briefly discussed in Ref.~\cite{Thrane:2013oya}, in an ideal case of autocorrelation,
we may use an  expression for the signal-to-noise ratio which is similar to cross-correlated cases.
In this paper we assume that this is indeed the case.
The signal-to-noise ratio in such cases reduces to 
\begin{align}
\left( \frac{S}{N} \right)^2
&=
T_{\rm obs}
\int_0^\infty df
~
\frac{\Gamma^2(f) S_h^2(f,\left\{ \hat{p} \right\})}
{\sigma^{{\rm (null)}2}(f)}.
\label{eq:SNSingle}
\end{align}
Here the label $I$ and $I'$ drop,
and also the factor of two drops compared to Eq.~(\ref{eq:SN}) 
because LISA has only one detector instead of two~\cite{Thrane:2013oya}.
The corresponding expression for the likelihood becomes
\begin{align}
\delta \chi^2 (\left\{ p \right \},\left\{ \hat{p} \right \})
&=
T_{\rm obs}
\sum_{(I,I')}
\int_0^\infty df
~
\frac{\Gamma^2(f)
\left[ S_h(f,\left\{ p \right\}) - S_h(f,\left\{ \hat{p} \right\}) \right]^2}
{\sigma^2(f)},
\label{eq:deltachi2Single}
\end{align}
with the denominator given by
\begin{align}
\sigma^2(f)
&=
\left[ S(f) + \Gamma(f)S_h(f,\left\{ \hat{p} \right\}) \right]^2.
\label{eq:sigmaSingle}
\end{align}
The procedure corresponding to Eqs.~(\ref{eq:Seff})--(\ref{eq:FabSeff}) is essentially the same.
We introduce the effective sensitivity by
\begin{align}
S_{\rm eff}(f)
&=
\left[
\Gamma^2(f)/\sigma^{{\rm (null)}2}(f)
\right]^{-1/2},
\label{eq:SeffSingle}
\end{align}
and write the expressions for $\delta \chi^2$ and ${\mathcal F}_{ab}$ as
\begin{align}
\delta \chi^2 (\left\{ p \right \},\left\{ \hat{p} \right \})
&=
T_{\rm obs}
\int_0^\infty df
~
\frac{\left[ S_h(f,\left\{ p \right\}) - S_h(f,\left\{ \hat{p} \right\}) \right]^2}
{\left[ S_{\rm eff}(f) + S_h(f,\left\{ \hat{p} \right\}) \right]^2},
\label{eq:deltachi2SeffSingle}
\end{align}
and
\begin{align}
{\mathcal F}_{ab}
&=
T_{\rm obs}
\int_0^\infty df
~
\frac{\partial_{p_a} S_h(f,\left\{ \hat{p} \right\}) \partial_{p_b} S_h(f,\left\{ \hat{p} \right\})}
{\left[ S_{\rm eff}(f) + S_h(f,\left\{ \hat{p} \right\}) \right]^2}.
\label{eq:FabSeffSingle}
\end{align}
The resulting approximate expression for $\delta \chi^2$ reduces to Eq.~(\ref{eq:deltachi2Approx}).

In Secs.~\ref{sec:General}--\ref{sec:Model} we sometimes show expected constraints in two-dimensional planes.
When the number of fundamental parameters is more than two, the results are obtained after marginalizing over the parameters
other than those shown in the figures by following the procedure below.
Denoting the marginalized parameters $\left\{ p_\perp \right \}$ collectively,
we first construct marginalized likelihood $\tilde{\mathcal L}$ by integrating out $\left\{ p_\perp \right \}$:
\begin{align}
\tilde{\mathcal L} (\left\{ p \right \},\left\{ \hat{p} \right \})
&=
\left( \prod \int dp_\perp \right)
{\mathcal L} (\left\{ p \right \},\left\{ \hat{p} \right \}).
\end{align}
It is understood that $\left\{ p \right \}$ in the L.H.S. does not contain $\left\{ p_\perp \right \}$.
Then the marginalized $\delta \chi^2$ is given by the likelihood ratio as
\begin{align}
\delta \chi^2 (\left\{ p \right \},\left\{ \hat{p} \right \})
&=
-2 \ln 
\frac{\tilde{\mathcal L} (\left\{ p \right \},\left\{ \hat{p} \right \})}
{\tilde{\mathcal L} (\left\{ \hat{p} \right \},\left\{ \hat{p} \right \})}.
\end{align}
%%

%%%%%%%%%%%%%%%%%%%%%%%%%%%%%%%%%%%%%%%%%%%%%%%%%%
\subsection{Effective sensitivity and foregrounds}
\label{subsec:SeffFG}
%%%%%%%%%%%%%%%%%%%%%%%%%%%%%%%%%%%%%%%%%%%%%%%%%%

In this subsection we clarify our assumptions on the effective sensitivity $S_{\rm eff}$
in Eqs.~(\ref{eq:Seff}) and (\ref{eq:SeffSingle}),
and also explain astrophysical foregrounds which enter $S_h$ in addition to the signal we would like to observe.

%%%%%%%%%%%%%%%%%%%%%%%%%%%%%%%%%%%%%%%%%%%%%%%%%%
\subsubsection{Effective sensitivity}
%%%%%%%%%%%%%%%%%%%%%%%%%%%%%%%%%%%%%%%%%%%%%%%%%%

We use the fitting formulas in Ref.~\cite{Klein:2015hvg} for LISA\footnote{
For a more recent sensitivity curve, see Ref.~\cite{Cornish:2018dyw}.
}
and the ones in Ref.~\cite{Yagi:2011wg} for DECIGO and BBO-like experiments.
\begin{itemize}
\item
LISA
\begin{align}
S_{\rm eff}(f)
&=
\frac{20}{3}
\frac{4S_{\rm acc}(f) + S_{\rm sn}(f) + S_{\rm omn}(f)}{L^2}
\left[
1 + \left( \frac{f}{0.41c/2L} \right)^2 
\right],
\label{eq:SeffLISA}
\end{align}
with $L = 5 \times 10^9$~m and
\begin{align}
S_{\rm acc}(f)
&=
9 \times 10^{-30} \frac{1}{(2\pi f / 1{\rm Hz})^4}
\left( 1 + \frac{10^{-4}}{f / 1{\rm Hz}} \right)~{\rm m^2Hz^{-1}}, \\
S_{\rm sn}(f)
&=
2.96 \times 10^{-23}~{\rm m^2Hz^{-1}}, \\
S_{\rm omn}(f)
&=
2.65 \times 10^{-23}~{\rm m^2Hz^{-1}}.
\end{align}
Here each denotes the acceleration noise, shot noise and other measurement noise, respectively.
\item
DECIGO
\begin{align}
S_{\rm eff}(f) 
&=
\left[
\frac{}{}
7.05 \times 10^{-48} 
\left[
1 + (f / f_p)^2
\right]
\right.
\nonumber \\
&~~~~
\left.
+ 
4.8 \times 10^{-51} 
\frac{(f / 1{\rm Hz})^{-4}}{1 + (f / f_p)^2} 
+
5.33 \times 10^{-52}
(f / 1{\rm Hz})^{-4}
\right]
~{\rm Hz^{-1}},
\label{eq:SeffDECIGO}
\end{align}
with $f_p = 7.36$~Hz.
\item
BBO 
\begin{align}
S_{\rm eff}(f)
&=
\left[
2.00 \times 10^{-49} 
(f / 1{\rm Hz})^2
+ 
4.58 \times 10^{-49}
+
1.26 \times 10^{-52}
(f / 1{\rm Hz})^{-4}
\right]
~{\rm Hz^{-1}}.
\label{eq:SeffBBO}
\end{align}
\end{itemize}
%%

%%%%%%%%%%%%%%%%%%%%%%%%%%%%%%%%%%%%%%%%%%%%%%%%%%
\subsubsection{Foregrounds}
%%%%%%%%%%%%%%%%%%%%%%%%%%%%%%%%%%%%%%%%%%%%%%%%%%

It is known that GWs from astrophysical sources form unresolvable foregrounds.
In this paper, we incorporate their effects by including the following power spectrum to $S_h$
in addition to the signal from first-order phase transitions.\footnote{
It should be noted that these astrophysical foregrounds are correlated among the detectors 
and their treatment might be modified in a more realistic situation.
}
In the analysis in Sec.~\ref{sec:General}--\ref{sec:Model} 
we assume that the spectral form of these foregrounds are already known from other studies and do not consider their uncertainties.

One of such foregrounds comes from compact white dwarf binaries in our Galaxy in the millihertz regime. 
The noise spectrum adopted in Ref.~\cite{Klein:2015hvg} is
\begin{align}
S'_{\rm WD}(f)
&=
\left\{
\begin{matrix*}[l]
(20/3) (f / 1~{\rm Hz})^{-2.3} \times 10^{-44.62}~{\rm Hz^{-1}}
&\equiv S^{(1)}_{\rm WD}(f)
~~~~
(10^{-5}~{\rm Hz} < f < 10^{-3}~{\rm Hz}), \\
(20/3) (f / 1~{\rm Hz})^{-4.4} \times 10^{-50.92}~{\rm Hz^{-1}}
&\equiv S^{(2)}_{\rm WD}(f)
~~~~
(10^{-3}~{\rm Hz} < f < 10^{-2.7}~{\rm Hz}), \\
(20/3) (f / 1~{\rm Hz})^{-8.8} \times 10^{-62.8}~{\rm Hz^{-1}}
&\equiv S^{(3)}_{\rm WD}(f)
~~~~
(10^{-2.7}~{\rm Hz} < f < 10^{-2.4}~{\rm Hz}), \\
(20/3) (f / 1~{\rm Hz})^{-20.0} \times 10^{-89.68}~{\rm Hz^{-1}}
&\equiv S^{(4)}_{\rm WD}(f)
~~~~
(10^{-2.4}~{\rm Hz} < f < 10^{-2}~{\rm Hz}).
\end{matrix*}
\right. 
\end{align}
In our analysis, we use the following smoothened noise spectrum:
\begin{align}
S_{\rm WD}(f)
&=
\frac{1}{1/S^{(1)}_{\rm WD}(f) + 1/S^{(2)}_{\rm WD}(f) + 1/S^{(3)}_{\rm WD}(f) + 1/S^{(4)}_{\rm WD}(f)}.
\label{eq:SWD}
\end{align}
Note that this is a smooth function since $S_{\rm WD}^{(1,2,3,4)}$ are smooth
and effectively works as $S_{\rm WD} \simeq \max (S_{\rm WD}^{(1)}, S_{\rm WD}^{(2)}, S_{\rm WD}^{(3)}, S_{\rm WD}^{(4)})$. 
Also note that $S'_{\rm WD}$ above corresponds to the foreground to the N2A5 configuration of LISA in Ref.~\cite{Klein:2015hvg},
and therefore might not be applicable to DECIGO and BBO in a strict sense.
However, we adopt this expression also for these detectors as a reference value.

Another source of foreground is binary neutron stars and binary black holes.
As recently discussed in Ref.~\cite{Abbott:2017xzg},
the merger rate of neutron stars and black holes
inferred from the detections of GWs by LIGO and Virgo collaboration 
might lead to a significant amount of foreground to stochastic GWs.
However, since there are still large uncertainties in this foreground,
we do not take this into account in the results presented 
in Secs.~\ref{sec:General}--\ref{sec:Model}.
However, 
in Appendix~\ref{app:Other} we show the results 
including this foreground 
by adopting the following 
function given in Ref.~\cite{Abbott:2017xzg}: 
\begin{align}
S_{\rm NSBH} (f)
&= 
\frac{3H_0^2}{2\pi^2}
\frac{1}{f^3}
\times
1.8 \times 10^{-8}
\left(
\frac{f}{25~{\rm Hz}}
\right)^\frac{2}{3}
\label{eq:SNSBH}
\end{align}
for $1~{\rm Hz} < f < 10^3~{\rm Hz}$.
Though in Ref.~\cite{Abbott:2017xzg} the foreground is shown only for 
$10~{\rm Hz} < f < 10^3~{\rm Hz}$,
we have slightly extrapolated it down to $1~{\rm Hz}$ to make conservative estimates.

%%%%%%%%%%%%%%%%%%%%%%%%%%%%%%%%%%%%%%%%%%%%%%%%%%
\subsection{GW spectrum from first-order phase transitions}
%%%%%%%%%%%%%%%%%%%%%%%%%%%%%%%%%%%%%%%%%%%%%%%%%%

In this subsection we summarize the spectral form of GW signal from first-order phase transitions.
In cosmological first-order phase transitions,
bubbles of true vacuum first nucleate at some temperature,
and then they expand due to the pressure difference between the true and false vacua.
They eventually collide and merge with each other, and during this phase 
GWs are sourced by the energy-momentum tensor of the system.
The dynamics is mainly determined by the following parameters:
\begin{align}
T_*,
~~~~
\eta,
~~~~
\alpha,
~~~~
\frac{\beta}{H_*}.
\label{eq:TransitionParam1}
\end{align}
Here $T_*$ is the temperature of the Universe just after the phase transition,
$\eta$ is the (symbolically denoted) coupling of the scalar field to the surrounding plasma,
and $\alpha \equiv \rho_0/\rho_{\rm rad}$ is the ratio between the released latent heat $ \rho_0$ and 
the background plasma energy density $\rho_{\rm rad}$  at the time of transition.
Also, $\beta/H_* = d(S_3/T)/d\ln T|_{T = T_N}$ is the logarithmic temperature derivative 
of the three-dimensional bounce action
with $H_*$ being the Hubble parameter at the time of the transition.
This quantity determines the bubble nucleation rate.
The $\eta$ dependence can be translated to the dependence on bubble wall velocity $v_w$
through the relation in Ref.~\cite{Espinosa:2010hh},\footnote{
In deflagration case, 
the released energy heats up the plasma in front of the bubble walls.
This heating back-reacts on the walls and decreases the pressure exerted on them,
and as a result the wall velocity can change as the transition proceeds (see e.g. Ref.~\cite{Megevand:2017vtb}).
In this paper we do not consider such effects to make our analysis simple.
}
because the wall velocity is determined by the balance between the released energy and the friction on the walls.
Therefore, instead of the parameter set (\ref{eq:TransitionParam1}), 
we consider the following one in the analysis below:
\begin{align}
T_*,
~~~~
v_w,
~~~~
\alpha,
~~~~
\frac{\beta}{H_*},
\label{eq:TransitionParam}
\end{align}

As a result of the scalar and plasma dynamics mentioned above, 
three types of GW sources arise~\cite{Caprini:2015zlo}:
\begin{itemize}
\item
Bubble collisions
\item
Sound waves
\item
Turbulence
\end{itemize}

The first one comes from the collision of walls, i.e. scalar field configurations.
This contribution is well approximated by the envelope of the configurations with infinitely thin shells~\cite{Kosowsky:1991ua,Kosowsky:1992rz,Kosowsky:1992vn,Kamionkowski:1993fg}.
More recently the resulting GW spectrum has been calculated by 
many-bubble simulations~\cite{Huber:2008hg,Weir:2016tov,Konstandin:2017sat}
and also by an analytic approach~\cite{Jinno:2016vai}.\footnote{
This approach also gives a rough estimate on the dependence of the GW spectrum on the nucleation rate, 
which can be used to distinguish particle physics models once we observe GWs from first-order phase transitions~\cite{Jinno:2017ixd}.
}
This scalar field contribution becomes significant when the bubble walls run away~\cite{Bodeker:2009qy},
which occurs when the friction from the thermal plasma on the walls cannot stop the acceleration of the walls.
However, it has recently been pointed out that such runaway bubbles are unlikely 
after taking into account particle splitting processes around the walls~\cite{Bodeker:2017cim}.
Therefore, this scalar contribution now is not considered to be a dominant source of GWs.

The second contribution arises from the dynamics of the fluid, in contrast to bubble collisions.
During bubble expansion, a significant fraction of the released energy is converted to 
the bulk motion of plasma surrounding the walls.
This plasma motion is launched into free propagation after bubbles collide with each other,
and it propagates as sound waves at the level of linear approximation.
These sound waves have been found to continuously source GWs with wavenumbers 
corresponding to the thickness of the bulk fluid~\cite{Hindmarsh:2013xza,Hindmarsh:2015qta,Hindmarsh:2017gnf},
and it has been proposed to model this GW production by sound shells~\cite{Hindmarsh:2016lnk}.
The resulting GW spectrum is~\cite{Hindmarsh:2017gnf}
\begin{align}
\Omega_{\rm sw} h^2
&=
2.65 \times 10^{-6}
\left( \frac{H_*}{\beta} \right)
\left( \frac{\kappa_{\rm sw} \alpha}{1 + \alpha} \right)^2 
\left( \frac{100}{g_*} \right)^{1/3} 
v_w
S_{\rm sw}(f),
\label{eq:OmegaSW}
\end{align}
where 
\begin{align}
&S_{\rm sw}(f)
= 
(f / f_{\rm sw})^3
\left(
\frac{7}{4 + 3 (f / f_{\rm sw})^2}
\right)^{7/2},
\label{eq:SSW}
\\[1ex]
&f_{\rm sw}
= 
1.9 \times 10^{-7}{\rm Hz}
\left( \frac{1}{v_w} \right)
\left( \frac{\beta}{H_*} \right)
\left( \frac{T_*}{1{\rm GeV}} \right)
\left( \frac{g_*}{100} \right)^{1/6}.
\label{eq:fSW}
\end{align}
Here $g_*$ is the number of relativistic degrees of freedom, 
which we take to be $106.75$ throughout the paper.
Also, $\kappa_{\rm sw}$ is the fraction of the released latent heat 
which goes into the plasma bulk motion and contributes to sound-wave formation. 
The peak frequency $f_{\rm sw}$ comes from the aforementioned thickness of the sound shell.

The last one, turbulence contribution, arises when the sound waves develop into 
nonlinear regime at late times. 
In this paper we adopt the spectral form given in Ref.~\cite{Caprini:2009yp,Binetruy:2012ze},
based on the Kolmogorov-type turbulence proposed in Ref.~\cite{Kosowsky:2001xp}:
\begin{align}
\Omega_{\rm turb} h^2
&=
3.35 \times 10^{-4}
\left( \frac{H_*}{\beta} \right)
\left( \frac{\kappa_{\rm turb} \alpha}{1 + \alpha} \right)^{3/2}
\left( \frac{100}{g_*} \right)^{1/3} 
v_w
S_{\rm turb}(f),
\label{eq:OmegaTurb}
\end{align}
where
\begin{align}
&S_{\rm turb}(f)
= 
\frac{(f / f_{\rm turb})^3}
{(1 + (f / f_{\rm turb}))^{11/3}(1 + 8\pi f / h_*)},
\label{eq:STurb}
\\[1ex]
&f_{\rm turb}
= 
2.7 \times 10^{-7}{\rm Hz}
\left( \frac{1}{v_w} \right)
\left( \frac{\beta}{H_*} \right)
\left( \frac{T_*}{1{\rm GeV}} \right)
\left( \frac{g_*}{100} \right)^{1/6},
\label{eq:fTurb}
\\[1ex]
&h_*
= 
1.65 \times 10^{-7}{\rm Hz}
\left( \frac{T_*}{1{\rm GeV}} \right)
\left( \frac{g_*}{100} \right)^{1/6}.
\end{align}
Here $\kappa_{\rm turb}$ is the fraction of the released latent heat which goes into turbulent motion of the plasma.
In numerical simulations it is found that $\kappa_{\rm turb} \simeq (0.05-0.1)\kappa_{\rm sw}$.
In our analysis we fix $\kappa_{\rm turb} = 0.1\kappa_{\rm sw}$.

We note in passing that the estimation of GW spectrum resulting from a first-order phase transition is an ongoing hot topic
(e.g. Refs.~\cite{Hindmarsh:2016lnk,Hindmarsh:2017gnf,Jinno:2017fby,Jinno:2017ixd,Konstandin:2017sat,
Cutting:2018tjt,Jackson:2018maa,Niksa:2018ofa}),
and therefore the above spectra might not be exact.
However, an important point is that the phase transition dynamics is determined by a few parameters,
and the GW spectrum is determined by such parameters accordingly.
Therefore, it is interesting to ask what kind of information we can obtain from the observation of GWs 
if we know the exact form of the GW spectrum, which depends on a few parameters related to phase transition.
In this paper we illustrate this point by using the expressions (\ref{eq:OmegaSW}) and (\ref{eq:OmegaTurb}).

%%%%%%%%%%%%%%%%%%%%%%%%%%%%%%%%%%%%%%%%%%%%%%%%%%
\section{Fisher analysis on general spectrum}
\label{sec:General}
\setcounter{equation}{0}
%%%%%%%%%%%%%%%%%%%%%%%%%%%%%%%%%%%%%%%%%%%%%%%%%%

In this section we first perform a Fisher analysis on a general peaky GW spectrum,
taking the peak amplitude, peak frequency and spectral slopes as free parameters.\footnote{
For a recent study on more general spectral shapes, see Ref.~\cite{Kuroyanagi:2018csn}.
}
We assume that the signal takes the following form
\begin{align}
\Omega_{\rm GW}(f)
&= 
\Omega_{\rm GW, peak}
\times
\left[ 
(f / f_{\rm peak})^{-n_L}
+
(f / f_{\rm peak})^{-n_R}
\right]^{-1}
\nonumber \\
&\simeq
\Omega_{\rm GW, peak}
\times
\left\{
\begin{matrix*}[l]
(f / f_{\rm peak})^{n_L}
&(f < f_{\rm peak}), \\
(f / f_{\rm peak})^{n_R}
&(f > f_{\rm peak}).
\end{matrix*}
\right. 
\label{eq:OmegaGeneral}
\end{align}
We also assume $n_L > 0$ and $n_R < 0$.

We first show the result of a Fisher analysis using 
$\delta \chi^2$ given in Eqs.~(\ref{eq:deltachi2Seff}), (\ref{eq:deltachi2SeffSingle})
with the effective sensitivities (\ref{eq:SeffLISA})--(\ref{eq:SeffBBO}).
We take several fiducial values for the parameters $(f_{\rm peak}, \Omega_{\rm GW,peak}, n_L, n_R)$ as examples.
The sample points we consider are
\begin{itemize}
\item
Point 1:
$(f_{\rm peak}, \Omega_{\rm peak}) = (10^{-2}~{\rm Hz}, 10^{-7})$,
\item
Point 2:
$(f_{\rm peak}, \Omega_{\rm peak}) = (10^{-1}~{\rm Hz}, 10^{-10})$,
\item
Point 3:
$(f_{\rm peak}, \Omega_{\rm peak}) = (10~{\rm Hz}, 10^{-10})$,
\item
Point 4:
$(f_{\rm peak}, \Omega_{\rm peak}) = (10^{-1}~{\rm Hz}, 10^{-14})$.
\end{itemize}
For the spectral slope, we consider two cases with $(n_L,n_R) = (3,-4)$ and $(1,-3)$.
The former corresponds to the sound-wave form given in Eqs.~(\ref{eq:OmegaSW})--(\ref{eq:fSW}),
while the latter corresponds to the one coming from the bubble-like structure mentioned at the end of the previous section.
In this section we only show the results for $(n_L,n_R) = (3,-4)$,
and the ones for $(n_L,n_R) = (1,-3)$ are shown in Appendix~\ref{app:Other}.

\begin{figure}
\begin{center}
\includegraphics[width=0.95\columnwidth]{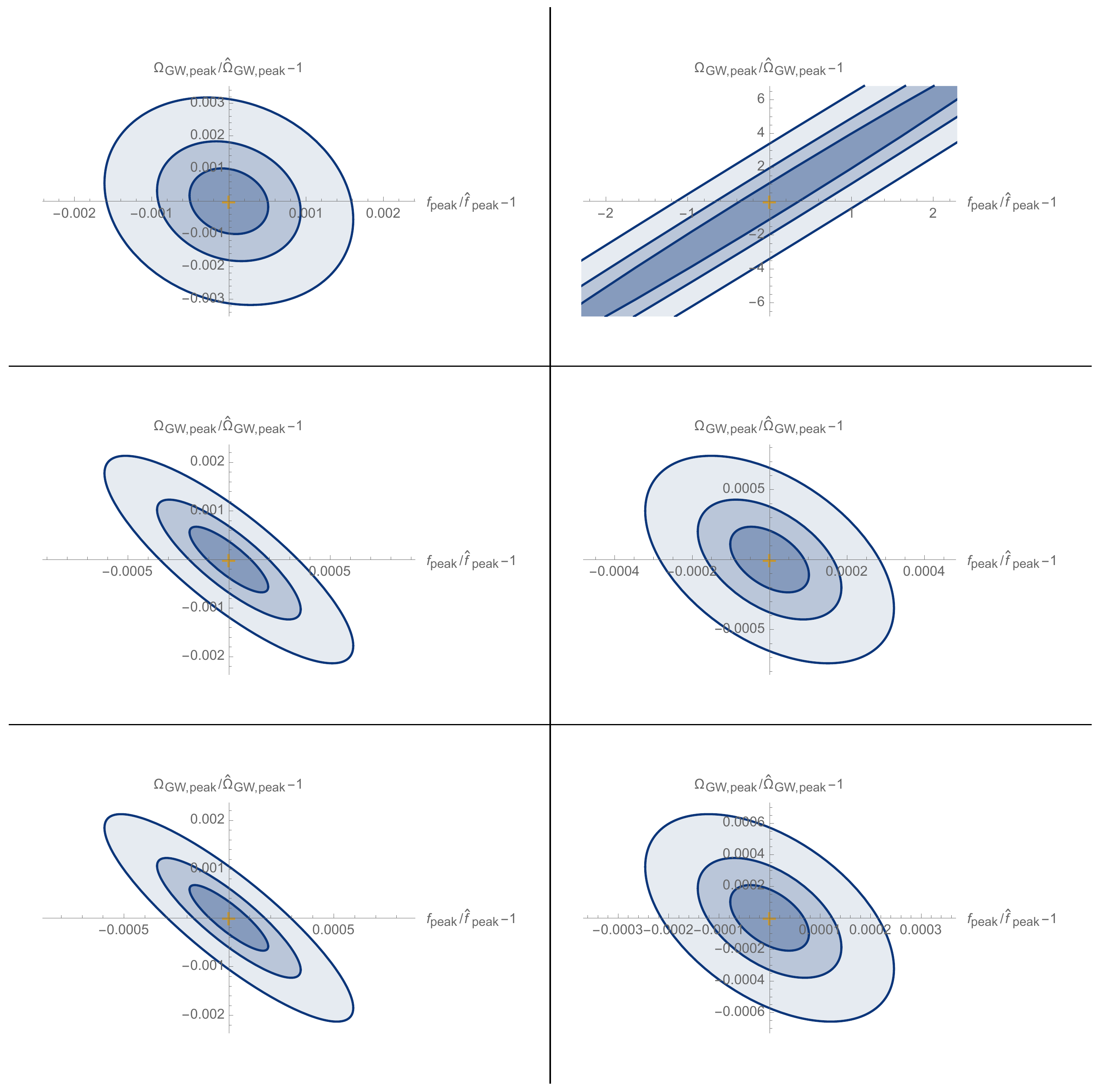} 
\caption{\small
$1~\sigma$ contours for Point 1 (left column) and 2 (right column)
for LISA (top), DECIGO (middle) and BBO (bottom).
Three contours in each panel correspond to $T_{\rm obs} = 1$, $3$ and $10$ years.
The spectral slopes are taken to be $(n_L,n_R) = (3,-4)$.
}
\label{fig:FisherGeneral1}
\end{center}
\end{figure}

\begin{figure}
\begin{center}
\includegraphics[width=0.95\columnwidth]{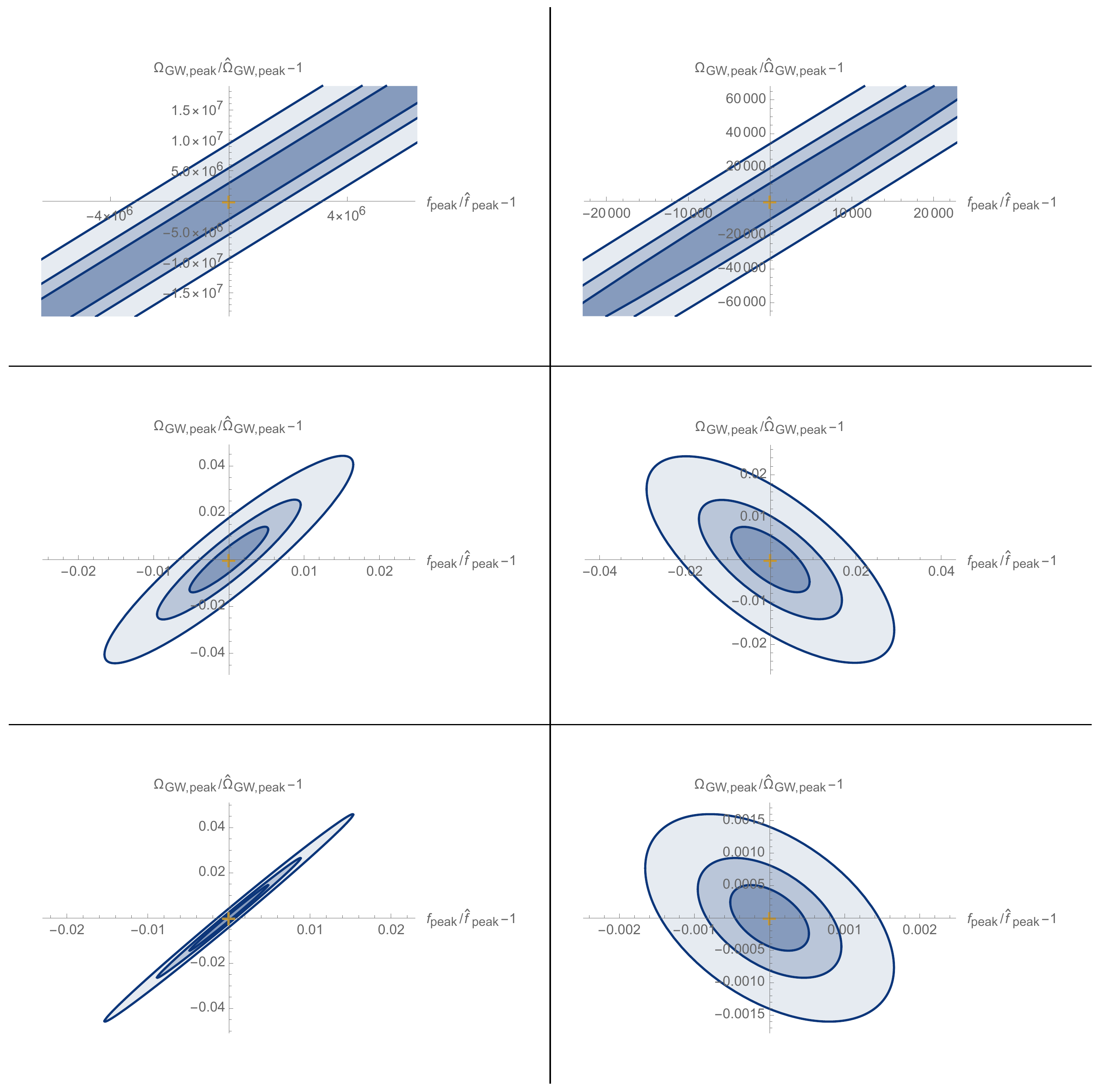}
\caption{\small
$1~\sigma$ contours for Point 3 (left column) and 4 (right column).
Otherwise the same as Fig.~\ref{fig:FisherGeneral1}.
}
\label{fig:FisherGeneral2}
\end{center}
\end{figure}

First, in Fig.~\ref{fig:fOmegaGeneral},
the sensitivity curves for LISA, DECIGO and BBO-like experiments (\ref{eq:SeffLISA}), (\ref{eq:SeffDECIGO}) and (\ref{eq:SeffBBO}),  
the foreground from white dwarfs (\ref{eq:SWD}),
and the signals for Point 1--4 are shown. 
The results of a Fisher analysis for Point 1--4 are shown in 
Fig.~\ref{fig:FisherGeneral1} (Point 1 and 2) and 
Fig.~\ref{fig:FisherGeneral2} (Point 3 and 4).
In these figures, we marginalize the two spectral indices $n_L$ and $n_R$ following the procedure in Sec.~\ref{subsec:Stat}
and show contours of $f_{\rm peak} /\hat{f}_{\rm peak} -1$ 
and $\Omega_{\rm GW, peak} /\hat{\Omega}_{\rm GW, peak} -1$ for $\delta \chi^2 = 2.3$,
which corresponds to 1 $\sigma$ in the two dimensional plane.
The three contours in each panel correspond to $T_{\rm obs} = 1$, $3$ and $10$ years.
Also, the panel is enlarged when the size of outermost ellipse 
(which corresponds to $T_{\rm obs} = 1$ year) far exceeds unity.
For Point 1 we expect parameter determination with a good precision 
with all the three detectors.
For Point 2--4 parameter determination by LISA is challenging 
but we still expect a good sensitivity for DECIGO and BBO.
It is seen that for Point 4 DECIGO can perform well even though the signal is below the sensitivity curve.
This is understood through Eq.~(\ref{eq:deltachi2Seff}): 
even if the signal $S_h$ is below the sensitivity of the detector $S_{\rm eff}$,
we have an additional factor $(T_{\rm obs} \times f_{\rm typ}) \sim ({\mathcal O}(1)~{\rm years} \times 0.1~{\rm Hz})$ 
with $f_{\rm typ}$ being the typical signal frequency 
for the fiducial values of $\hat{f}_{\rm peak}$ and $\hat{\Omega}_{\rm GW, peak}$.

\begin{figure}
\begin{center}
\includegraphics[width=\columnwidth]{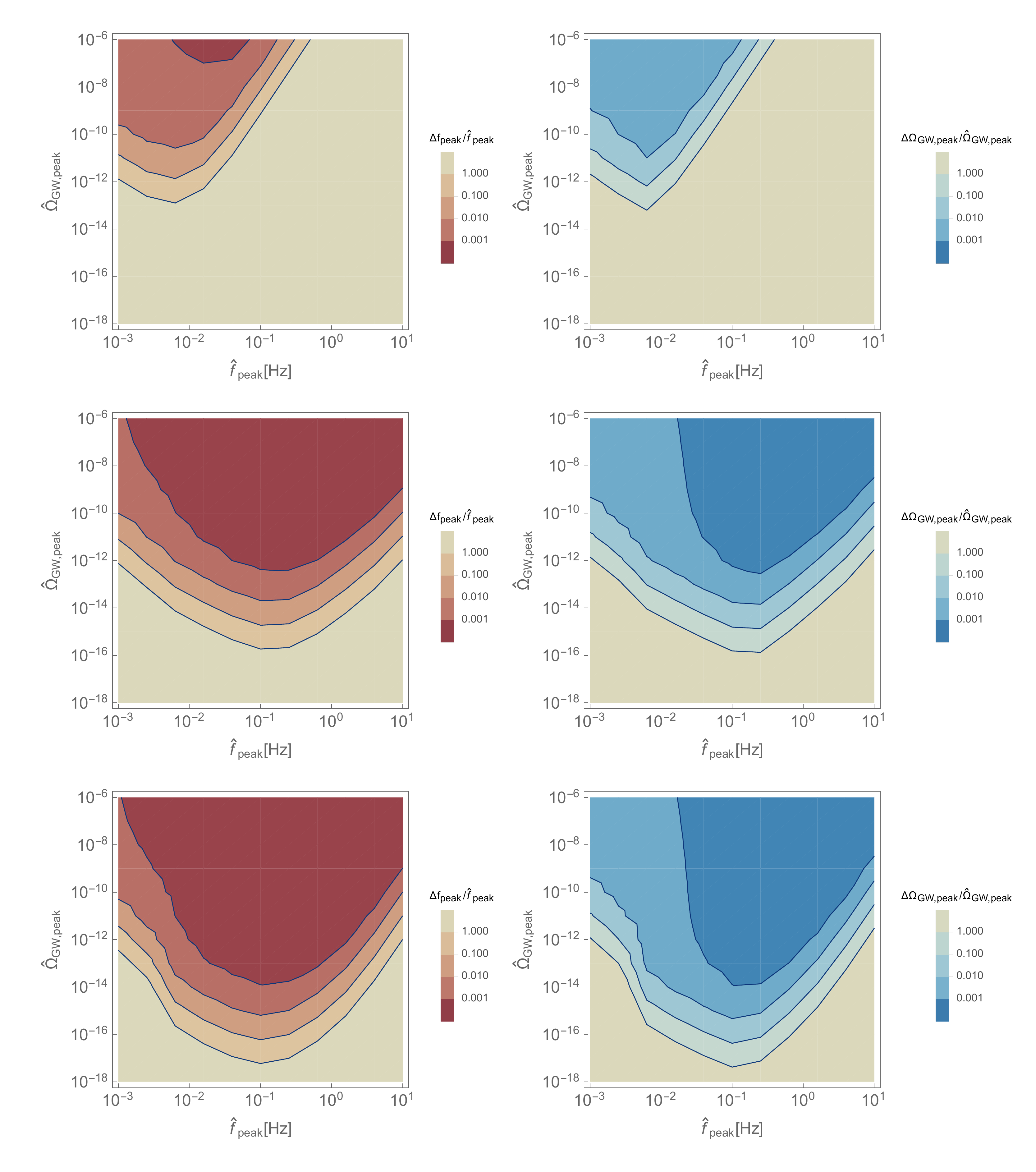} 
\caption{\small
$1~\sigma$ fractional error $\Delta f_{\rm peak}/\hat{f}_{\rm peak}$ (left) 
and $\Delta \Omega_{\rm GW,peak}/\hat{\Omega}_{\rm GW,peak}$ (right)
for the fiducial values $\hat{f}_{\rm peak}$ and $\hat{\Omega}_{\rm GW, peak}$
for LISA (top), DECIGO (middle) and BBO (bottom).
The spectral slopes and observational period are taken to be $(n_L,n_R) = (3,-4)$ and $T_{\rm obs} = 1$~year.
}
\label{fig:DelfDelOmega}
\end{center}
\end{figure}

In Fig.~\ref{fig:DelfDelOmega} we show contours for 
$1~\sigma$ fractional error $\Delta f_{\rm peak}/\hat{f}_{\rm peak}$ 
and $\Delta \Omega_{\rm GW,peak}/\hat{\Omega}_{\rm GW,peak}$
(where $\Delta f_{\rm peak}$ and $\Delta \Omega_{\rm GW,peak}$ correspond to 
$\delta \chi^2 = 1$ for one degree of freedom)
after marginalizing the other parameters.

%%%%%%%%%%%%%%%
\begin{figure}
\begin{center}
\includegraphics[width=0.6\columnwidth]{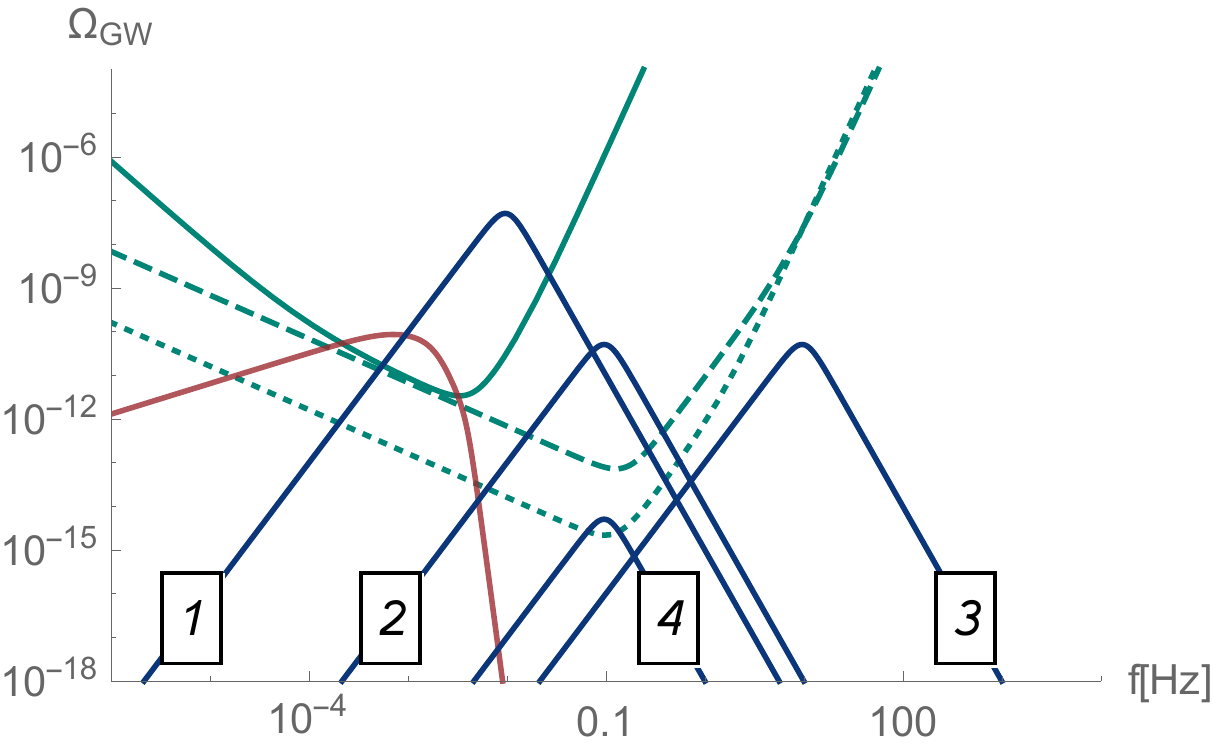} 
\caption{\small
Sensitivity curves for LISA~(green-solid), DECIGO~(green-dashed) and BBO~(green-dotted).
Blue curves correspond to the GW spectra for the sample points 1-4 in the main text.
Red lines show the contribution from compact white dwarf binaries $S_{\rm WD}$.
}
\label{fig:fOmegaGeneral}
\end{center}
\end{figure}
%%%%%%%%%%%%%%%

%%%%%%%%%%%%%%%
%%%%%%%%%%%%%%%

%%%%%%%%%%%%%%%
%%%%%%%%%%%%%%%

%%%%%%%%%%%%%%%
%%%%%%%%%%%%%%%

\clearpage

%%%%%%%%%%%%%%%%%%%%%%%%%%%%%%%%%%%%%%%%%%%%%%%%%%
\section{Fisher analysis on transition parameters}
\label{sec:Transition}
\setcounter{equation}{0}
%%%%%%%%%%%%%%%%%%%%%%%%%%%%%%%%%%%%%%%%%%%%%%%%%%

\begin{figure}
\begin{center}
\includegraphics[width=\columnwidth]{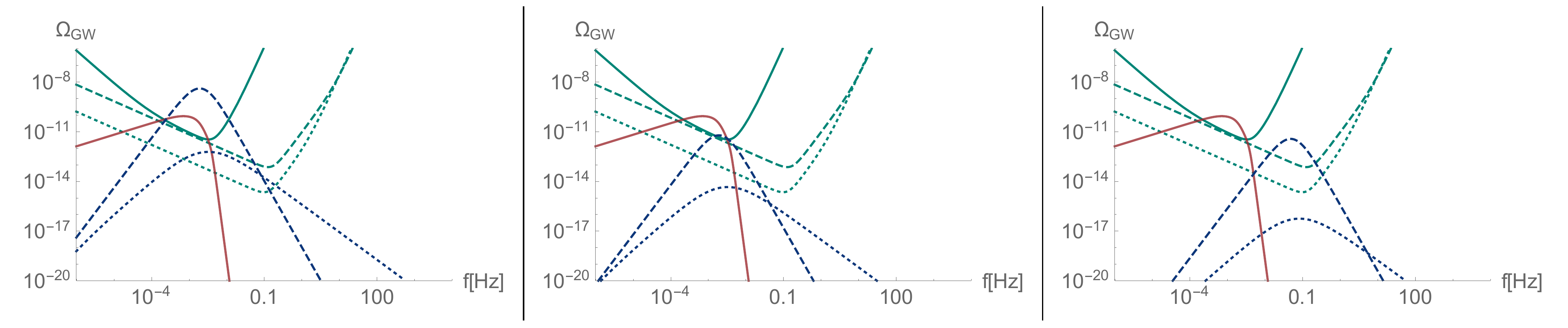}
\caption{\small
Sensitivity curves for LISA (green-solid), DECIGO (green-dashed) and BBO (green-dotted).
Red line shows the foreground from compact white dwarf binaries $S_{\rm WD}$.
Each panel corresponds to Point A--C in the main text from left to right.
}
\label{fig:fOmegaTransition}
\end{center}
\end{figure}

In this section we perform a Fisher analysis on the transition parameters (\ref{eq:TransitionParam})
with the spectrum provided in Eqs.~(\ref{eq:OmegaSW}) and (\ref{eq:OmegaTurb}).
When we show contours of constant likelihood below, we assume three fiducial points:
\begin{itemize}
\item
Point A:
$(\alpha, \beta/H_*, v_w, T_*) = (1, 100, 1, 100~{\rm GeV})$,
\item
Point B:
$(\alpha, \beta/H_*, v_w, T_*) = (0.1, 100, 1, 100~{\rm GeV})$,
\item
Point C:
$(\alpha, \beta/H_*, v_w, T_*) = (0.3, 500, 0.2, 100~{\rm GeV})$.
\end{itemize}
In Figs.~\ref{fig:fOmegaTransition} we show the signal with these parameter points 
as well as the sensitivity curves and the foreground from white dwarfs.

Before showing the results, it should be mentioned that, 
if we take all four parameters in Eq.~(\ref{eq:TransitionParam}) completely free, 
it is generically difficult to determine their values at the same time.
This is because of the following reason.
Suppose that the detector see only the sound-wave peak, Eq.~(\ref{eq:OmegaSW}).
(Notice that the sound-wave peak amplitude is typically much larger than that of turbulence.)
For the spectral shape given by Eq.~(\ref{eq:OmegaSW}), the information the detectors can obtain 
is the position (i.e. frequency and amplitude) of the peak,
which is not enough to determine all the four parameters.
Therefore in the analysis below we limit the number of free parameters
to two ($\alpha$ and $\beta/H_*$) or to three ($\alpha$, $\beta/H_*$ and $T_*$).
When we show two-parameter planes in three-parameter analysis,
we marginalize over $T_*$ following the procedure in Sec.~\ref{subsec:Stat}.

\begin{figure}
\begin{center}
\includegraphics[width=0.95\columnwidth]{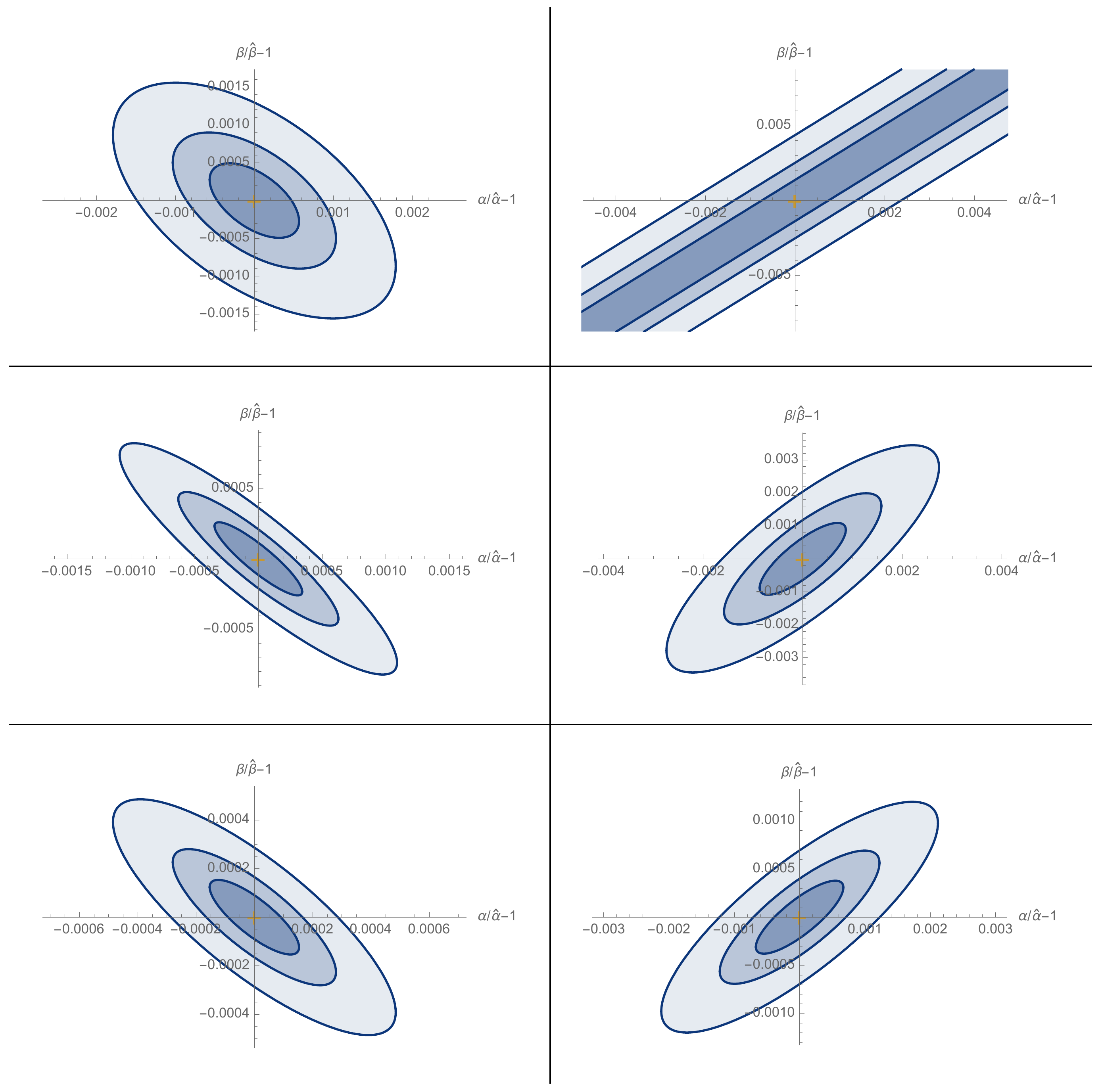}
\caption{\small
$1~\sigma$ contours for Point A for LISA (top), DECIGO (middle) and BBO (bottom).
Left and right columns correspond to 2- and 3-parameter analysis, respectively.
Three contours in each panel correspond to $T_{\rm obs} = 1$, $3$ and $10$ years.
}
\label{fig:FisherTransition1}
\end{center}
\end{figure}

\begin{figure}
\begin{center}
\includegraphics[width=0.95\columnwidth]{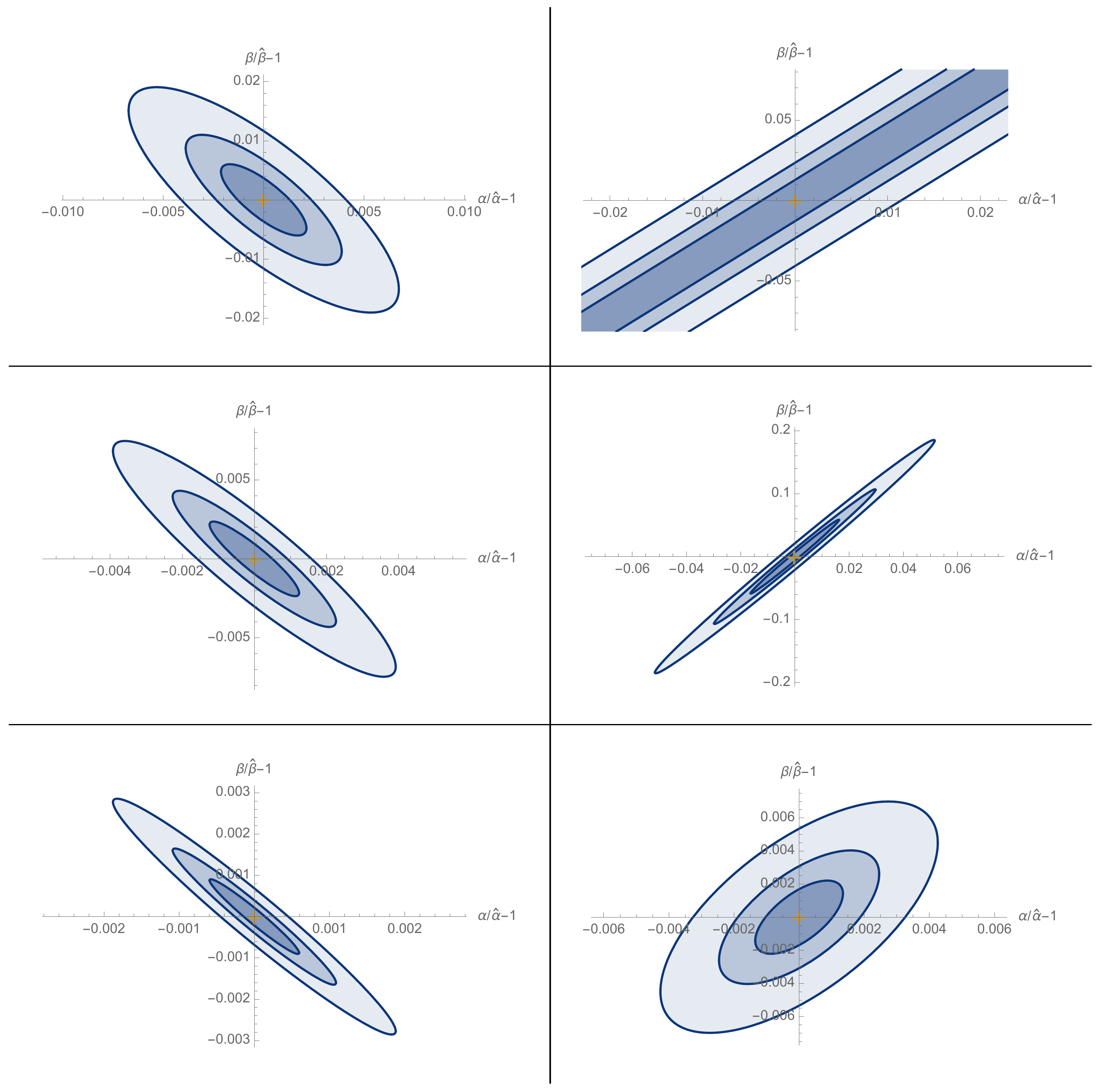}
\caption{\small
$1~\sigma$ contours for Point B.
Otherwise the same as Fig.~\ref{fig:FisherTransition1}.
}
\label{fig:FisherTransition2}
\end{center}
\end{figure}

\begin{figure}
\begin{center}
\includegraphics[width=0.95\columnwidth]{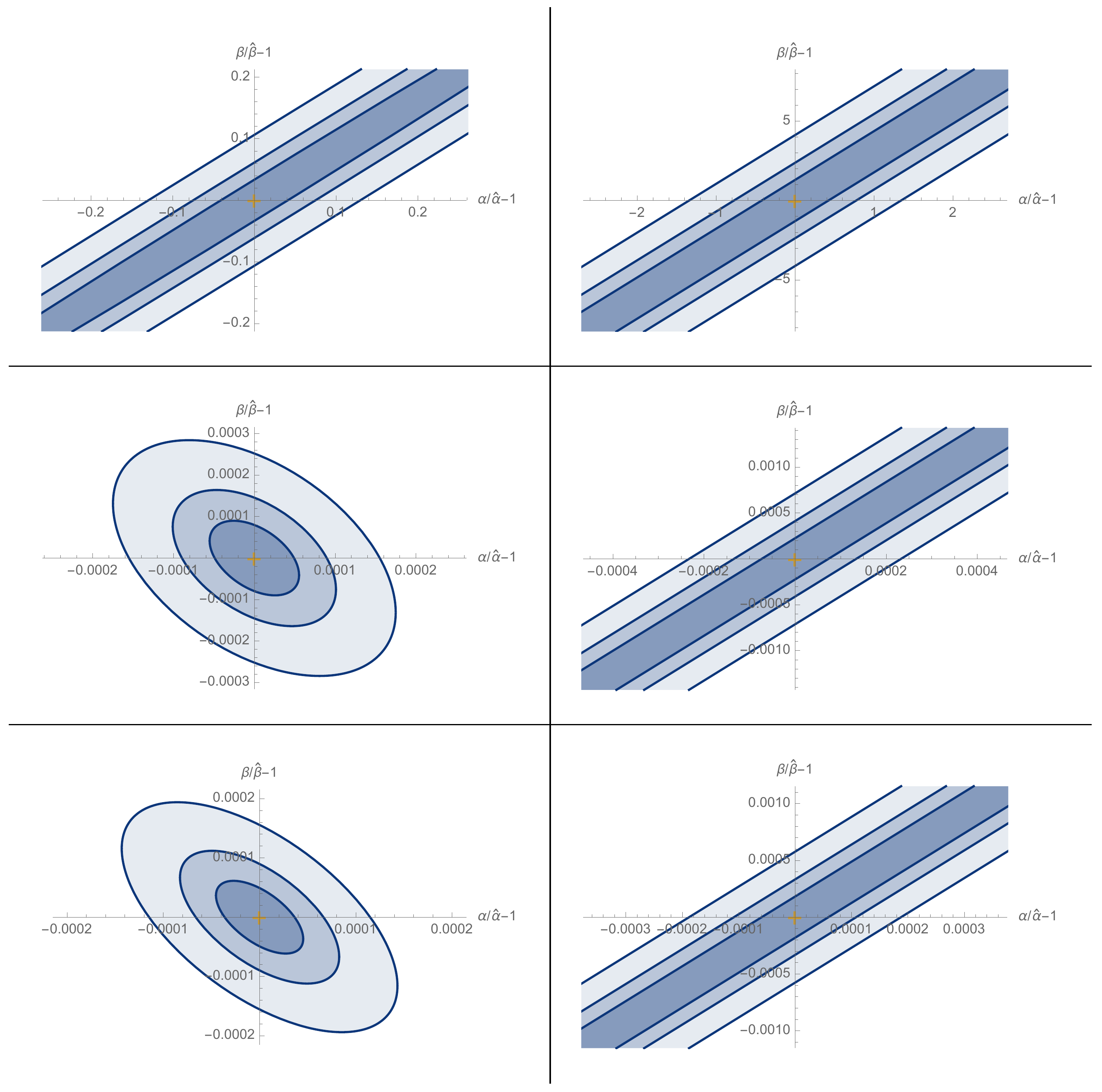}
\caption{\small
$1~\sigma$ contours for Point C.
Otherwise the same as Fig.~\ref{fig:FisherTransition1}.
}
\label{fig:FisherTransition3}
\end{center}
\end{figure}

Figs.~\ref{fig:FisherTransition1}--\ref{fig:FisherTransition3} are the results of a Fisher analysis
for the three fiducial points above.
In these figures the left and right columns correspond to two- and three-parameter analysis, respectively,
and the three contours in each panel correspond to the analysis with $T_{\rm obs} = 1, 3$ and $10$ years.
Also, the panel is enlarged when the size of outermost ellipse 
(which corresponds to $T_{\rm obs} = 1$ year) far exceeds unity.
First, for two-parameter analysis (left columns of Figs.~\ref{fig:FisherTransition1}--\ref{fig:FisherTransition3},
it is seen that the parameters are well determined (except for Point C for LISA).
This reflects the fact that those detectors indeed see the spectral peak from sound waves.
Also, even for LISA with Point C, one combination of $\alpha$ and $\beta/H_*$ is well determined,
even though the spectrum do not hit the sensitivity curve in the right panel of Fig.~\ref{fig:fOmegaTransition}.
This is because of the same reason as Sec.~\ref{sec:General}:
we have an additional factor $(T_{\rm obs} \times f_{\rm typ}) \sim ({\mathcal O}(1)~{\rm years} \times 0.1~{\rm Hz})$ 
which boosts the sensitivity (compared to na\"ive sensitivity curve argument).
Second, for the three-parameter analysis, 
it is seen that DECIGO and BBO still perform well. 
This is because they can see the spectral shape coming from turbulence in addition to sound waves.
On the other hand, for LISA, there appears a strong degeneracy 
in Figs.~\ref{fig:FisherTransition1}--\ref{fig:FisherTransition3}.
This degeneracy arises from the fact that LISA cannot see the spectrum from turbulence
and cannot determine three (or more) parameters at the same time.
However, it should be noted that LISA is still able to determine two parameters,
which means that it can significantly contribute to narrowing down parameters of underlying particle physics models.
We will return to this point in Sec.~\ref{sec:Model}.

\begin{figure}
\begin{center}
\includegraphics[width=\columnwidth]{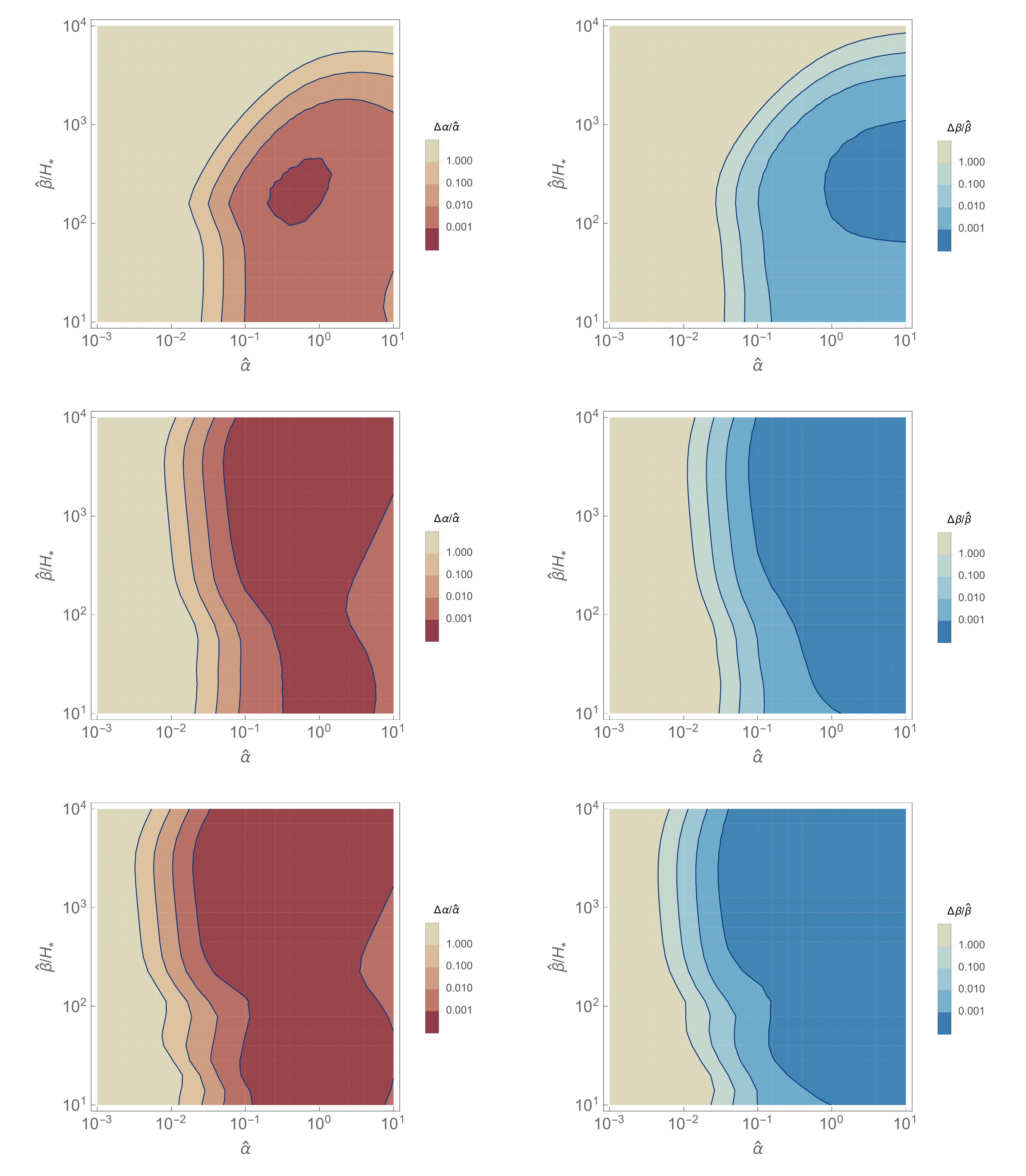} 
\caption{\small
1 $\sigma$ fractional error $\Delta \alpha / \hat{\alpha}$ (left) and $\Delta \beta / \hat{\beta}$ (right) 
for the fiducial values $\hat{\alpha}$ and $\hat{\beta}$ for 2-parameter analysis.
Each row corresponds to LISA (top), DECIGO (middle) and BBO (bottom).
The wall velocity is taken to be $v_w = 1$.
}
\label{fig:S4N2DelalphaDelbeta}
\end{center}
\end{figure}

\begin{figure}
\begin{center}
\includegraphics[width=\columnwidth]{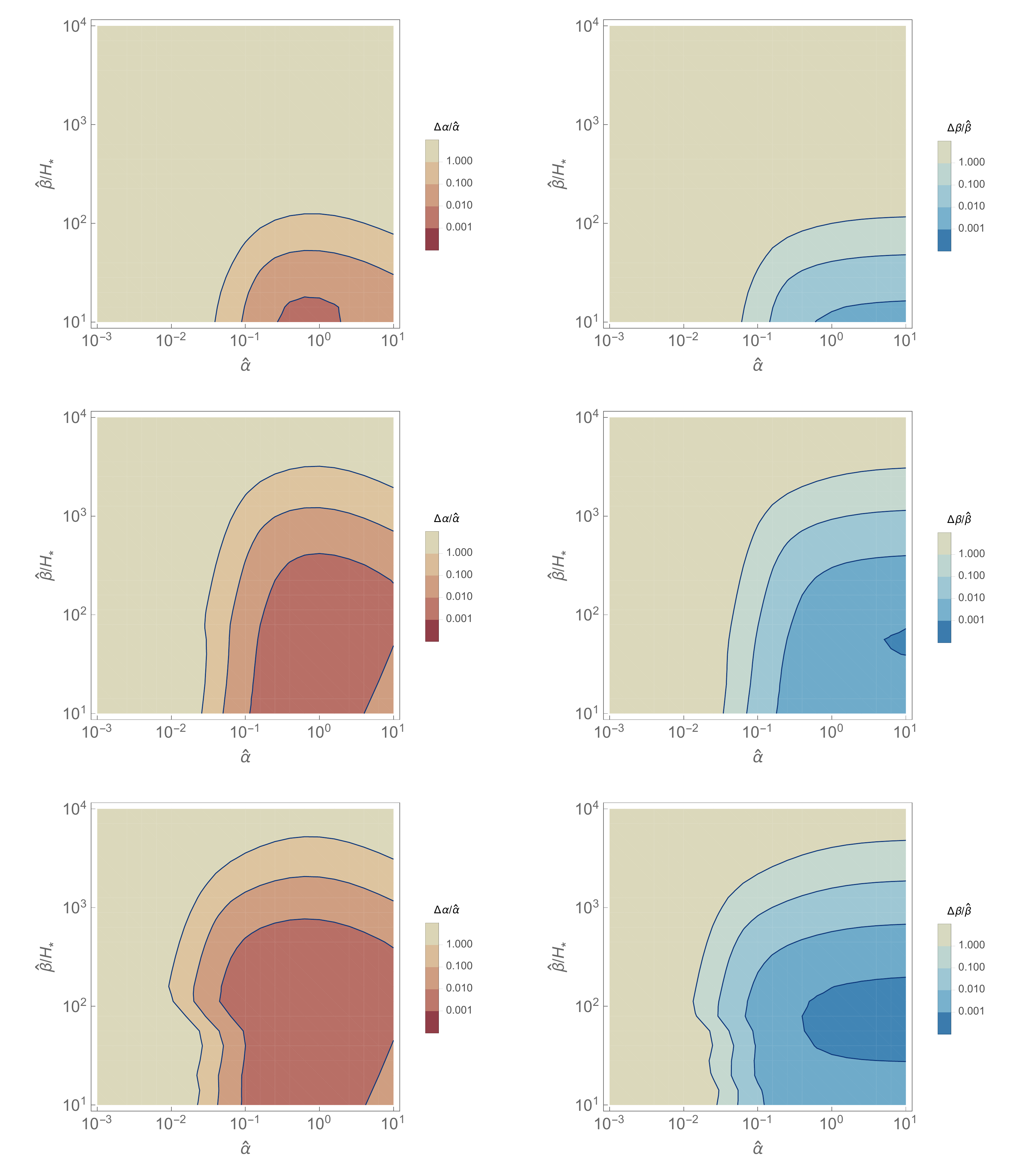} 
\caption{\small
1 $\sigma$ fractional error $\Delta \alpha / \hat{\alpha}$ (left) and $\Delta \beta / \hat{\beta}$ (right) for 3-parameter analysis.
Each row corresponds to LISA (top), DECIGO (middle) and BBO (bottom).
Otherwise the same as Fig.~\ref{fig:S4N2DelalphaDelbeta}.
}
\label{fig:S4N3DelalphaDelbeta}
\end{center}
\end{figure}

Figs.~\ref{fig:S4N2DelalphaDelbeta} and \ref{fig:S4N3DelalphaDelbeta}
show 1 $\sigma$ fractional error for $\alpha$ and $\beta/H_*$
for two- and three-parameter analysis, respectively.
In these figures we fixed $T_* = 100$~GeV.
It is seen that a high sensitivity spot appears in some of the panels.
This is because, if one fixes $\alpha$, 
there is a typical value of $\beta/H_*$ which makes the signal peak close to the frequency
at which the detector is most sensitive.
(Note that for too small $\beta/H_*$ the signal from sound waves starts to overlap with 
the foreground from white dwarfs.)
Also, it is seen that the sensitivity on $\alpha$ becomes worse as $\alpha$ increases for fixed $\beta/H_*$.
This is because the spectral shape, Eqs.~(\ref{eq:OmegaSW}) and (\ref{eq:OmegaTurb}),
becomes almost independent of $\alpha$ for $\alpha \gg 1$. 
Physically this means that the transition dynamics looks almost the same
when the released latent heat dominates the radiation energy density (i.e. $\alpha \gg 1$).

%%%%%%%%%%%%%%%
%%%%%%%%%%%%%%%

%%%%%%%%%%%%%%%
%%%%%%%%%%%%%%%

%%%%%%%%%%%%%%%
%%%%%%%%%%%%%%%

%%%%%%%%%%%%%%%
%%%%%%%%%%%%%%%

%%%%%%%%%%%%%%%
%%%%%%%%%%%%%%%

%%%%%%%%%%%%%%%
%%%%%%%%%%%%%%%

\clearpage

%%%%%%%%%%%%%%%%%%%%%%%%%%%%%%%%%%%%%%%%%%%%%%%%%%
\section{Fisher analysis on model parameters}
\label{sec:Model}
\setcounter{equation}{0}
%%%%%%%%%%%%%%%%%%%%%%%%%%%%%%%%%%%%%%%%%%%%%%%%%%

In this section we take some specific examples of particle physics models 
which give rise to a first-order phase transition in the early Universe,
and illustrate how the detection of GWs contributes to narrow down the fundamental model parameters.
Below we consider 
(1) models with additional isospin singlet scalar fields 
with and without the classical conformal invariance
(2) a model with an extra Higgs singlet field,
and 
(3) a classically conformal $B-L$ model,
respectively.

%%%%%%%%%%%%%%%%%%%%%%%%%%%%%%%%%%%%%%%%%%%%%%%%%%
\subsection{\texorpdfstring{$O(N)$}{Lg} singlet extensions of the SM}
\label{subsec:SingletON}
%%%%%%%%%%%%%%%%%%%%%%%%%%%%%%%%%%%%%%%%%%%%%%%%%%

We first consider extensions of the SM in which $N$ additional isospin singlet scalars 
$\vec{S}=(S_1,\dots, S_N)^T$ with a global $O(N)$ symmetry are added to the SM particle content.
There are two classes in such extensions: with or without classical conformal invariance (CCI). 

%%%%%%%%%%%%%%%%%%%%%%%%%%%%%%%%%%%%%%%%%%%%%%%%%%
\subsubsection*{Model}
%%%%%%%%%%%%%%%%%%%%%%%%%%%%%%%%%%%%%%%%%%%%%%%%%%

The tree-level potential of the $O(N)$ models with CCI is given by
\begin{align}
V_0 = \lambda_\Phi |\Phi|^4+\frac{\lambda_S}{4}|\vec{S}|^4+\frac{\lambda_{\Phi S}}{2}|\Phi|^2|\vec{S}|^2,
\label{eq:VwithCCI}
\end{align}
where $\Phi$ is the isospin doublet Higgs field.
In the CCI models, the electroweak symmetry breaking occurs by the Coleman-Weinberg mechanism~\cite{Coleman:1973jx}. 
This class of models has a distinctive phenomenological feature: 
the deviation in the $hhh$ coupling is universally about $70\%$~\cite{Hashino:2015nxa}.
Gravitational-wave production in this class of models has been studied in e.g. Refs.~\cite{Kakizaki:2015wua,Hashino:2016rvx}.
In the following analysis, we take the free parameters to be $N$ and $\lambda_S$.
This is reasonable because two of the original four parameters $\lambda_\Phi, \lambda_S, \lambda_{\Phi S}$ and $N$ are
fixed by the observed Higgs mass $m_h$ and its vacuum expectation value $v$.

On the other hand, the tree-level potential for the $O(N)$ models without CCI is given by
\begin{align}
V_0 
= -\mu^2|\Phi|^2 + \mu_S^2|\vec{S}|^2 + \lambda_\Phi |\Phi|^4 + \frac{\lambda_S}{4}|\vec{S}|^4 + \frac{\lambda_{\Phi S}}{2}|\Phi|^2|\vec{S}|^2.
\label{eq:VwithoutCCI}
\end{align}
Compared to the above models, we have two additional parameters $\mu^2$ and $\mu_S^2$.
This makes the number of the free parameters four instead of two.
In the following analysis we take the free parameters to be $N$, $\lambda_S$, $m_S$ and $\mu_S^2$,
where $m_S$ is the singlet mass after the transition.
In this class of models $m_S$ can be translated into the deviation in the triple Higgs coupling 
$\Delta\lambda_{hhh}/\lambda_{hhh}^{\rm SM} \equiv \lambda_{hhh}/\lambda_{hhh}^{\rm SM}-1$.

%%%%%%%%%%%%%%%%%%%%%%%%%%%%%%%%%%%%%%%%%%%%%%%%%%
\subsubsection*{Analysis}
%%%%%%%%%%%%%%%%%%%%%%%%%%%%%%%%%%%%%%%%%%%%%%%%%%

We take the following benchmark points for the models with and without CCI, respectively:
\begin{itemize}
\item
$(N,\lambda_S) = (2, 0.1)$ : with CCI
\item
$(N,\lambda_S,m_S~{\rm [GeV]},\mu_S^2~[{\rm GeV}^2]) 
= 
(8, 0.1, 385, 0)$ and $(12, 0.1, 385, 0)$ : without CCI
\end{itemize}
Several comments are in order before moving on to the results.
For the former model we perform 2-parameter analysis with $N$ and $\lambda_S$.
In this analysis we regard $N$ as a continuous parameter to make Eq.~(\ref{eq:FabSeffSingle}) directly applicable.
For the latter model, we fix $\mu_S$ to the fiducial value and 
perform 3-parameter analysis with $N$, $\lambda_S$ and $m_S$.
When showing the final figures, we marginalize over $N$ and translate $m_S$ 
into the triple Higgs coupling $\Delta\lambda_{hhh}/\lambda_{hhh}^{\rm SM}$ as mentioned above.
Finally, for the both models, we assume fixed values for $v_w$ 
since it is generically hard to calculate the wall velocity in a given model.
Therefore, the GW spectrum used in our analysis reflects the model parameters 
only through $\alpha$, $\beta/H_*$ and $T_*$.
The dependence through $v_w$ should be taken into account in more realistic analyses.

\begin{figure}
\begin{center}
\includegraphics[width=\columnwidth]{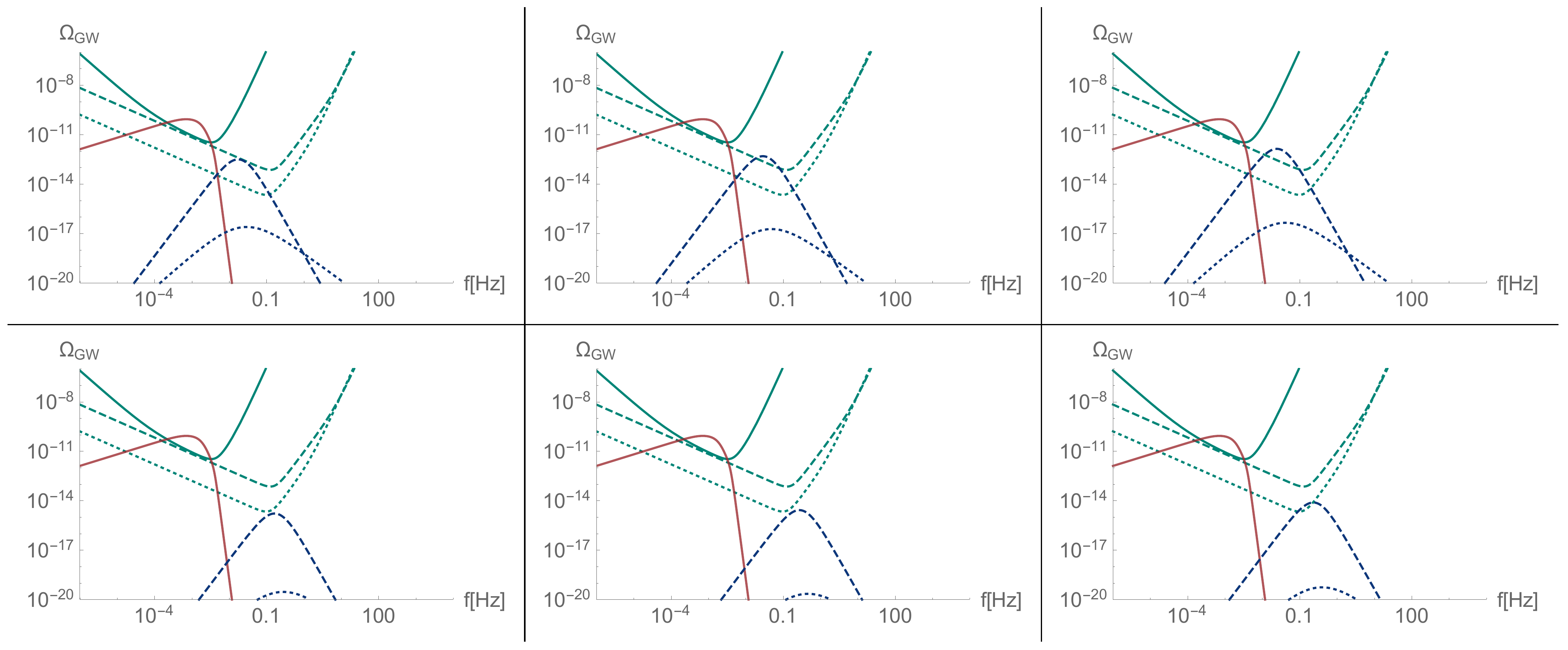} 
\caption{\small
Gravitational-wave spectra from sound waves (blue-dashed) and turbulence (blue-dotted) 
for the parameter point in Sec.~\ref{subsec:SingletON}.
The panels correspond to the model with CCI (left: $N = 2$) and without CCI (center: $N = 8$, right: $N = 12$), respectively.
The upper panels correspond to $v_w = 0.95$, while $v_w = 0.1$ for the lower panels.
}
\label{fig:S51fOmega}
\end{center}
\end{figure}

Fig.~\ref{fig:S51fOmega} is the GW spectrum realized in each model.
The corresponding electroweak phase transition parameters are
\begin{itemize}
\item
$(\alpha,\beta/H_*,T_*~{\rm [GeV]}) \simeq (0.080, 1000, 82)$ : with CCI
\item
$(\alpha,\beta/H_*,T_*~{\rm [GeV]}) \simeq (0.10, 1700, 83)$ and $(0.14, 1600, 77)$ : without CCI
\end{itemize}
In this figure we assumed $v_w = 0.95$ (top panels) and $v_w = 0.1$ (bottom panels).
We use the former value of $v_w$ for LISA and the latter for DECIGO and BBO, respectively. 

\begin{figure}
\begin{center}
\includegraphics[width=\columnwidth]{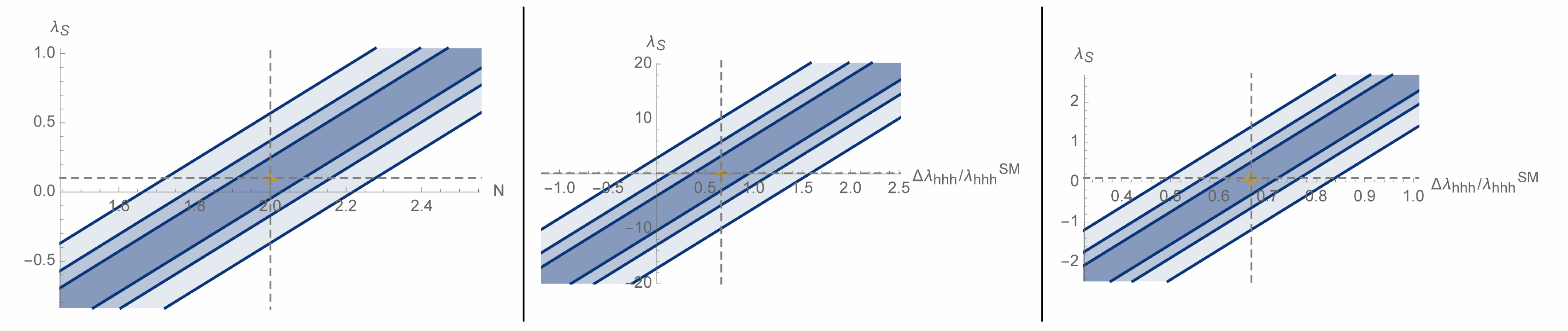} 
\caption{\small
(Left)
$1~\sigma$ contours for $O(N)$ singlet extensions of the SM with CCI for LISA.
(Center, Right)
$1~\sigma$ contours for $O(N)$ singlet extensions of the SM without CCI ($N = 8, 12$) for LISA.
}
\label{fig:S51D1}
\end{center}
\begin{center}
\includegraphics[width=\columnwidth]{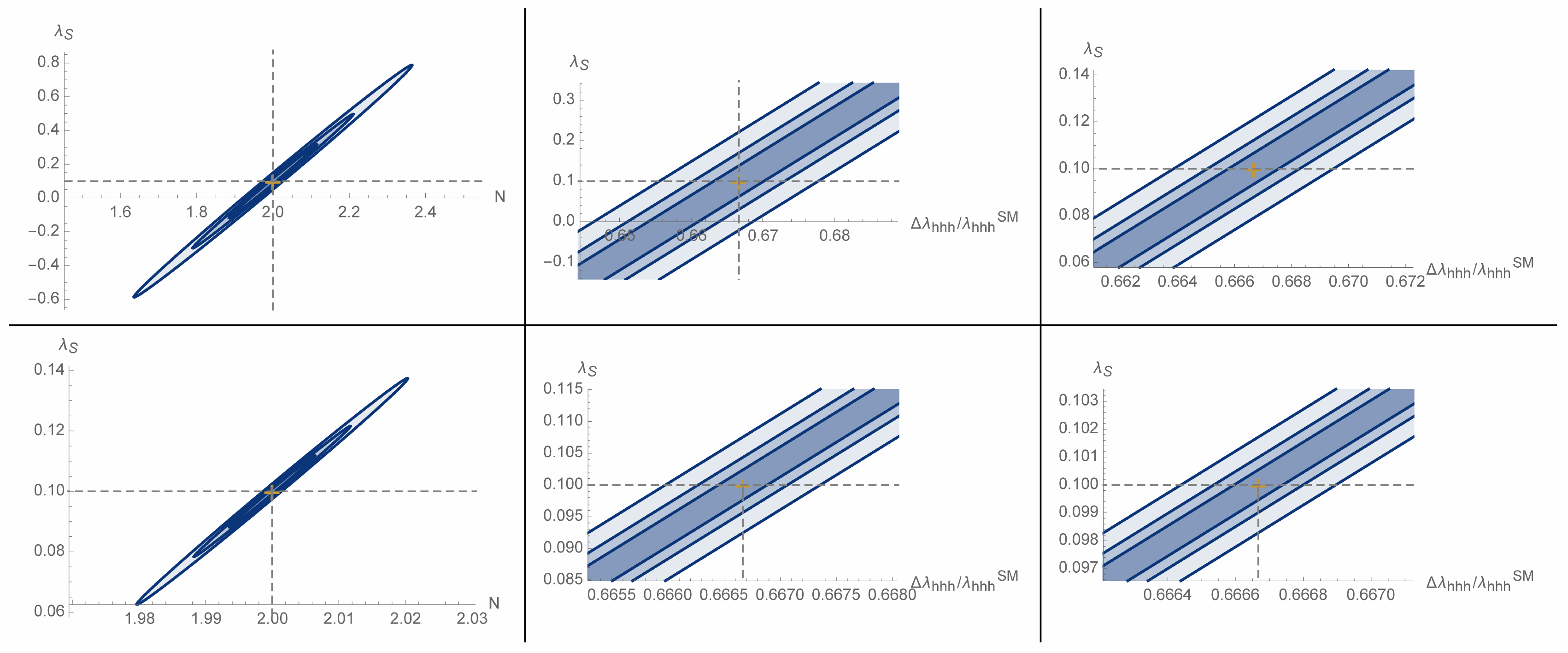} 
\caption{\small
(Left)
$1~\sigma$ contours for $O(N)$ singlet extensions of the SM with CCI
for DECIGO (top) and BBO (bottom).
(Center, Right)
$1~\sigma$ contours for $O(N)$ singlet extensions of the SM without CCI ($N = 8, 12$)
for DECIGO (top) and BBO (bottom).
}
\label{fig:S51D23}
\end{center}
\end{figure}

The results of a Fisher analysis is shown in Figs.~\ref{fig:S51D1}--\ref{fig:S51D23},
where Fig.~\ref{fig:S51D1} is for LISA while Fig.~\ref{fig:S51D23} is for DECIGO and BBO, respectively.
As seen in the left panel of Fig.~\ref{fig:S51D1}, 
LISA has the potential to contribute to narrow down the parameters for the model with CCI,
even if the spectrum is somewhat below the sensitivity curve.
This is because of the same reason as Sec.~\ref{sec:General}:
we have a factor of $(T_{\rm obs} \times f_{\rm typ}) \sim ({\mathcal O}(1)~{\rm years} \times 0.1~{\rm Hz})$ 
in Eq.~(\ref{eq:deltachi2Seff}) with $f_{\rm typ}$ being the typical frequency of the signal.
However, in passing it should be again noted that we have assumed the ideal case discussed in Ref.~\cite{Thrane:2013oya}.
Also, for the model without CCI, it is somewhat challenging to constrain the model parameters,
as seen in the right panel of the same figure.

On the other hand, 
DECIGO and BBO perform excellently to pin down the model parameters as shown in Fig.~\ref{fig:S51D23}.
For the model with CCI, both $N$ and $\lambda_S$ can be determined with a good precision even for $v_w$ as low as $0.1$.
Note that, after restricting $N$ to be an integer, the uncertainty in $\lambda_S$ becomes significantly small.
For the model without CCI as well,
though degeneracy appears because of the relatively large number of model parameters,
both detectors can contribute to determine a certain combination of the parameters.\footnote{
Even in parameter regions where only small amount of GWs are produced from the transition dynamics,
deformations in the primordial GW spectrum might also help pin down the model parameters:
see e.g. Refs.~\cite{Jinno:2011sw,Jinno:2013xqa}.
}

\begin{figure}
\begin{center}
\includegraphics[width=0.48\columnwidth]{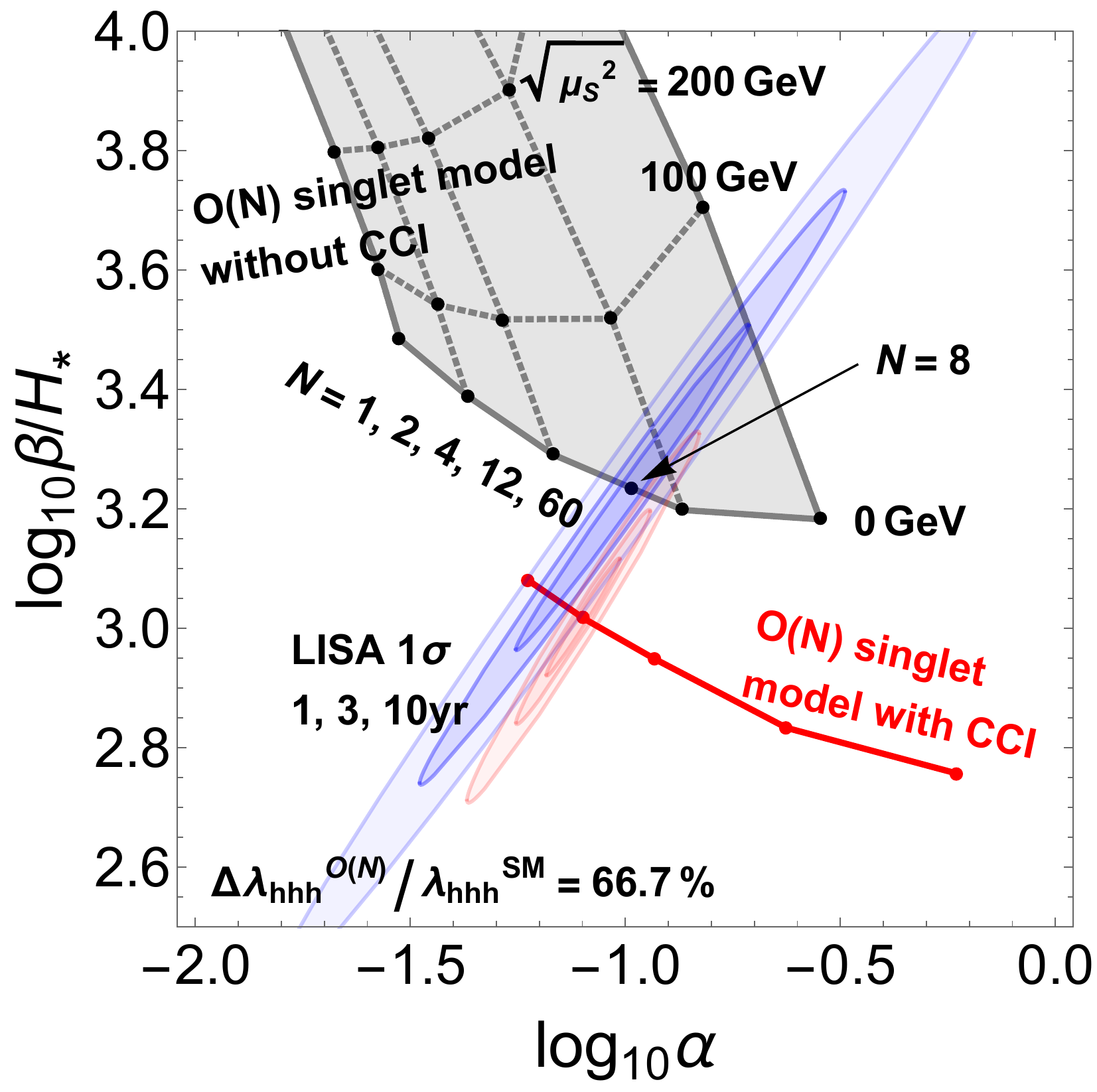} 
\includegraphics[width=0.48\columnwidth]{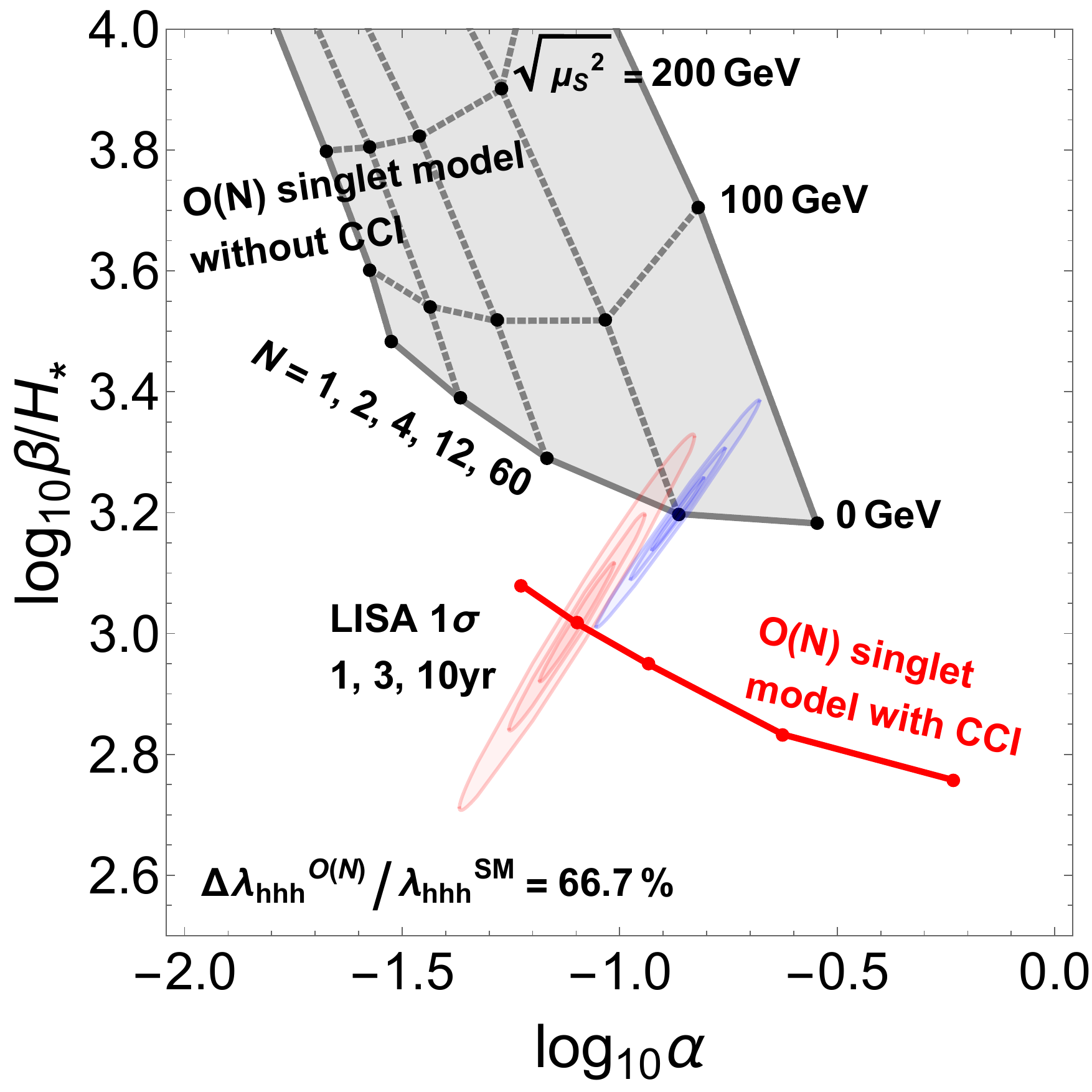} 
\caption{\small
LISA $1 \sigma$ contours for the $O(N)$ singlet models with and without CCI in $\alpha$-$\beta/H_*$ plane.
The red points correspond to $N=1, 2, 4, 12$ and $60$ for the $O(N)$ models with CCI from left to right,
while the gray points correspond to $N=1, 2, 4, 12$ and $60$ with $\mu_S^2 = 0$~GeV${}^2$
for the $O(N)$ models without CCI.
The fiducial values for the red contours correspond to the $O(N)$ models with CCI with $N=2$,
while the ones for the blue contours correspond to the $O(N)$ models without CCI with $N=8$ ($N=12$) and $\mu_S^2 = 0$~GeV${}^2$
in the left (right) panel.
Also, the three contours correspond to $T_{\rm obs}=1, 3$ and $10$ years. 
}
\label{fig:S51Synergy}
\end{center}
\end{figure}

Finally, we show in Fig.~\ref{fig:S51Synergy} the result of a Fisher analysis in $\alpha$-$\beta/H_*$ plane
in order to discuss the complementarity between collider and GW experiments.
In this figure, we perform a Fisher analysis on $\alpha$ and $\beta/H_*$ 
with $T_* \simeq 93$~GeV fixed at the fiducial value and also $v_w$ fixed to be $0.95$.
(In Appendix~\ref{app:Other} we show the results after marginalizing $T_*$.)
The resulting 1 $\sigma$ contours for LISA are shown as the red and blue lines for the $O(N)$ models with and without CCI, respectively.
The three contours correspond to $T_{\rm obs} = 1$, $3$ and $10$ years.
Both the left and right panels use the fiducial point $N = 2$ for the $O(N)$ models with CCI,
while they use $N = 8$ (left) and $N = 12$ (right) with $\sqrt{\mu_S^2}=0$~GeV for the $O(N)$ models without CCI.
In the latter models, we choose $m_S$ so that the triple Higgs coupling has the same value as the former:
$\Delta\lambda_{hhh}/\lambda_{hhh}^{\rm SM} = 66.7\%$.
As seen from the left panel, 
LISA may be able to distinguish $N=2$ with CCI from $N = 8$ without CCI for $T_{\rm obs} = 10$~years,
even when collider experiments cannot distinguish the two classes from the triple Higgs coupling.
On the other hand, as seen from the right panel,
LISA may differentiate $N = 2$ with CCI and $N = 12$ without CCI even in shorter observational periods.

%%%%%%%%%%%%%%%
%%%%%%%%%%%%%%%

%%%%%%%%%%%%%%%
%%%%%%%%%%%%%%%

%%%%%%%%%%%%%%%
%%%%%%%%%%%%%%%

%%%%%%%%%%%%%%%%%%%%%%%%%%%%%%%%%%%%%%%%%%%%%%%%%%
\subsection{Real Higgs singlet extension of the SM}
\label{subsec:RealHiggsSinglet}
%%%%%%%%%%%%%%%%%%%%%%%%%%%%%%%%%%%%%%%%%%%%%%%%%%

We next consider an extension of the SM with a real singlet scalar field $S$
which takes a nonzero expectation value at low temperatures.

%%%%%%%%%%%%%%%%%%%%%%%%%%%%%%%%%%%%%%%%%%%%%%%%%%
\subsubsection*{Model}
%%%%%%%%%%%%%%%%%%%%%%%%%%%%%%%%%%%%%%%%%%%%%%%%%%

The tree-level potential of this model is given by 
\begin{align}
V_0 = -\mu_\Phi^2|\Phi|^2 + \lambda_\Phi|\Phi|^4+\mu_{\Phi S}|\Phi|^2S +  \frac{\lambda_{\Phi S}}{2}|\Phi|^2S^2 + \mu_S^3 S + \frac{m_S^2}{2}S^2 + \frac{\mu_S'}{3}S^3  + \frac{\lambda_S}{4}S^4.
\label{eq:Vsinglet}
\end{align}
One of the eight parameters in the model can be removed by the redefinition of the singlet scalar field. 
In the following analysis, we take $\mu_S$ to be 0 by the field redefinition of $S$.

The electroweak phase transition in this model involves 
not only tree-level mixing effects between the scalar fields but also thermal loop effects.
If the transition is strongly first-order, the latter effects are imprinted in the resulting shape of the GWs.
Therefore we can test the model both by precision measurements of various Higgs boson couplings 
and by GW observations~\cite{Hashino:2016xoj}.

In the following analysis, we take the free parameters of the model to be 
the mass eigenvalue $m_H$ of the additional singlet scalar eigenstate $H$, 
the deviation $\kappa$ in the Higgs couplings to the gauge bosons and fermions, 
vacuum expectation value $v_S$ of singlet scalar field, $\mu_{\Phi S}$ and $\mu_S'$.

%%%%%%%%%%%%%%%%%%%%%%%%%%%%%%%%%%%%%%%%%%%%%%%%%%
\subsubsection*{Analysis}
%%%%%%%%%%%%%%%%%%%%%%%%%%%%%%%%%%%%%%%%%%%%%%%%%%

As explained above, there are five parameters in this model: $m_H$, $\kappa$, $v_S$, $\mu_{\Phi S}$ and $\mu_S'$.
However, in the following we fix $v_S$ and $\mu_S'$ at their fiducial values 
and include only the parameters related to the $H$ field ($m_H$, $\kappa$ and $\mu_{\Phi S}$) in the analysis.
This is because unremovable degeneracies appear if we take all the parameters as free.
In other words, observing the GW spectrum may not be enough to pin down the model parameters.
However, the detection of GWs indeed contributes to narrowing down the allowed parameter space, as we see below.
Also, for the wall velocity $v_w$, we fix its value as in Sec.~\ref{subsec:SingletON}.
We take the following benchmark point for all of LISA, DECIGO and BBO:
\begin{itemize}
\item
$(m_H~{\rm [GeV]}, \kappa, \mu_{\Phi S}~{\rm [GeV]}, v_S~{\rm [GeV]}, \mu_S'~[{\rm GeV}]) 
= (166, 0.96, -80, 90, -30)$.
\end{itemize}
The electroweak phase transition parameters for this fiducial point become
\begin{itemize}
\item
$(\alpha,\beta/H_*,T_*~{\rm [GeV]}) \simeq (0.085, 420, 93)$.
\end{itemize}

\begin{figure}
\begin{center}
\includegraphics[width=0.48\columnwidth]{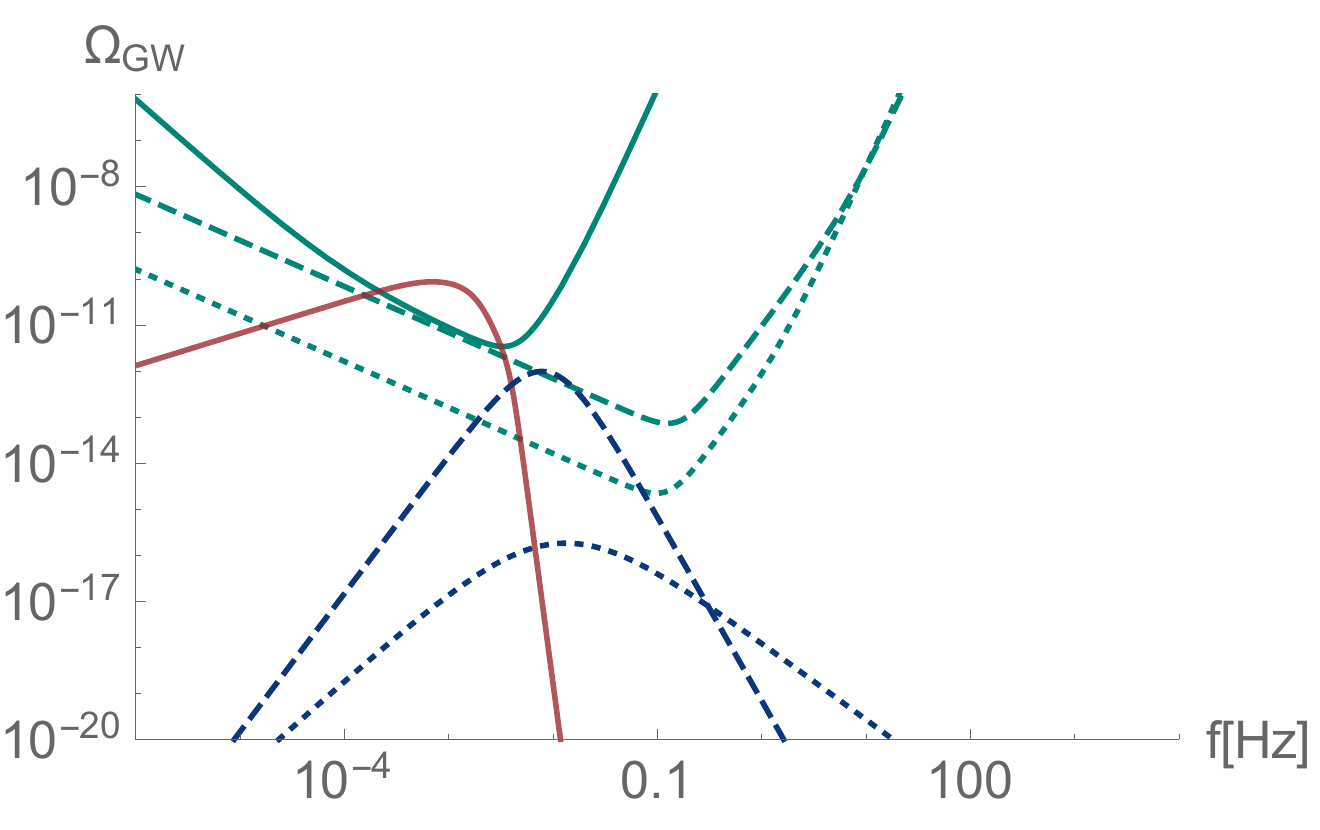} 
\includegraphics[width=0.48\columnwidth]{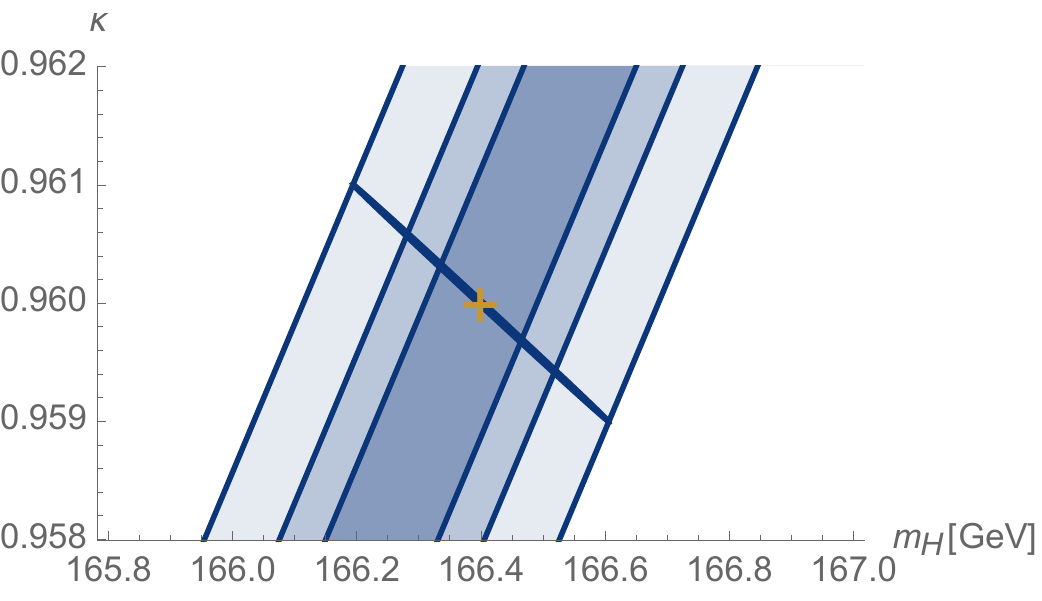} 
\caption{\small
(Left)
GW spectrum from sound waves (blue-dashed) and turbulence (blue-dotted) 
for the parameter point in Sec.~\ref{subsec:RealHiggsSinglet} with $v_w = 0.95$.
(Right)
$1~\sigma$ contours in the $m_H$--$\kappa$ plane for LISA.
The narrow contours correspond to fixed $\mu_{\Phi S}$,
while the wide contours correspond to marginalized $\mu_{\Phi S}$.
In drawing both contours, $v_S$ and $\mu_S'$ are fixed to be the fiducial values.
}
\label{fig:S52D1}
\end{center}
\begin{center}
\includegraphics[width=0.48\columnwidth]{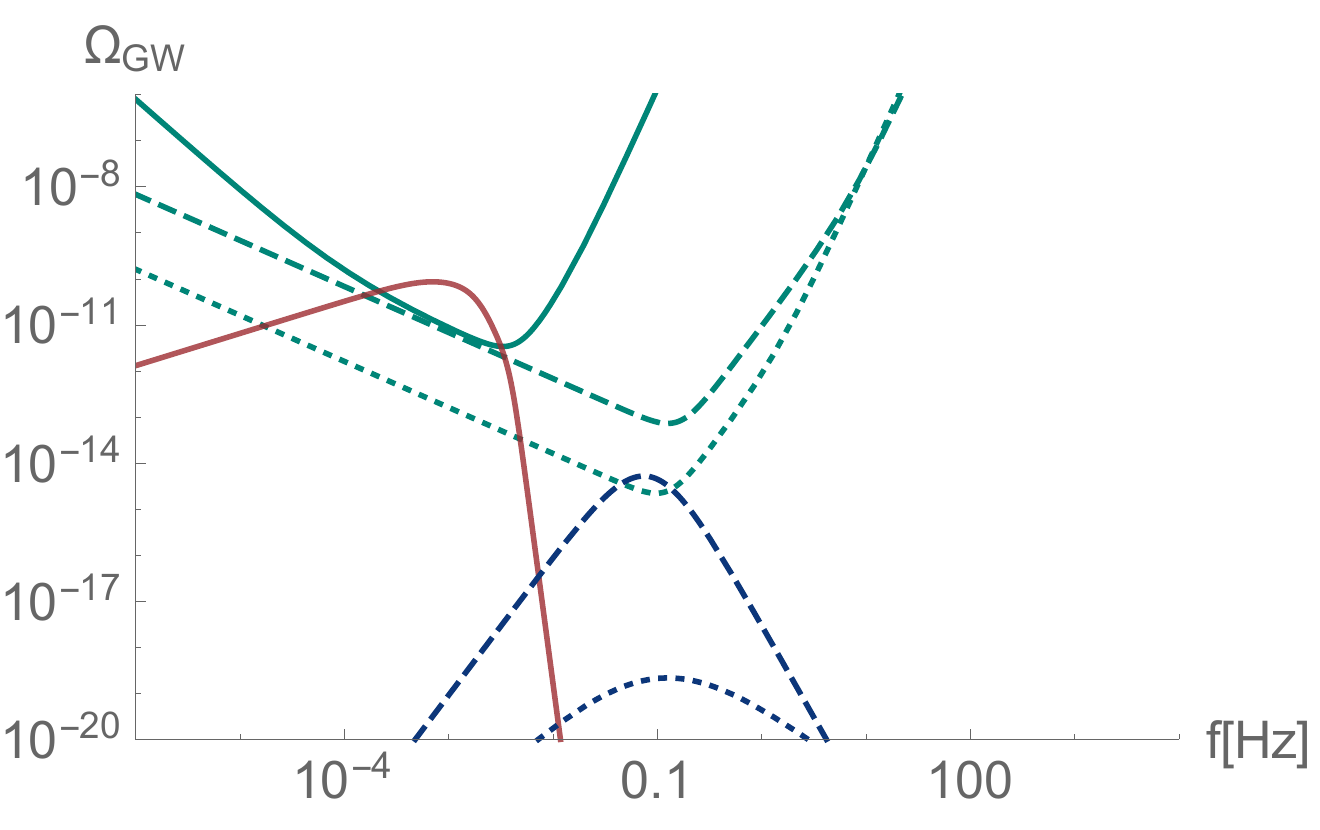} 
\includegraphics[width=\columnwidth]{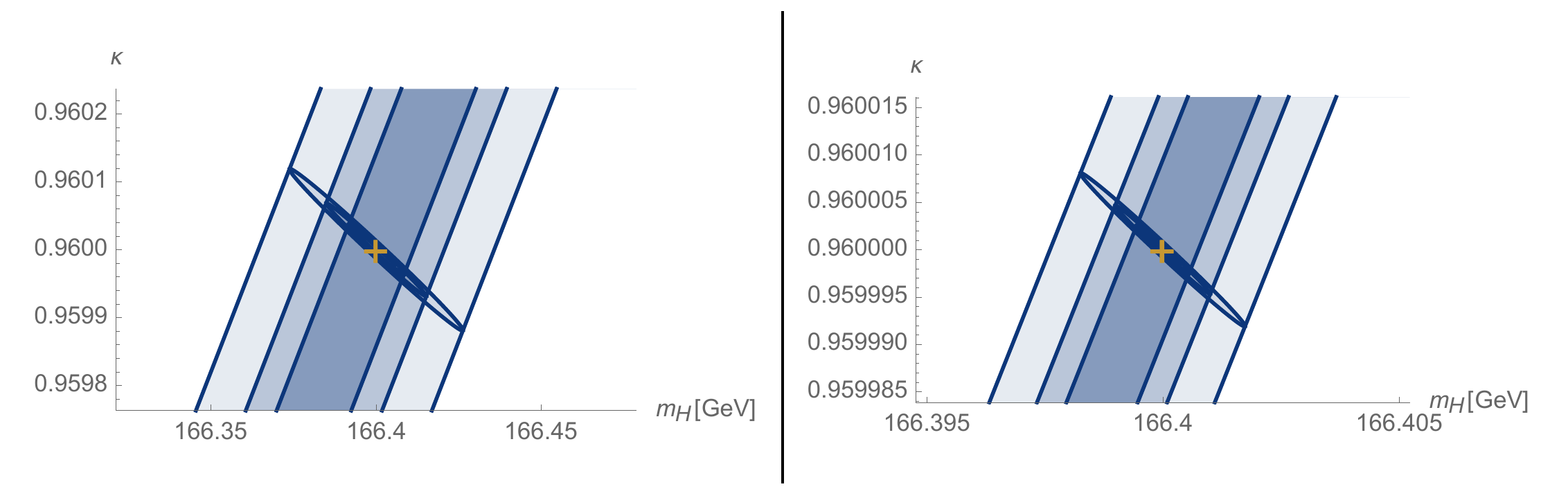} 
\caption{\small
(Top)
GW spectra from sound waves (blue-dashed) and turbulence (blue-dotted) 
for the parameter point in Sec.~\ref{subsec:RealHiggsSinglet} with $v_w = 0.1$.
(Bottom)
$1~\sigma$ contours in the $m_H$--$\kappa$ plane for DECIGO (left) and BBO (right).
Otherwise the same as the right panel of Fig.~\ref{fig:S52D1}.
}
\label{fig:S52D23}
\end{center}

\end{figure}

For LISA, we fix the wall velocity to be $v_w = 0.95$.
The GW spectrum realized at this parameter point is shown in the left panel of Fig.~\ref{fig:S52D1}.
The result of a Fisher analysis is shown in the right panel of the same figure.
The narrow contours are for fixed $\mu_{\Phi S}$
while the wide contours are obtained after marginalizing over $\mu_{\Phi S}$.
It is seen that LISA can contribute to constraining $m_H$ and $\kappa$ with a good accuracy.

For DECIGO and BBO, we fix the wall velocity to be $v_w = 0.1$.
The GW spectrum realized at this parameter point is shown in the top panel of Fig.~\ref{fig:S52D23}.
The results of a Fisher analysis is shown in the bottom panels of the same figure.
The narrow contours are the result for fixed $\mu_{\Phi S}$,
while the wide contours are the ones after marginalizing over $\mu_{\Phi S}$.
It is seen that, though the GW amplitude is much smaller than the previous parameter point,
DECIGO and BBO can perform excellently in constraining the model parameters.

\begin{figure}
\begin{center}
\includegraphics[width=0.6\columnwidth]{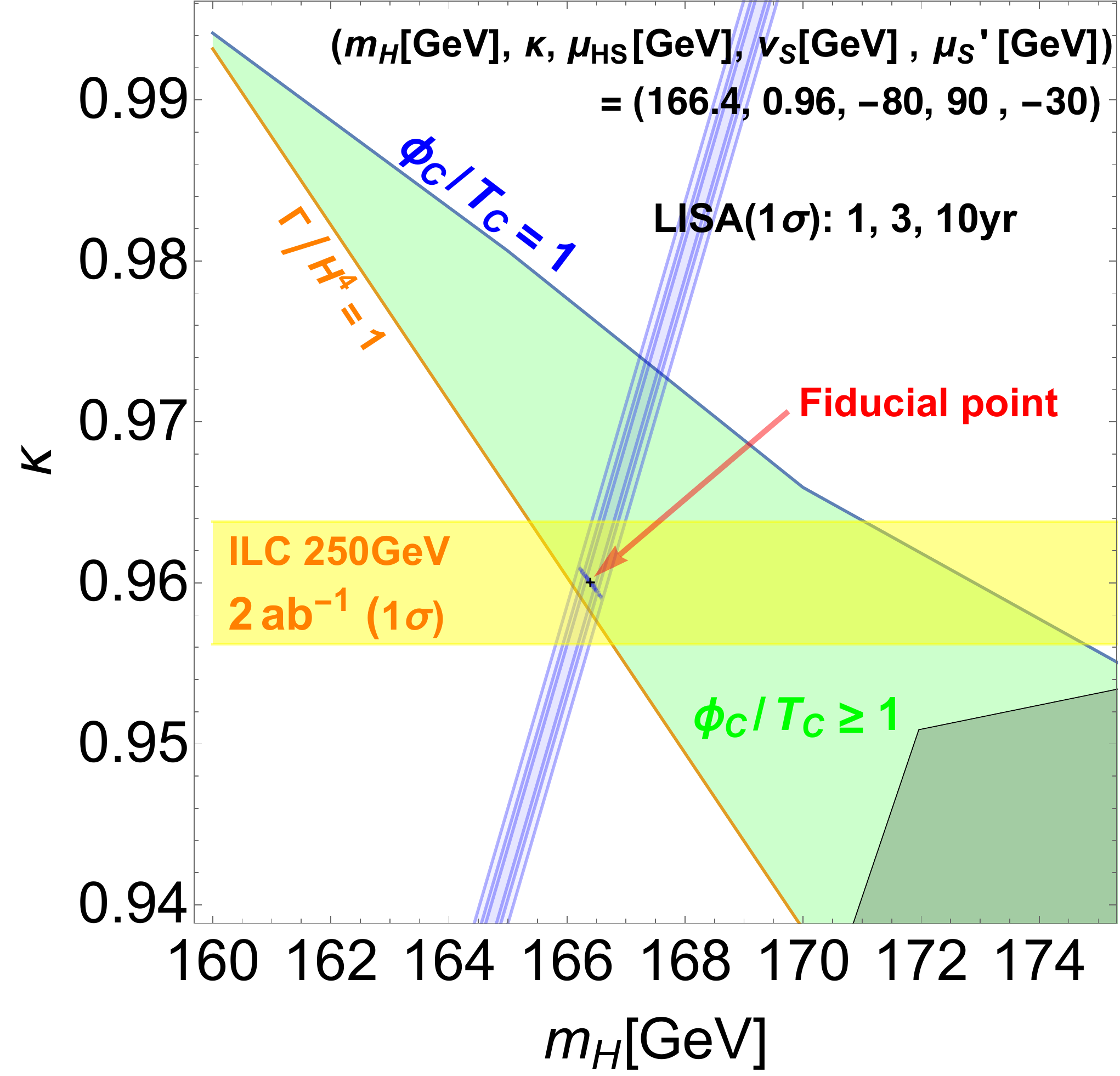} 
\caption{\small
(Blue) 
LISA $1 \sigma$ contours for the real Higgs singlet model with the fiducial values $m_H=166.4$~GeV and $\kappa=0.96$.  
Narrow and wide contours correspond to fixed and marginalized $\mu_{\Phi S}$, respectively.
The same as Fig.~\ref{fig:S52D1}.
(Green)
The region where the condition for a strongly first-order phase transition is satisfied.
(Yellow)
ILC $1 \sigma$ sensitivity region with $\sqrt{s} = 250$~GeV and ${\mathcal L} = 2$~ab$^{-1}$.
(Gray) The region excluded by the direct search for a heavy Higgs~\cite{Ilnicka:2018def}. 
}
\label{fig:S52rHSM}
\end{center}
\end{figure}

Finally we see the synergy between collider and GW experiments in Fig.~\ref{fig:S52rHSM}.
The condition for a strongly first-order electroweak phase transition is satisfied in the green-shaded region,
and the same result as the right panel of Fig.~\ref{fig:S52D1} is shown in blue.
The yellow region is the $1\sigma$ expected sensitivity of ILC 
with $\sqrt{s} = 250$ GeV and ${\cal L}=2$ ab${}^{-1}$~\cite{Fujii:2017vwa},
while the right-bottom shaded region is excluded by the current experimental data for the direct search for a heavy Higgs~\cite{Ilnicka:2018def}. 
It is seen that we can test the model both by future GW experiments and collider experiments. 

%%%%%%%%%%%%%%%
%%%%%%%%%%%%%%%

%%%%%%%%%%%%%%%
%%%%%%%%%%%%%%%

%%%%%%%%%%%%%%%%%%%%%%%%%%%%%%%%%%%%%%%%%%%%%%%%%%
\subsection{Classically conformal \texorpdfstring{$B-L$}{Lg} model}
\label{subsec:Clacon}
%%%%%%%%%%%%%%%%%%%%%%%%%%%%%%%%%%%%%%%%%%%%%%%%%%

We next consider the classically conformal $B-L$ model proposed in Refs.~\cite{Iso:2009ss,Iso:2009nw}
based on the argument on classical conformal theories~\cite{Bardeen:1995kv}.
It is known that, in nearly-conformal models, a large amount of GWs can be produced
due to huge supercooling and slow change of the nucleation rate 
(see e.g. Refs.~\cite{Konstandin:2011dr,Jinno:2016knw,Iso:2017uuu,Bai:2018vik,Bruggisser:2018mus,Bruggisser:2018mrt}).
Gravitational-wave production in the classically conformal $B-L$ model was studied in 
Ref.~\cite{Iso:2017uuu} and Ref.~\cite{Jinno:2016knw} for relatively small and $B-L$ gauge coupling, respectively.
In the following analysis we consider the former parameter region.

%%%%%%%%%%%%%%%%%%%%%%%%%%%%%%%%%%%%%%%%%%%%%%%%%%
\subsubsection*{Model}
%%%%%%%%%%%%%%%%%%%%%%%%%%%%%%%%%%%%%%%%%%%%%%%%%%

The relevant part of the model is the scalar sector, whose tree-level potential is given by
\begin{align}
V
&= 
\lambda_\Phi |\Phi|^4 + \lambda_X |X|^4 - \lambda_{\Phi X} |\Phi|^2|X|^2,
\end{align}
where only four-point couplings appear due to the assumption of the classical conformal symmetry.
Here $\Phi$ is the SM Higgs doublet and $X$ is the $B-L$ breaking scalar with $B-L$ charge $+2$.\footnote{
In this model, there are also right handed neutrinos. In this paper, we neglect their effects
assuming that their Yukawa couplings are small enough. 
}
The $B-L$ scalar field $X$ develops the vacuum expectation value 
$M \equiv \sqrt{2} \left< X \right>$ due to the running of the coupling $\lambda_X$. 
The mixing term $\lambda_{\Phi X}|\Phi|^2|X|^2$ generates the negative mass term for the SM Higgs
and electroweak symmetry breaking is realized at zero temperature.
We consider the parameter space where $M$ is relatively larger than the electroweak scale.
In such cases, the mixing coupling $\lambda_{\Phi X}$
becomes negligible and the potential for $X$ field is mainly determined by the $B-L$ gauge interaction. 

In this scenario, the phase transition in $X$ direction occurs
in the early Universe and produce large amount of GWs.
To understand this, first note that the finite-temperature effective potential for $X$ roughly 
consists of the thermal mass term and the energy-dependent quartic coupling:
\begin{align}
V_{\rm eff}
&\sim
\frac{g_{B-L}(T)^2}{2} T^2 \chi^2 + \frac{\lambda_X(T)}{4} \chi^4,
\end{align}
with $\chi = \sqrt{2}{\rm Re}[X]$ parametrizing the transition direction.
Also, both the $B-L$ gauge coupling $g_{B-L}$ and the quartic coupling $\lambda_X$
are understood as dependent on the typical energy scale of the system,
which is the temperature $T$ of the Universe.
For low enough temperature, the effective quartic coupling becomes negative
and the origin $X = 0$ becomes the false vacuum.
Then, the resulting tunneling rate can be written just by the combination of the couplings
because there is no scale other than the temperature $T$:
\begin{align}
\frac{S_3}{T}
&\sim
\frac{g_{B-L}(T)}{|\lambda_X(T)|}.
\end{align}
Note that
it is only logarithmically dependent on the temperature.
As a result,
the parameter $\beta/H_*=d(S_3/T)/d\ln T$ becomes relatively small 
and we expect large amount of GW production.

The number of the free parameters in this scenario is just two:
the vacuum expectation value of the $B-L$ breaking scalar $M \equiv \sqrt{2} \left< X \right>$
and the $B-L$ gauge coupling $g_{B-L}$ (or equivalently $\alpha_{B-L} = g_{B-L}^2/4\pi$) at scale $M$.
The allowed parameter space is shown in Figs.~\ref{fig:S53TstarTR} and \ref{fig:S53alphabeta}.
The regions shaded in red, green and yellow correspond to 
\begin{itemize}
\item
Red:
Landau pole develops below the Planck scale,
\item
Green:
Excluded by $Z'$ search (see Refs.~\cite{Okada:2016gsh}),
\item
Yellow:
Phase transition does not complete in sufficiently large regions (see Refs.~\cite{Jinno:2016knw,Iso:2017uuu}).
\end{itemize}
It is seen that a significant supercooling occurs in this model ($\alpha\gg1$).
In such cases the combustion mode of the walls is likely to be very strong detonation, where the wall velocity approaches almost unity.
(Note that even in this case most of the released energy is still carried by the fluid motion~\cite{Bodeker:2017cim}.)
Therefore, in the following analysis we fix $v_w = 1$.

Below we see that, for particle models with such a small number of free parameters,
the detection of GWs significantly contributes to pin down the model parameters.

%%%%%%%%%%%%%%%%%%%%%%%%%%%%%%%%%%%%%%%%%%%%%%%%%%
\subsubsection*{Analysis}
%%%%%%%%%%%%%%%%%%%%%%%%%%%%%%%%%%%%%%%%%%%%%%%%%%

\begin{figure}
\begin{center}
\includegraphics[width=\columnwidth]{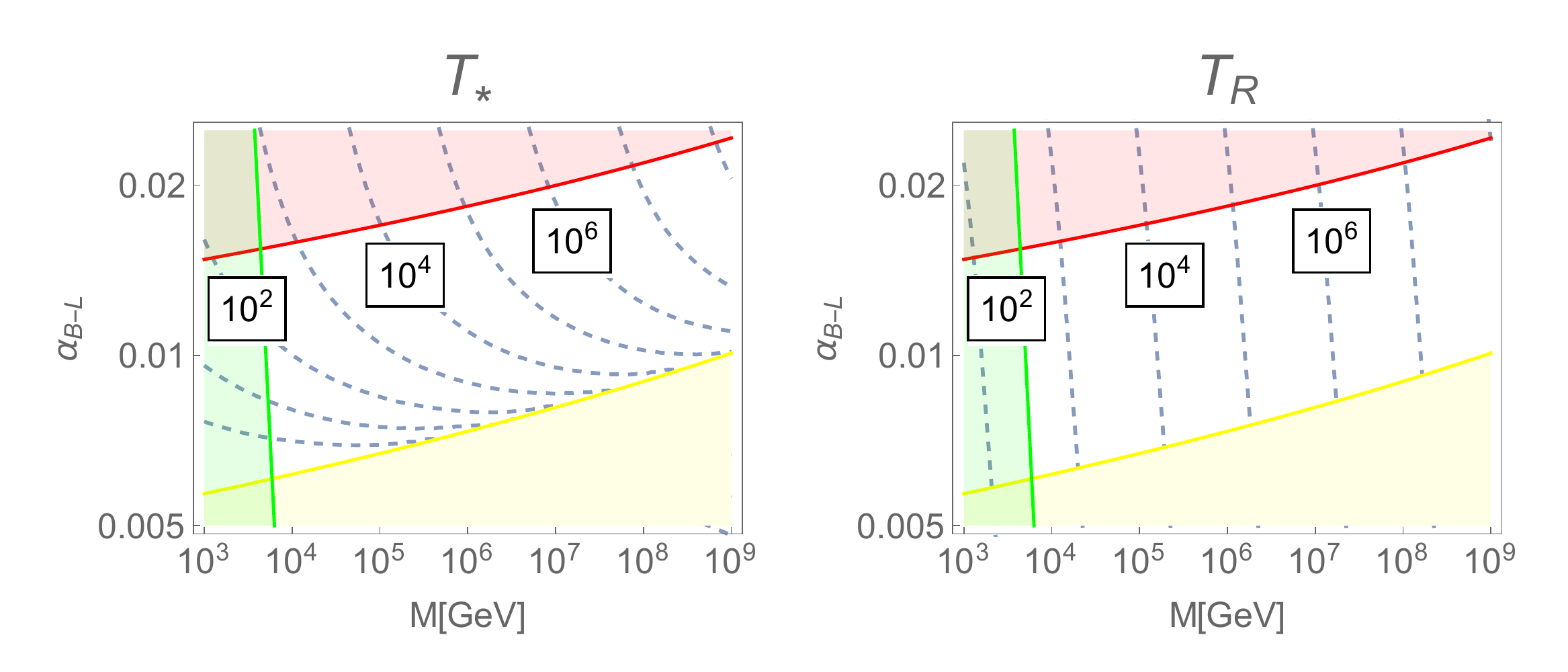} 
\caption{\small
Contours of the temperature just before the transition $T_*$ (left) 
and the temperature just after the transition $T_R$ (right)
in unit of GeV.
The regions shaded in red, green and yellow show 
the region which develops Landau pole below the Planck scale,
the region excluded by $Z'$ search,
and the region where the transition does not complete, respectively.
}
\label{fig:S53TstarTR}
\end{center}
\begin{center}
\includegraphics[width=\columnwidth]{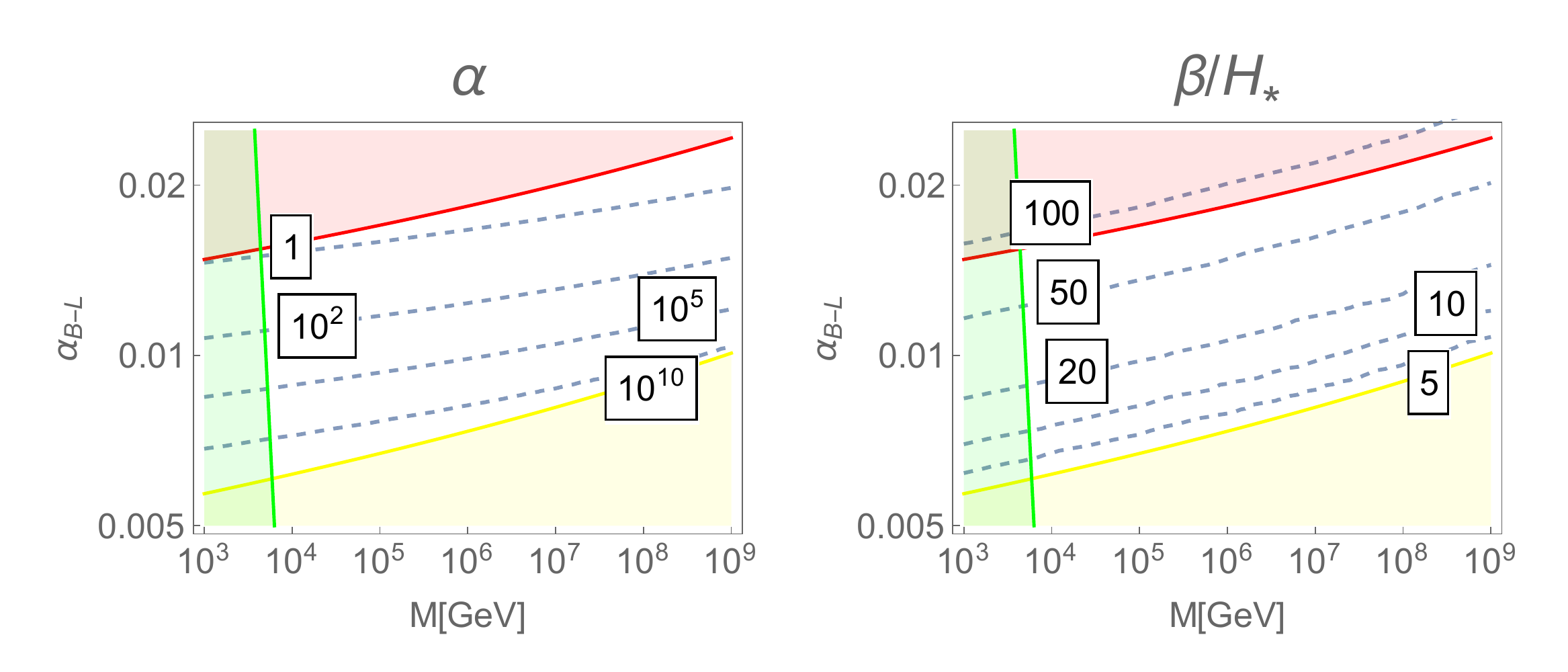} 
\caption{\small
Contours of the latent heat fraction $\alpha$ (left)
and bubble nucleation speed $\beta/H_*$ (right).
The regions shaded in red, green and yellow are the same as Fig.~\ref{fig:S53TstarTR}.
}
\label{fig:S53alphabeta}
\end{center}
\end{figure}

\begin{figure}
\begin{center}
\includegraphics[width=\columnwidth]{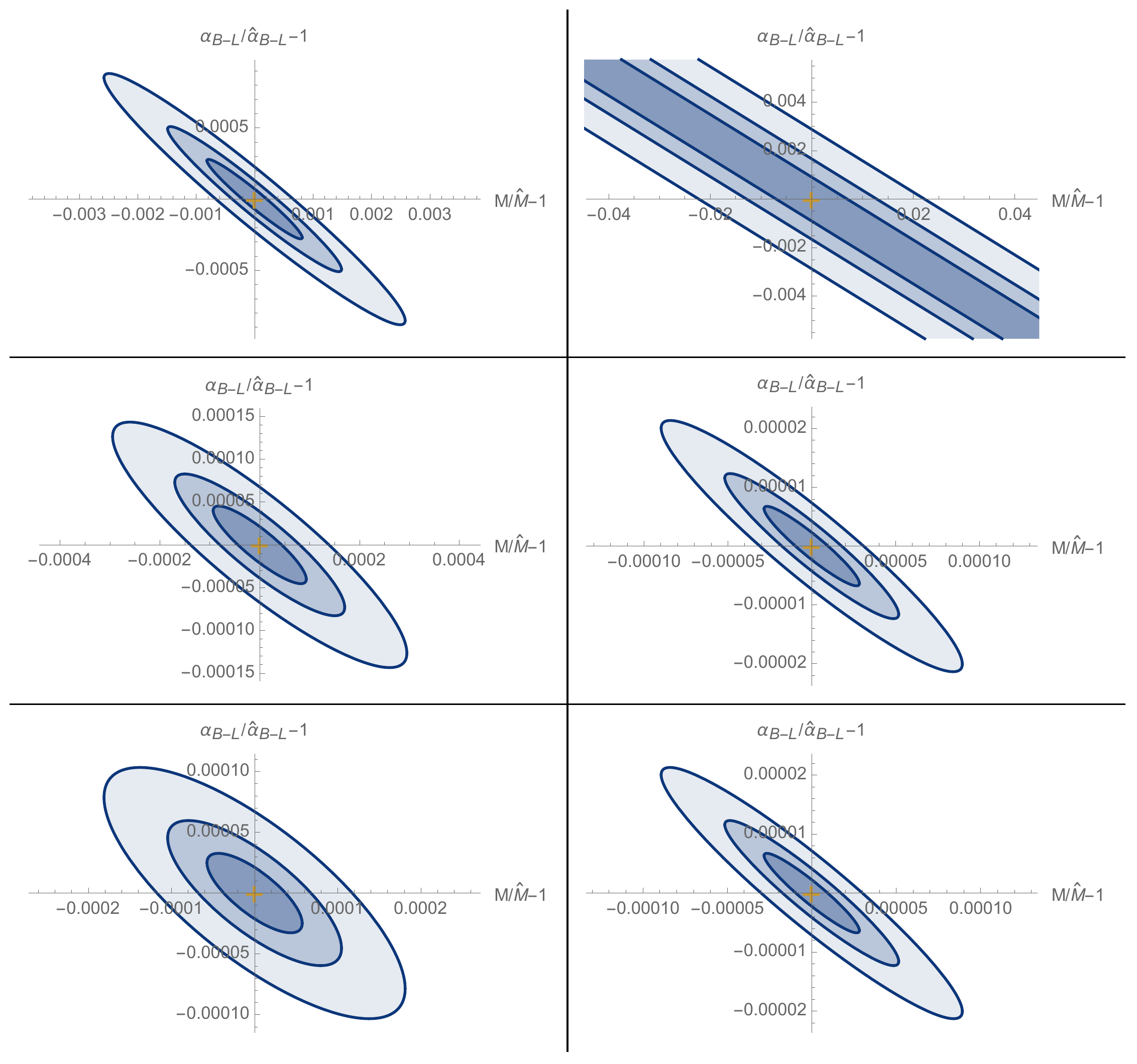} 
\caption{\small
$1~\sigma$ contours for Point 1 for LISA (top), DECIGO (middle) and BBO (bottom).
Left and right columns correspond to Point 1 and 2, respectively.
Three contours in each panel correspond to $T_{\rm obs} = 1$, $3$ and $10$ years.
}
\label{fig:S53P12}
\end{center}
\end{figure}

We first take two fiducial points:
\begin{itemize}
\item
Point 1:
$(M, \alpha_{B-L}) = (10^4~{\rm GeV}, 0.01)$
\item
Point 2:
$(M, \alpha_{B-L}) = (10^7~{\rm GeV}, 0.01)$
\end{itemize}
The GW spectra realized for these parameter points are shown in Fig.~\ref{fig:S53P12}.\footnote{
For the gauge dependence of the GW production in classically conformal models, see Ref.~\cite{Chiang:2017zbz}.
}
It is seen that the resulting GW amplitude is extremely large 
due to the behavior of $\alpha$ and $\beta/H_*$ shown in Fig.~\ref{fig:S53alphabeta}.
We show the result of a Fisher analysis for these parameter points in Fig.~\ref{fig:S53P12}.
It is seen that the model parameters are precisely determined except for Point 2 with LISA,
in which case the peak frequency of the GW spectrum becomes relatively high.
However, even in such a case, GW detection still contributes to constraining the parameters
as we see in the top-right panel of Fig.~\ref{fig:S53P12}.

\begin{figure}
\begin{center}
\includegraphics[width=\columnwidth]{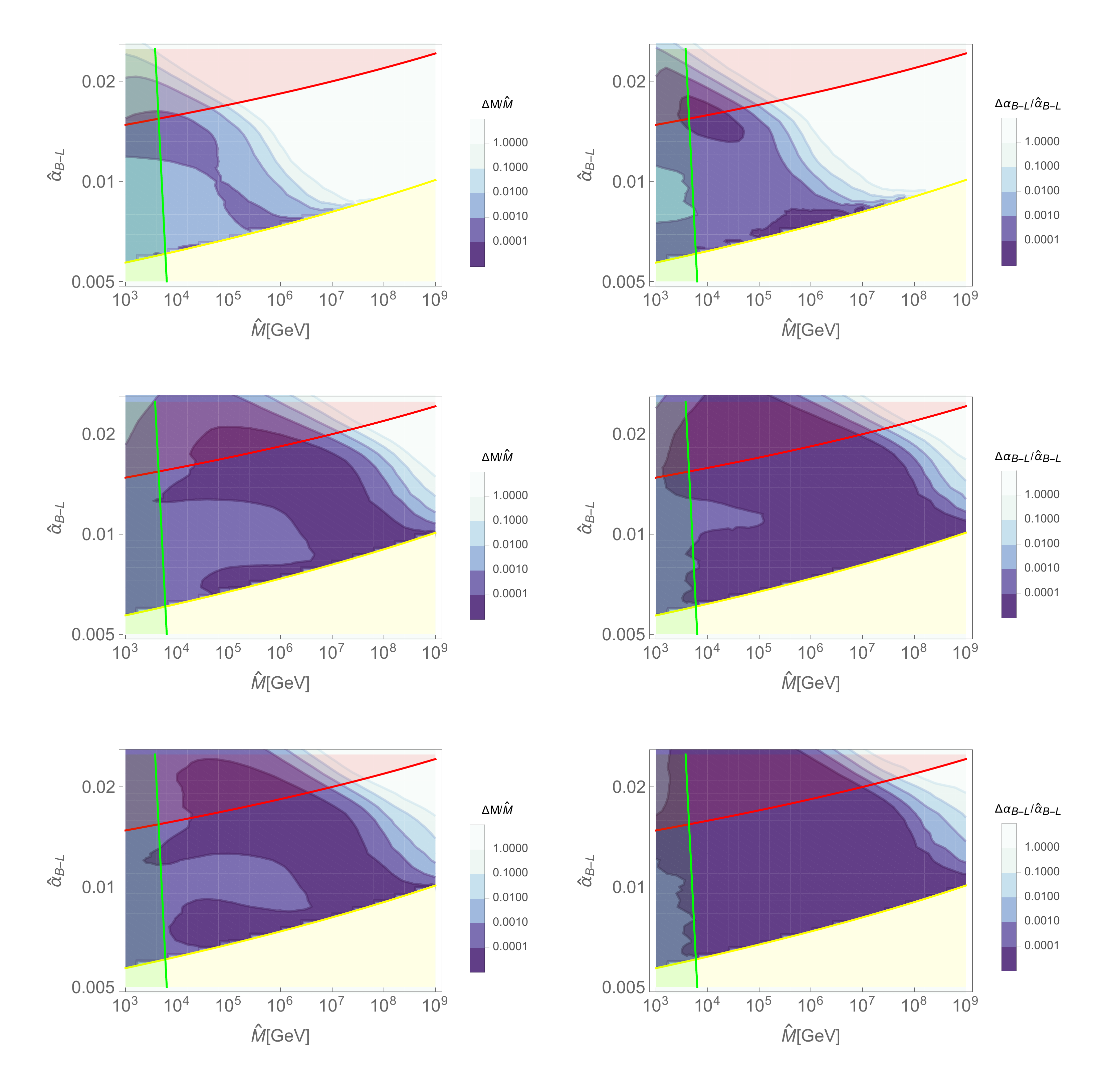} 
\caption{\small
1 $\sigma$ fractional error for $\Delta M / \hat{M}$ (left) and $\Delta \alpha_{B-L} / \hat{\alpha}_{B-L}$ (right) 
for the fiducial values $\hat{M}$ and $\hat{\alpha}_{B-L}$.
Each row corresponds to LISA (top), DECIGO (middle) and BBO (bottom).
Regions shaded in red, green and yellow are the same as Figs.~\ref{fig:S53TstarTR}--\ref{fig:S53alphabeta}.
}
\label{fig:S53DelMDelalpha}
\end{center}
\end{figure}

Next we show contour plots for $\Delta M / \hat{M}$ and $\Delta \alpha_{B-L} / \hat{\alpha}_{B-L}$ 
for different fiducial values for $\hat{M}$ and $\hat{\alpha}_{B-L}$ in Fig.~\ref{fig:S53DelMDelalpha}.
The three rows show LISA, DECIGO and BBO from top to bottom, respectively.
It is seen that LISA can pin down the model parameters in a wide range of the parameter space,
while such a parameter space becomes much wider for DECIGO and BBO.
Also note that $Z'$ searches can corner the parameter space from lower values of $M$,
which is favored from the viewpoint of naturalness.

%%%%%%%%%%%%%%%
%%%%%%%%%%%%%%%

%%%%%%%%%%%%%%%
\begin{figure}
\begin{center}
\includegraphics[width=\columnwidth]{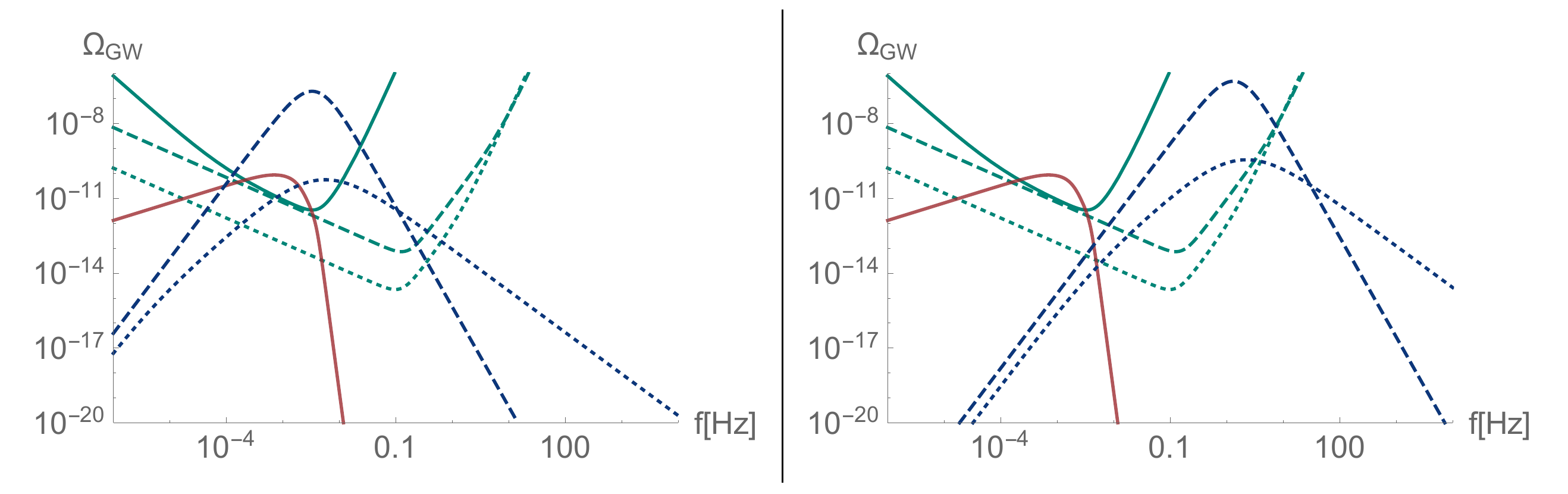} 
\caption{\small
(Left)
GW spectrum from sound waves (blue-dashed) and turbulence (blue-dotted) 
for Point 1 in Sec.~\ref{subsec:Clacon}.
(Right)
GW spectrum for Point 2. Otherwise the same as the left panel.
}
\label{fig:S5fOmega}
\end{center}
\end{figure}
%%%%%%%%%%%%%%%

%%%%%%%%%%%%%%%
%%%%%%%%%%%%%%%

%%%%%%%%%%%%%%%
%%%%%%%%%%%%%%%

\clearpage

%%%%%%%%%%%%%%%%%%%%%%%%%%%%%%%%%%%%%%%%%%%%%%%%%%
\section{Discussion and conclusions}
\label{sec:Conc}
\setcounter{equation}{0}
%%%%%%%%%%%%%%%%%%%%%%%%%%%%%%%%%%%%%%%%%%%%%%%%%%

In this paper, we investigated 
to what extent future space-based gravitational wave (GW) detectors such as LISA, DECIGO and BBO(-like one)
can contribute to pin down new physics beyond the standard model
through the detection of GWs from first-order phase transition.
In order to go beyond na\"ive comparison between the GW signal and the sensitivity curves
and quantify the attainable precisions, we adopted the method of Fisher analysis in this paper.

First, in Sec.~\ref{sec:General} (and Appendix~\ref{app:General}),
we studied the sensitivity of the detectors to the parameters which characterize a general peaky spectrum.
We parameterized the spectrum with the peak frequency, peak amplitude and spectral indices.
We performed a Fisher analysis to see the attainable uncertainties,
and it was found that, not only ultimately sensitive detectors such as DECIGO and BBO 
but also LISA, a relatively near-future detector, can significantly contribute to study GW spectral shapes.

Next, in Sec.~\ref{sec:Transition} (and Appendix~\ref{app:Transition}),
we performed a Fisher analysis on the parameters which characterize the phase transition
such as the the latent heat fraction $\alpha$, 
time dependence of the bubble nucleation rate $\beta/H_*$
and the transition temperature $T_*$.
We adopted a classification of GW sources in first-order phase transition in the literature
(i.e. bubble collisions, sound waves and turbulence) and used the spectral shapes provided there.
Though the classification and determination of the spectral shapes 
realized in first-order phase transition is a still ongoing hot topic 
(e.g. Refs.~\cite{Hindmarsh:2016lnk,Hindmarsh:2017gnf,Jinno:2017fby,Jinno:2017ixd,Konstandin:2017sat,
Cutting:2018tjt,Jackson:2018maa,Niksa:2018ofa}),
we illustrated in this paper the procedure to determine the transition parameters 
by detecting one or several spectral shapes which have different parameter dependences
by adopting expressions in the literature.
As a result, it was found that, though the detection of single spectral shape is indeed helpful,
the degeneracies in the parameters are resolved and their precise determination is possible 
if more than one spectral shapes are detected.

Finally, in Sec.~\ref{sec:Model} (and Appendix~\ref{app:Model}),
we studied how the detection of GWs contribute to the determination of fundamental model parameters.
This is possible because the transition parameters above are determined by the parameters of the particle physics model
which drive a first-order phase transition.
We illustrated this point by taking three examples:
(1) models with additional isospin singlet scalars
(2) a model with an extra real Higgs singlet,
and 
(3) a classically conformal $B-L$ model.
We found that the detection of the GW spectrum is indeed extremely powerful in pinning down the model parameters.
However, the exploration of new physics becomes truly interesting 
when GW searches are combined with collider experiments.
We also illustrated this point by taking the above three examples.
For the first models, the determination of the triple Higgs coupling helps to identify the existence of the classical scale invariance.
However, even if its value takes similar values both for the cases with and without the classical scale invariance, 
GW detection can distinguish the two (Fig.~\ref{fig:S51Synergy}).
In this sense, we can narrow down the model candidates and finally identify one by using two different experimental methods.
For the second model as well, colliders such as ILC give different constraints than GW observations (Fig.~\ref{fig:S52rHSM}).
For the last model, GW observations can corner the model with the help of $Z'$ searches (Fig.~\ref{fig:S53DelMDelalpha}).

To summarize, we found that future gravitational wave observations can play complementary roles to future collider experiments.
Fortunately, the LISA project and precision measurements of the Higgs boson couplings 
come around the same time in the future:
a great synergy between GW observations and collider experiments is awaiting us!

%%%%%%%%%%%%%%%%%%%%%%%%%%%%%%%%%%%%%%%%%%%%%%%%%%%%%%%
\section*{Acknowledgments}
%%%%%%%%%%%%%%%%%%%%%%%%%%%%%%%%%%%%%%%%%%%%%%%%%%%%%%%

RJ is grateful to M.~Hindmarsh and A.~J.~Long for helpful discussions.
The work of KH was supported by the Sasakawa Scientific Research Grant from The Japan Science Society.
The work of RJ was supported by IBS under the project code, IBS-R018-D1.
The work of MK was supported in part by Grant-in-Aid for Scientific Research on Innovative Areas, 
the Ministry of Education, Culture, Sports, Science and Technology, 
No.~16H01093 (MK), No.~17H05400 (MK).
The work of SK was supported in part by Grant-in-Aid for Scientific Research on Innovative Areas, 
the Ministry of Education, Culture, Sports, Science and Technology, 
No.~16H06492 and No.~18H04587, Grant H2020-MSCA-RISE-2014 No.~645722 (Non-Minimal Higgs).
The work of TT was supported by JSPS KAKENHI Grant Number 15K05084~(TT),  17H01131~(TT) 
and MEXT KAKENHI Grant Number 15H05888~(TT).
The work of MT was supported by JSPS Research Fellowships for Young Scientists.

\clearpage

\appendix

%%%%%%%%%%%%%%%%%%%%%%%%%%%%%%%%%%%%%%%%%%%%%%%%%%%%%%%
\section{Numerical results for other parameter sets}
\label{app:Other}
\setcounter{equation}{0}
%%%%%%%%%%%%%%%%%%%%%%%%%%%%%%%%%%%%%%%%%%%%%%%%%%%%%%%

In this appendix we show numerical results for different parameter sets
or different noise assumptions from the ones in the main text.

%%%%%%%%%%%%%%%%%%%%%%%%%%%%%%%%%%%%%%%%%%%%%%%%%%%%%%%
\subsection{Fisher analysis on general spectrum}
\label{app:General}
%%%%%%%%%%%%%%%%%%%%%%%%%%%%%%%%%%%%%%%%%%%%%%%%%%%%%%%

This subsection supplements the results in Sec.~\ref{sec:General}.
We see how they change depending on the spectral indices $(n_L,n_R)$ and the foregrounds.

%%%%%%%%%%%%%%%%%%%%%%%%%%%%%%%%%%%%%%%%%%%%%%%%%%%%%%%
\subsubsection*{\texorpdfstring{Case 1: $(n_L,n_R) = (3,-4)$}{Lg}}
%%%%%%%%%%%%%%%%%%%%%%%%%%%%%%%%%%%%%%%%%%%%%%%%%%%%%%%

As mentioned in Sec.~\ref{subsec:SeffFG},
unresolvable foregrounds from neutron stars and black holes can have significant effects on our results. 
Below we show how the results change if we include $S_{\rm NSBH}$ in Eq.~(\ref{eq:SNSBH}) in our analysis.

\begin{figure}
\begin{center}
\includegraphics[width=\columnwidth]{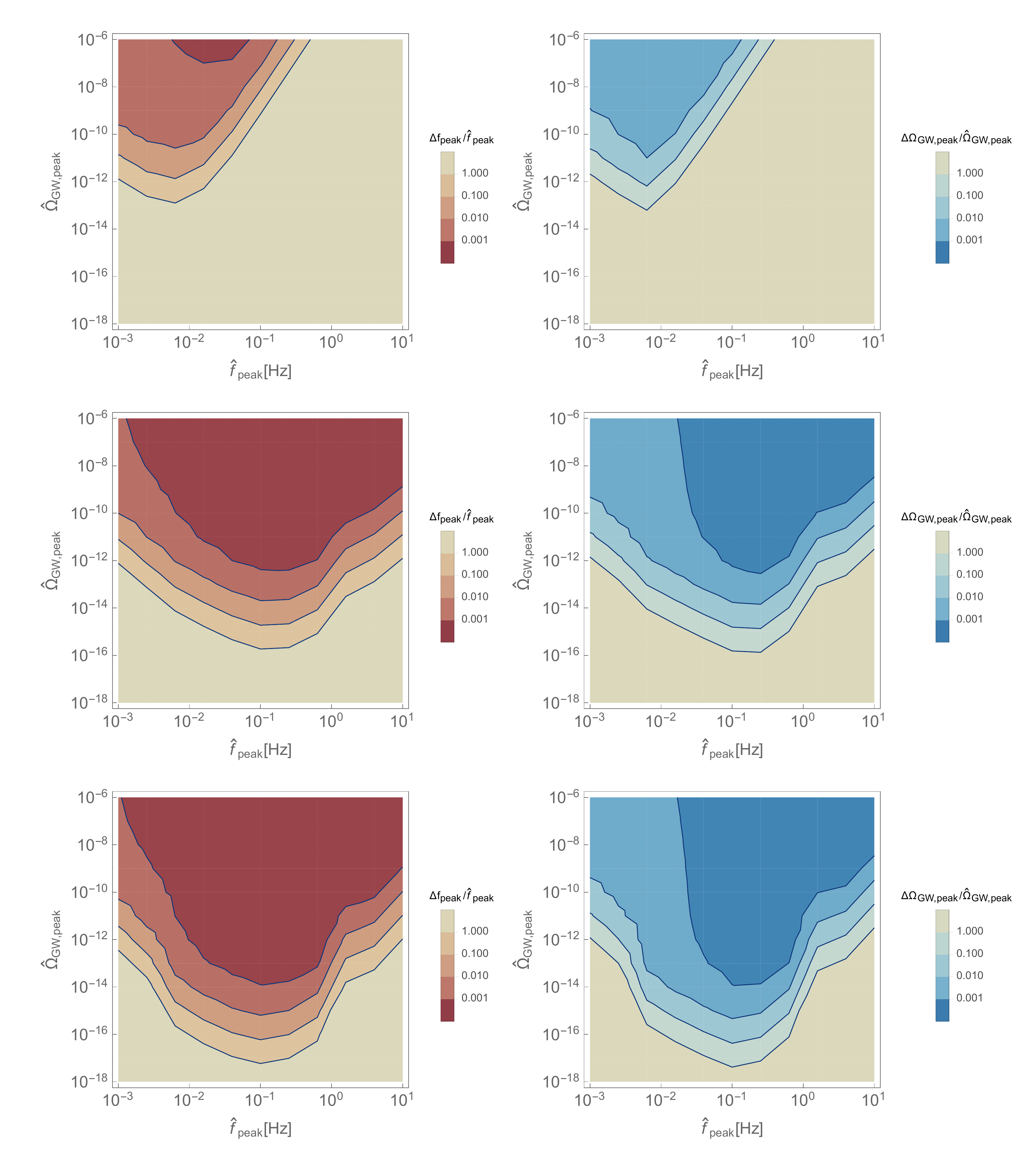} 
\caption{\small
$1~\sigma$ fractional error $\Delta f_{\rm peak}/\hat{f}_{\rm peak}$ (left) 
and $\Delta \Omega_{\rm GW,peak}/\hat{\Omega}_{\rm GW,peak}$ (right)
for the fiducial values $\hat{f}_{\rm peak}$ and $\hat{\Omega}_{\rm GW, peak}$
for LISA (top), DECIGO (middle) and BBO (bottom).
The spectral slopes and foreground are taken to be $(n_L,n_R) = (3,-4)$ and $S_{\rm WD} + S_{\rm NSBH}$.
}
\label{fig:A11DelfDelOmega}
\end{center}
\end{figure}

Fig.~\ref{fig:A11DelfDelOmega} is the result of a Fisher analysis for $(n_L,n_R) = (3,-4)$ 
and $S_h \supset S_{\rm WD} + S_{\rm NSBH}$.
This figure corresponds to Fig.~\ref{fig:DelfDelOmega}.
It is seen that for the frequency $1~{\rm Hz} < f < 10^3~{\rm Hz}$ the precision becomes worse,
but still there are possibilities of parameter determination for $\Omega_{\rm GW,peak} \gtrsim 10^{-12}$.

%%%%%%%%%%%%%%%
%%%%%%%%%%%%%%%

%%%%%%%%%%%%%%%%%%%%%%%%%%%%%%%%%%%%%%%%%%%%%%%%%%%%%%%
\subsubsection*{\texorpdfstring{Case 2: $(n_L,n_R) = (1,-3)$}{Lg}}
%%%%%%%%%%%%%%%%%%%%%%%%%%%%%%%%%%%%%%%%%%%%%%%%%%%%%%%

\begin{figure}
\begin{center}
\includegraphics[width=0.6\columnwidth]{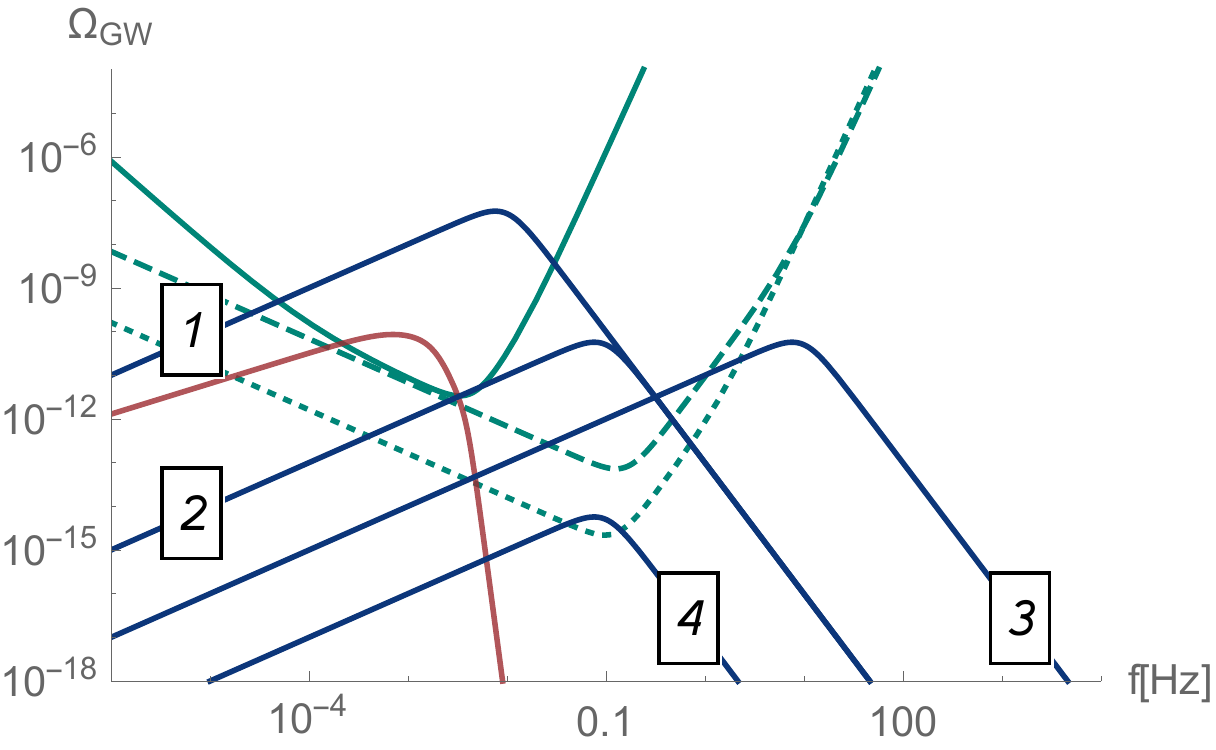} 
\caption{\small
Sensitivity curves for LISA~(green-solid), DECIGO~(green-dashed) and BBO~(green-dotted).
Blue curves correspond to the sample points 1-4 in Sec.~\ref{sec:General} in the main text.
Red lines show the contribution from compact white dwarf binaries $S_{\rm WD}$.
}
\label{fig:A11fOmegaGeneral}
\end{center}
\end{figure}

We next change the spectral indices in Eq.~(\ref{eq:OmegaGeneral}) to $(n_L,n_R) = (1,-3)$,
and see how the result changes.
In fact, these spectral indices are suggested in the study of GW production from thin bubbles~\cite{Jinno:2017fby,Konstandin:2017sat}.
First we show the sensitivity curves and GW signals in Fig.~\ref{fig:A11fOmegaGeneral}. 
The signals now have broader peaks compared to Fig.~\ref{fig:fOmegaGeneral}.

\begin{figure}
\begin{center}
\includegraphics[width=\columnwidth]{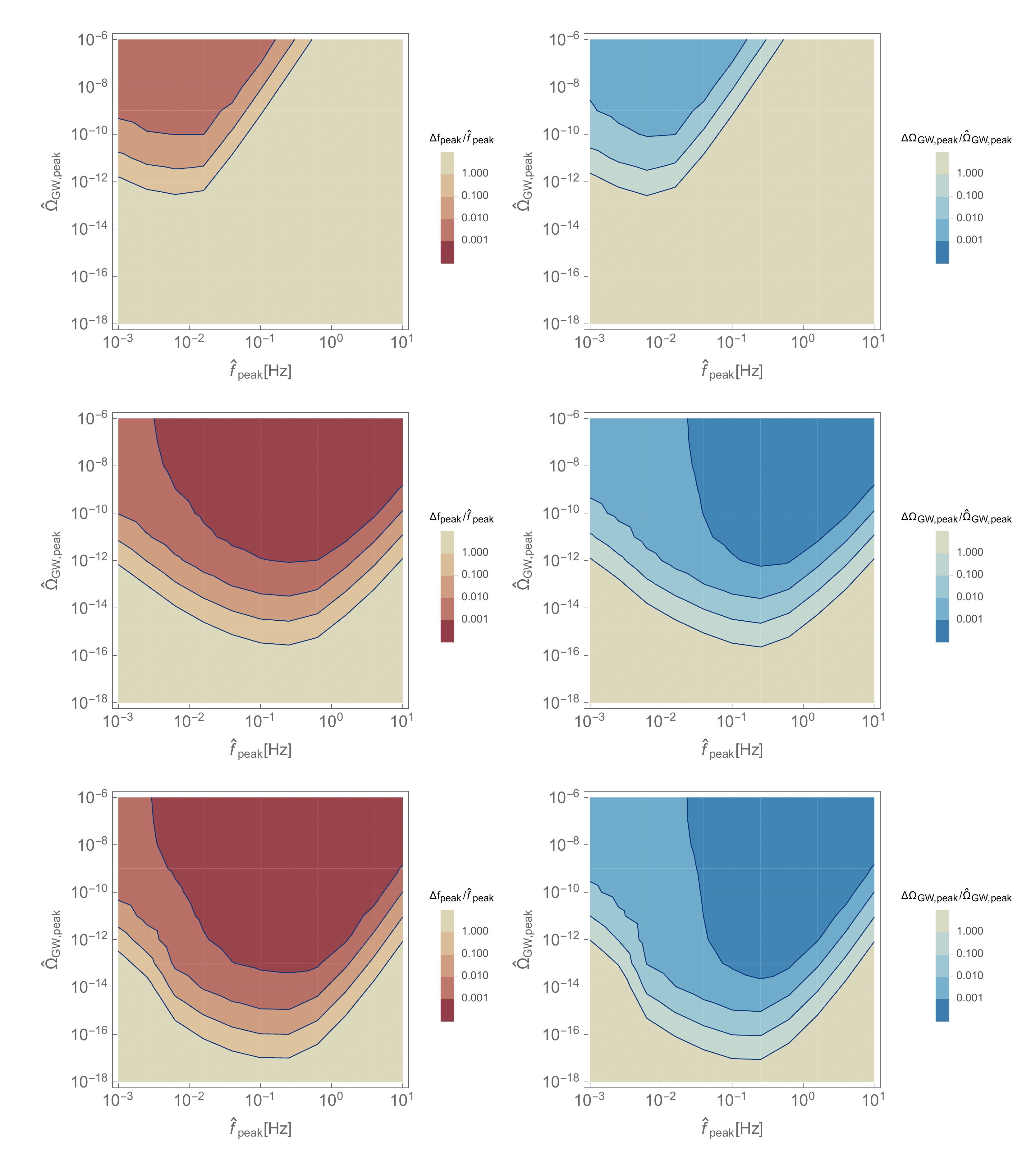} 
\caption{\small
$1~\sigma$ fractional error $\Delta f_{\rm peak}/\hat{f}_{\rm peak}$ (left) 
and $\Delta \Omega_{\rm GW,peak}/\hat{\Omega}_{\rm GW,peak}$ (right)
for the fiducial values $\hat{f}_{\rm peak}$ and $\hat{\Omega}_{\rm GW, peak}$
for LISA (top), DECIGO (middle) and BBO (bottom).
The spectral slopes and foreground are taken to be $(n_L,n_R) = (1,-3)$ and $S_{\rm WD}$.
}
\label{fig:A121DelfDelOmega}
\end{center}
\end{figure}

\begin{figure}
\begin{center}
\includegraphics[width=\columnwidth]{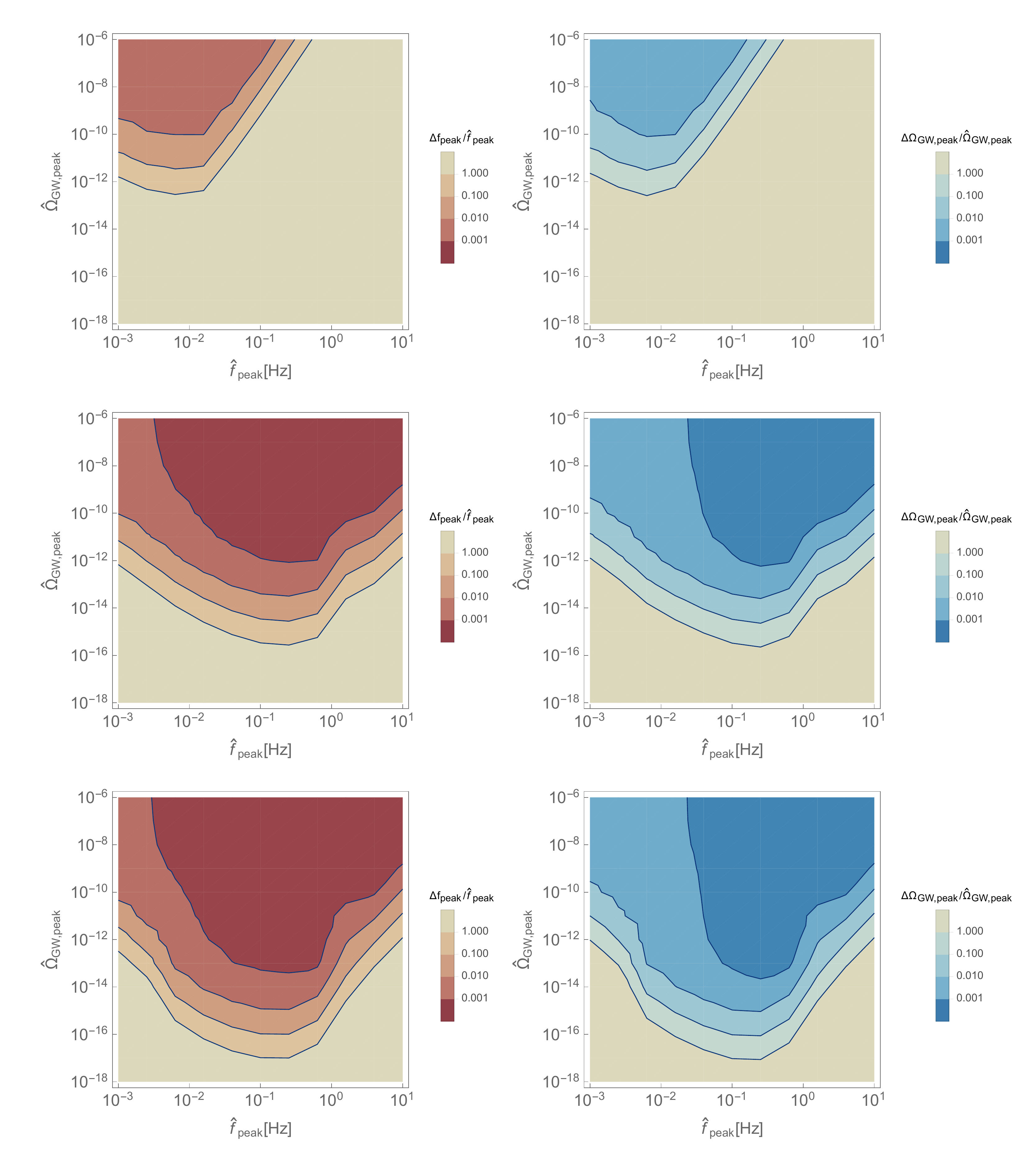} 
\caption{\small
$1~\sigma$ fractional error $\Delta f_{\rm peak}/\hat{f}_{\rm peak}$ (left) 
and $\Delta \Omega_{\rm GW,peak}/\hat{\Omega}_{\rm GW,peak}$ (right)
for the fiducial values $\hat{f}_{\rm peak}$ and $\hat{\Omega}_{\rm GW, peak}$
for LISA (top), DECIGO (middle) and BBO (bottom).
The spectral slopes and foreground are taken to be $(n_L,n_R) = (1,-3)$ and $S_{\rm WD} + S_{\rm NSBH}$.
}
\label{fig:A122DelfDelOmega}
\end{center}
\end{figure}

We show the results of a Fisher analysis in Figs.~\ref{fig:A121DelfDelOmega}--\ref{fig:A122DelfDelOmega}.
Fig.~\ref{fig:A121DelfDelOmega} corresponds to the case with $S_h \supset S_{\rm WD}$ only,
while Fig.~\ref{fig:A122DelfDelOmega} correspond to the one with $S_h \supset S_{\rm WD} + S_{\rm NSBH}$.

%%%%%%%%%%%%%%%
%%%%%%%%%%%%%%%

%%%%%%%%%%%%%%%
%%%%%%%%%%%%%%%

%%%%%%%%%%%%%%%
%%%%%%%%%%%%%%%

\clearpage

%%%%%%%%%%%%%%%%%%%%%%%%%%%%%%%%%%%%%%%%%%%%%%%%%%%%%%%
\subsection{Fisher analysis on transition parameters}
\label{app:Transition}
%%%%%%%%%%%%%%%%%%%%%%%%%%%%%%%%%%%%%%%%%%%%%%%%%%%%%%%

This subsection supplements the results in Sec.~\ref{sec:Transition}.
We see how they change depending on the wall velocity $v_w$ and the foregrounds.

%%%%%%%%%%%%%%%%%%%%%%%%%%%%%%%%%%%%%%%%%%%%%%%%%%%%%%%
\subsubsection{\texorpdfstring{$v_w = 1$}{Lg}}
%%%%%%%%%%%%%%%%%%%%%%%%%%%%%%%%%%%%%%%%%%%%%%%%%%%%%%%

\begin{figure}
\begin{center}
\includegraphics[width=\columnwidth]{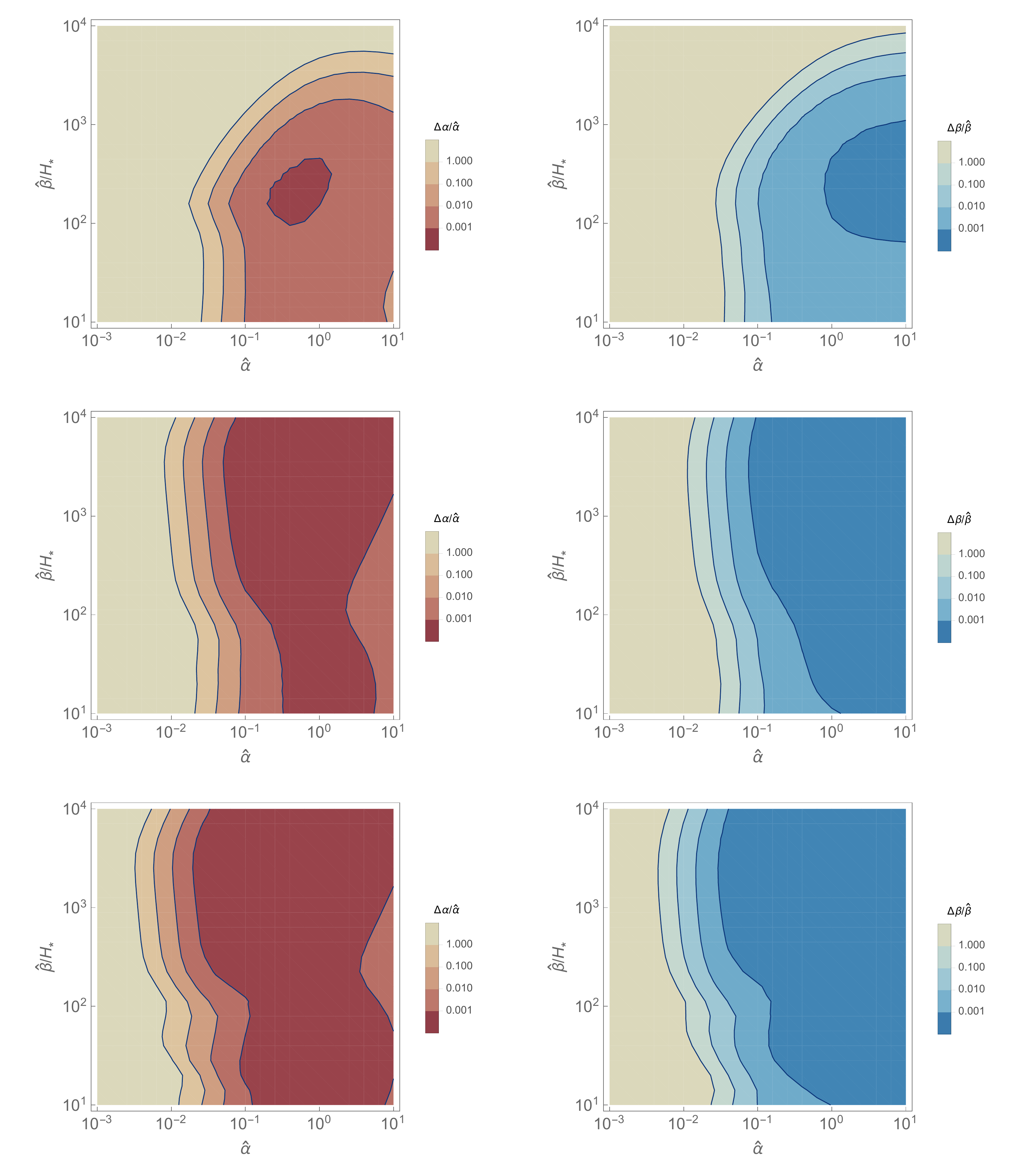} 
\caption{\small
1 $\sigma$ fractional error $\Delta \alpha / \hat{\alpha}$ (left) and $\Delta \beta / \hat{\beta}$ (right)
for the fiducial values $\hat{\alpha}$ and $\hat{\beta}$ for 2-parameter analysis.
Each row corresponds to LISA (top), DECIGO (middle) and BBO (bottom).
The wall velocity and foreground are taken to be $v_w = 1$ and $S_{\rm WD} + S_{\rm NSBH}$.
}
\label{fig:A21N2DeltaalphaDeltabeta}
\end{center}
\end{figure}

\begin{figure}
\begin{center}
\includegraphics[width=\columnwidth]{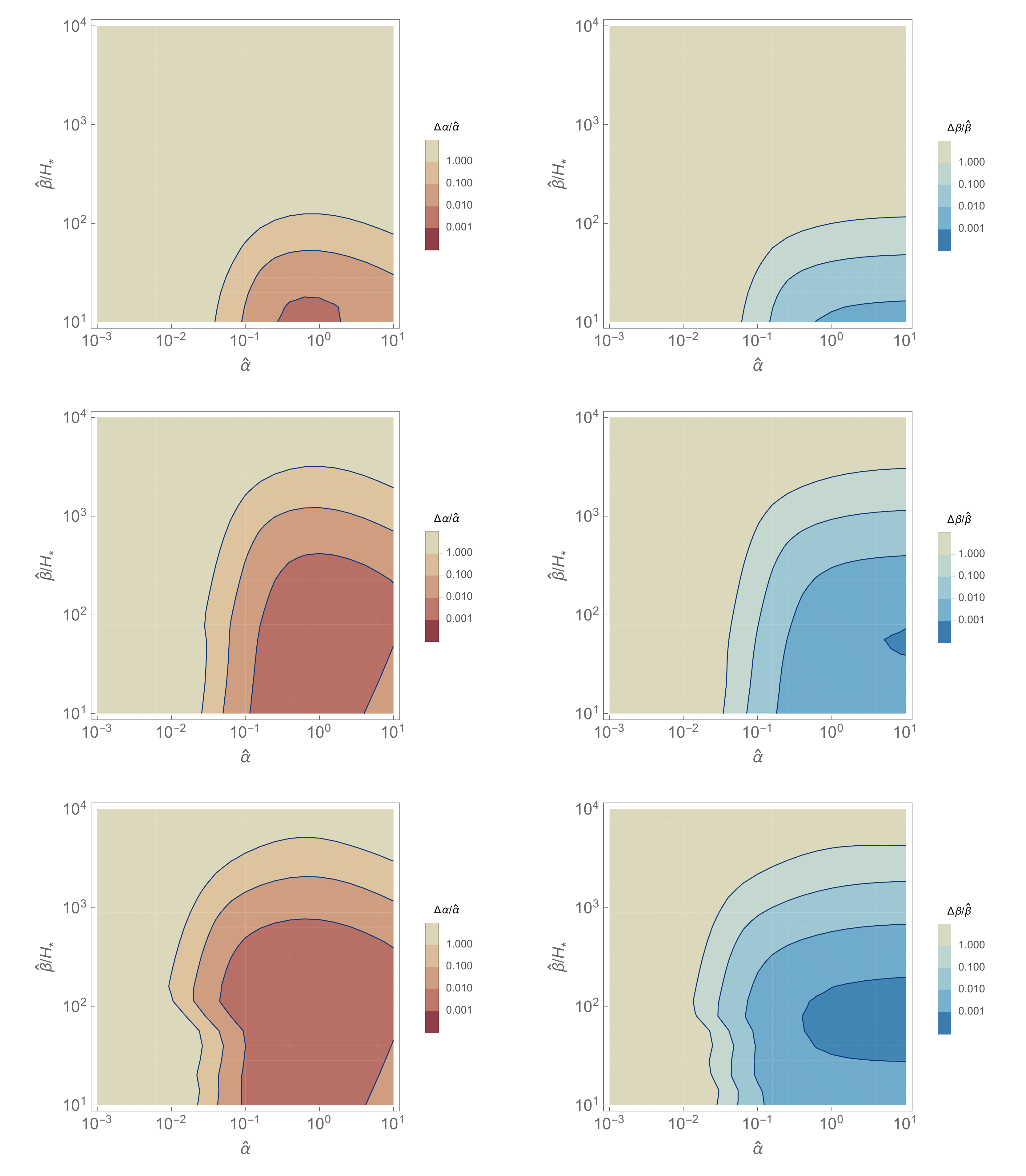} 
\caption{\small
1 $\sigma$ fractional error $\Delta \alpha / \hat{\alpha}$ (left) and $\Delta \beta / \hat{\beta}$ (right) for 3-parameter analysis.
Each row corresponds to LISA (top), DECIGO (middle) and BBO (bottom).
Otherwise the same as Fig.~\ref{fig:A21N2DeltaalphaDeltabeta}.
}
\label{fig:A21N3DeltaalphaDeltabeta}
\end{center}
\end{figure}

We first show how the result change if we include $S_{\rm NSBH}$, i.e. $S_h \supset S_{\rm WD} + S_{\rm NSBH}$,
keeping $v_w = 1$ as in Sec.~\ref{sec:Transition}.
Figs.~\ref{fig:A21N2DeltaalphaDeltabeta} and \ref{fig:A21N3DeltaalphaDeltabeta} are 
two- and three-parameter analyses, respectively.
These figures are practically the same as Figs.~\ref{fig:S4N2DelalphaDelbeta} and \ref{fig:S4N3DelalphaDelbeta}.
This is because, for $T_*$ fixed at $100$~GeV, the GW spectrum tend to have its peak below $1$~Hz,
and therefore the foreground $S_{\rm NSBH}$ in Eq.~(\ref{eq:SNSBH}) becomes almost irrelevant.

%%%%%%%%%%%%%%%
%%%%%%%%%%%%%%%

%%%%%%%%%%%%%%%
%%%%%%%%%%%%%%%

%%%%%%%%%%%%%%%%%%%%%%%%%%%%%%%%%%%%%%%%%%%%%%%%%%%%%%%
\subsubsection{\texorpdfstring{$v_w = 0.3$}{Lg}}
%%%%%%%%%%%%%%%%%%%%%%%%%%%%%%%%%%%%%%%%%%%%%%%%%%%%%%%

\begin{figure}
\begin{center}
\includegraphics[width=\columnwidth]{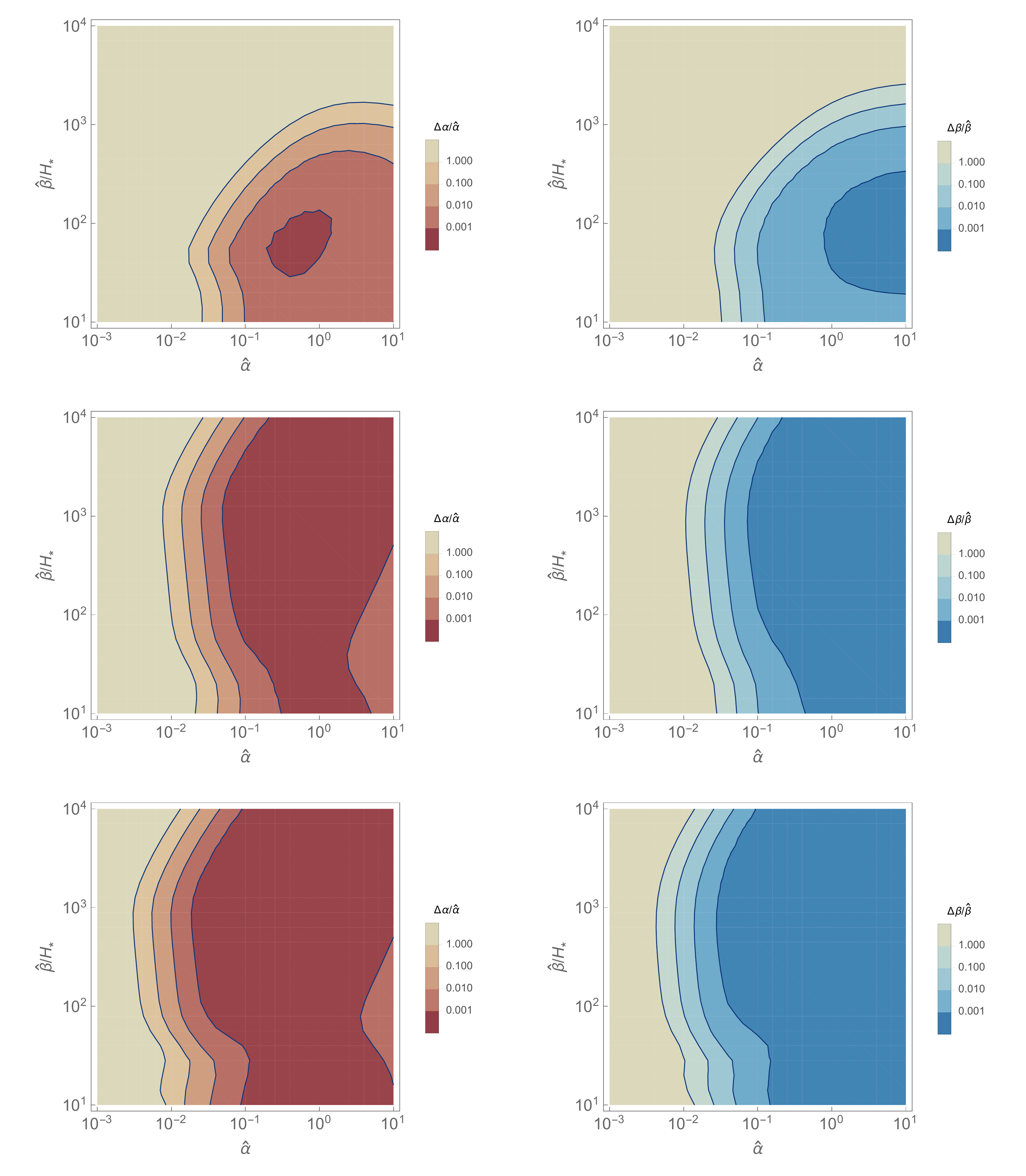} 
\caption{\small
1 $\sigma$ fractional error $\Delta \alpha / \hat{\alpha}$ (left) and $\Delta \beta / \hat{\beta}$ (right)
for the fiducial values $\hat{\alpha}$ and $\hat{\beta}$ for 2-parameter analysis.
Each row corresponds to LISA (top), DECIGO (middle) and BBO (bottom).
The wall velocity and foreground are taken to be $v_w = 0.3$ and $S_{\rm WD}$.
}
\label{fig:A221N2DeltaalphaDeltabeta}
\end{center}
\end{figure}

\begin{figure}
\begin{center}
\includegraphics[width=\columnwidth]{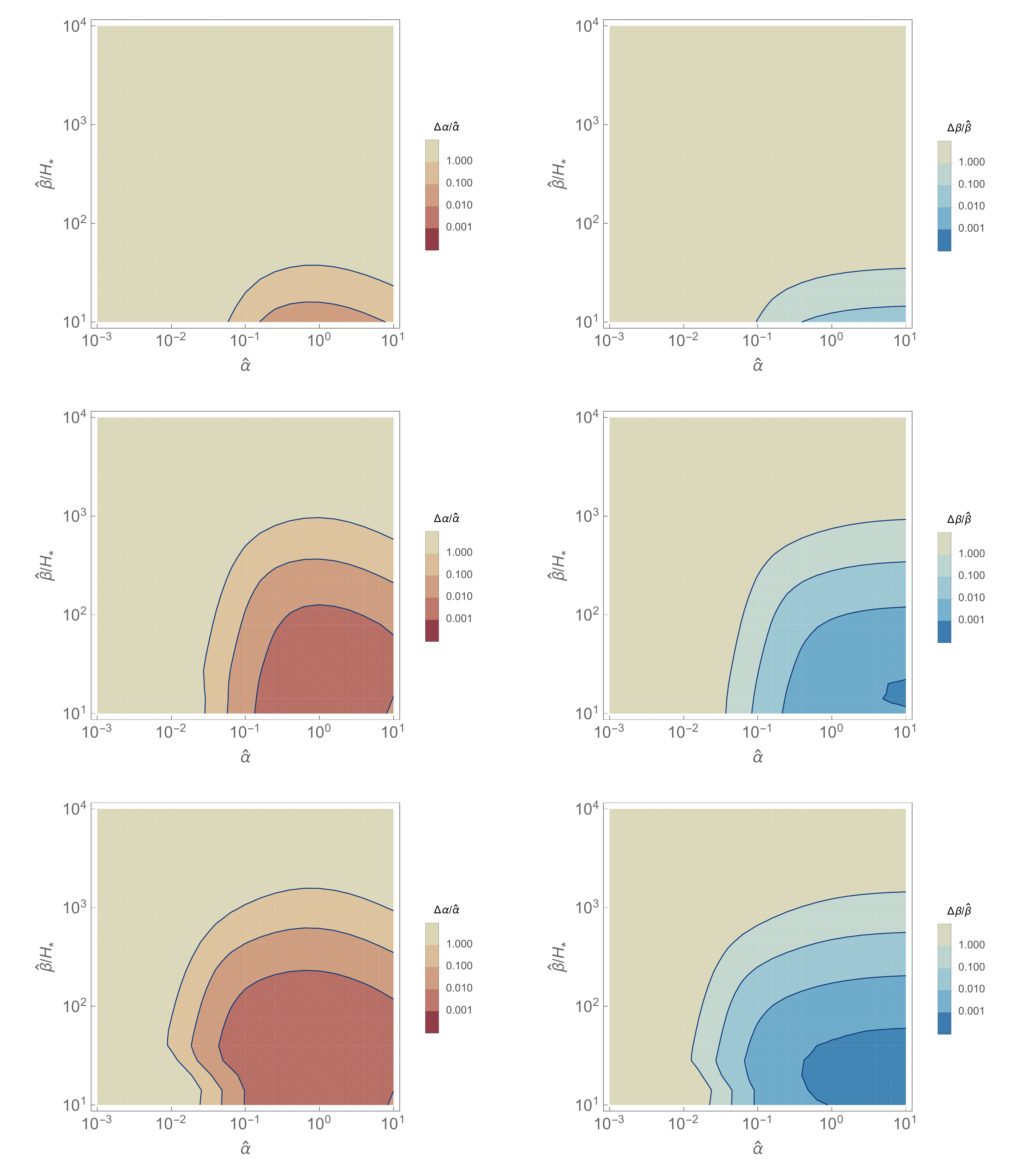} 
\caption{\small
1 $\sigma$ fractional error $\Delta \alpha / \hat{\alpha}$ (left) and $\Delta \beta / \hat{\beta}$ (right) for 3-parameter analysis.
Each row corresponds to LISA (top), DECIGO (middle) and BBO (bottom).
Otherwise the same as Fig.~\ref{fig:A221N2DeltaalphaDeltabeta}.
}
\label{fig:A221N3DeltaalphaDeltabeta}
\end{center}
\end{figure}

\begin{figure}
\begin{center}
\includegraphics[width=\columnwidth]{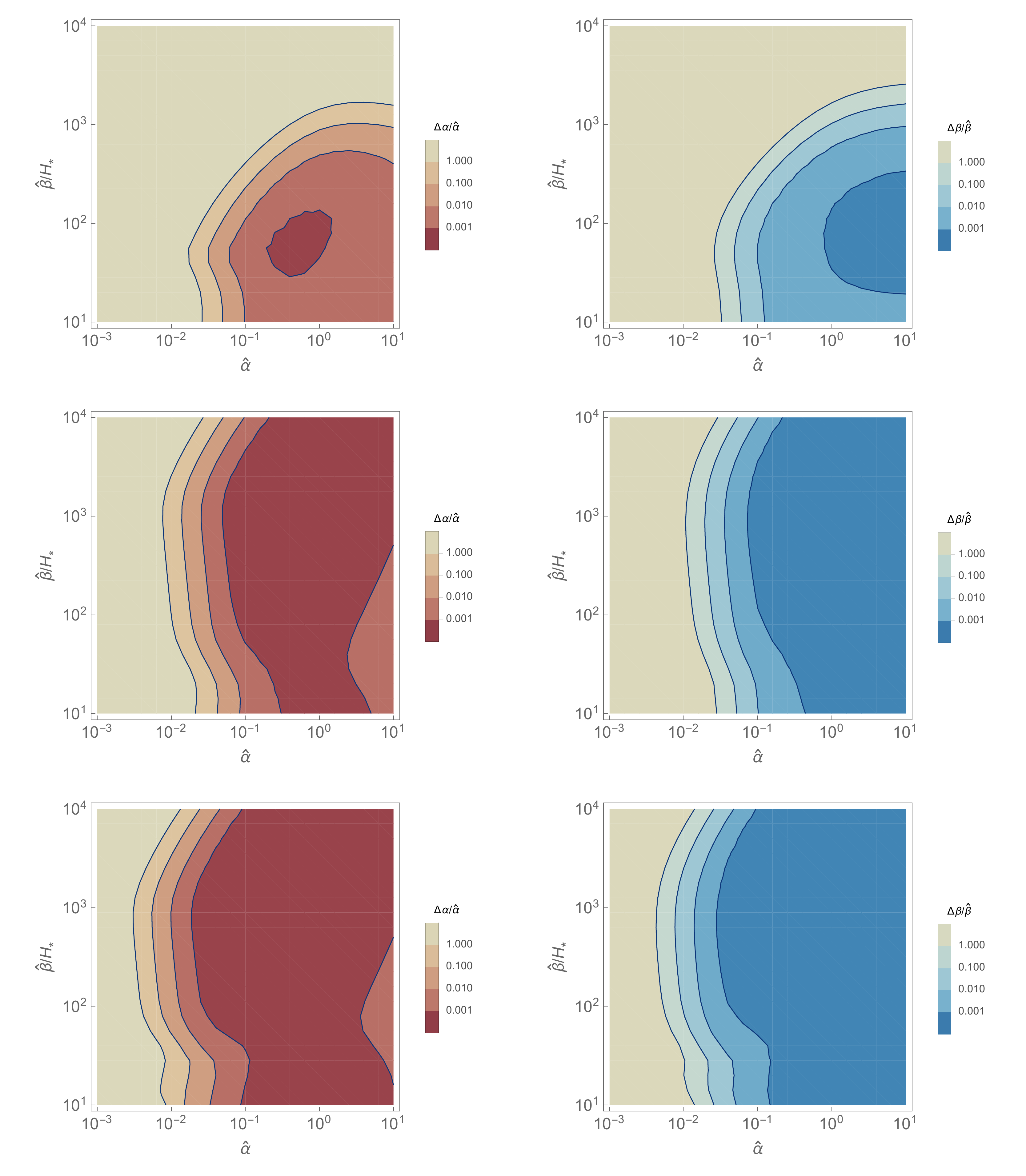} 
\caption{\small
1 $\sigma$ fractional error $\Delta \alpha / \hat{\alpha}$ (left) and $\Delta \beta / \hat{\beta}$ (right)
for the fiducial values $\hat{\alpha}$ and $\hat{\beta}$ for 2-parameter analysis.
Each row corresponds to LISA (top), DECIGO (middle) and BBO (bottom).
The wall velocity and foreground are taken to be $v_w = 0.3$ and $S_{\rm WD} + S_{\rm NSBH}$.
}
\label{fig:A222N2DeltaalphaDeltabeta}
\end{center}
\end{figure}

\begin{figure}
\begin{center}
\includegraphics[width=\columnwidth]{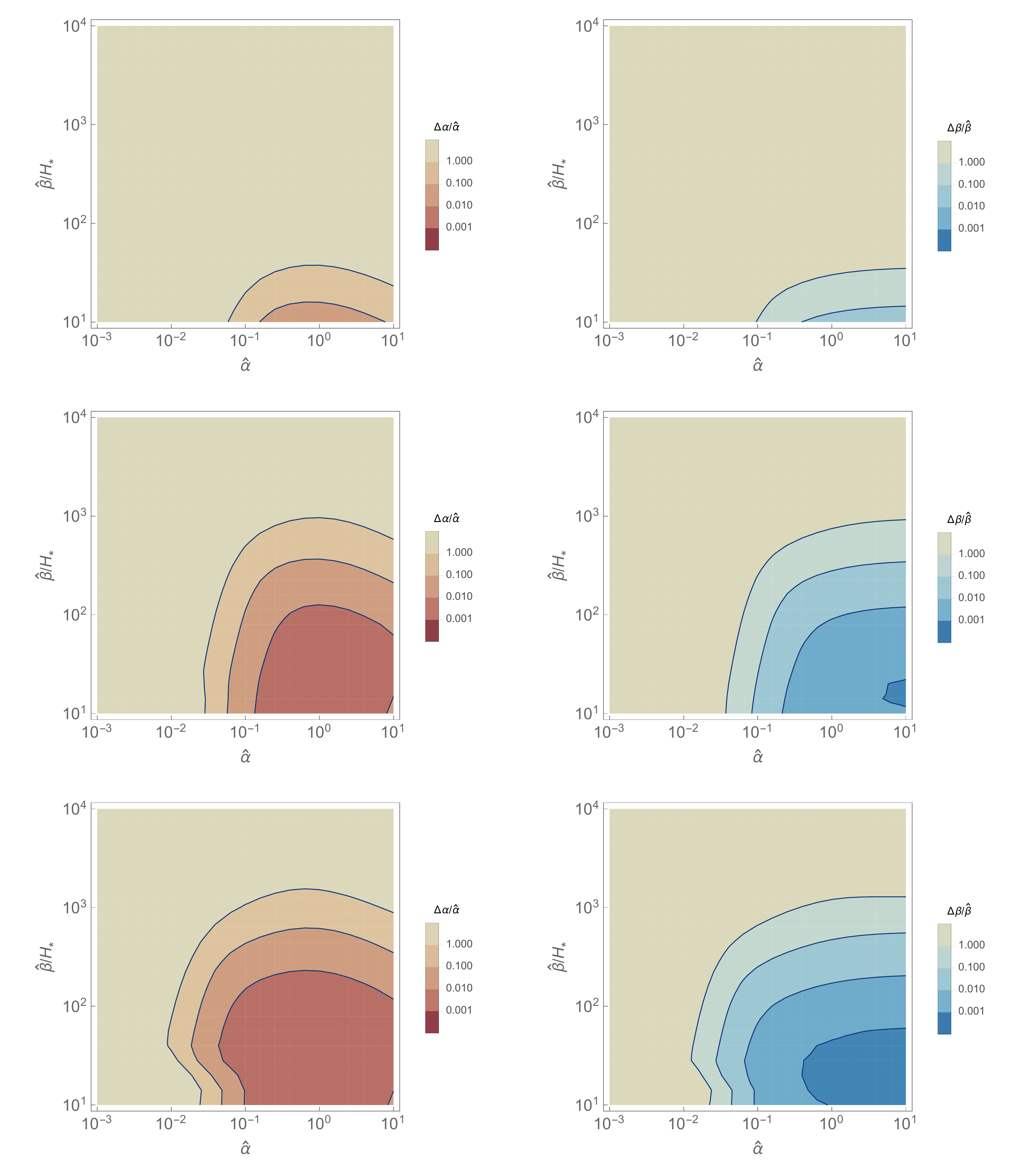} 
\caption{\small
1 $\sigma$ fractional error $\Delta \alpha / \hat{\alpha}$ (left) and $\Delta \beta / \hat{\beta}$ (right) for 3-parameter analysis.
Each row corresponds to LISA (top), DECIGO (middle) and BBO (bottom).
Otherwise the same as Fig.~\ref{fig:A222N2DeltaalphaDeltabeta}.
}
\label{fig:A222N3DeltaalphaDeltabeta}
\end{center}
\end{figure}

We next discuss how the result change if we take $v_w = 0.3$.
Fig.~\ref{fig:A221N2DeltaalphaDeltabeta} and \ref{fig:A221N3DeltaalphaDeltabeta} are the results of 
two- and three-parameter analyses, respectively,
with the foreground from white dwarfs only.
On the other hand,
Fig.~\ref{fig:A222N2DeltaalphaDeltabeta} and \ref{fig:A222N3DeltaalphaDeltabeta} are the results of 
two- and three-parameter analyses with the foreground from neutron stars and black holes also included.
It is seen that the parameter region shifts towards lower $\beta/H_*$
compared to $v_w = 1$ case.
This is because 
lower $v_w$ makes the peak frequency higher,
while lower $\beta/H_*$ compensates that by shifting the peak to lower frequency.
It is also seen that 
Fig.~\ref{fig:A221N2DeltaalphaDeltabeta}--\ref{fig:A221N3DeltaalphaDeltabeta}
and
Fig.~\ref{fig:A222N2DeltaalphaDeltabeta}--\ref{fig:A222N3DeltaalphaDeltabeta}
are almost the same.
This is because of the same reason as the previous subsection:
for $T_*$ fixed around $100$~GeV, there is essentially no effect of $S_{\rm NSBH}$.

%%%%%%%%%%%%%%%
%%%%%%%%%%%%%%%

%%%%%%%%%%%%%%%
%%%%%%%%%%%%%%%

%%%%%%%%%%%%%%%
%%%%%%%%%%%%%%%

%%%%%%%%%%%%%%%
%%%%%%%%%%%%%%%

%%%%%%%%%%%%%%%%%%%%%%%%%%%%%%%%%%%%%%%%%%%%%%%%%%%%%%%
\subsection{Fisher analysis on model parameters}
\label{app:Model}
%%%%%%%%%%%%%%%%%%%%%%%%%%%%%%%%%%%%%%%%%%%%%%%%%%%%%%%

This subsection supplements the results in Sec.~\ref{sec:Model}.

%%%%%%%%%%%%%%%%%%%%%%%%%%%%%%%%%%%%%%%%%%%%%%%%%%%%%%%
\subsubsection{\texorpdfstring{$O(N)$}{Lg} singlet extension of the SM}
%%%%%%%%%%%%%%%%%%%%%%%%%%%%%%%%%%%%%%%%%%%%%%%%%%%%%%%

\begin{figure}
\begin{center}
\includegraphics[width=0.32\columnwidth]{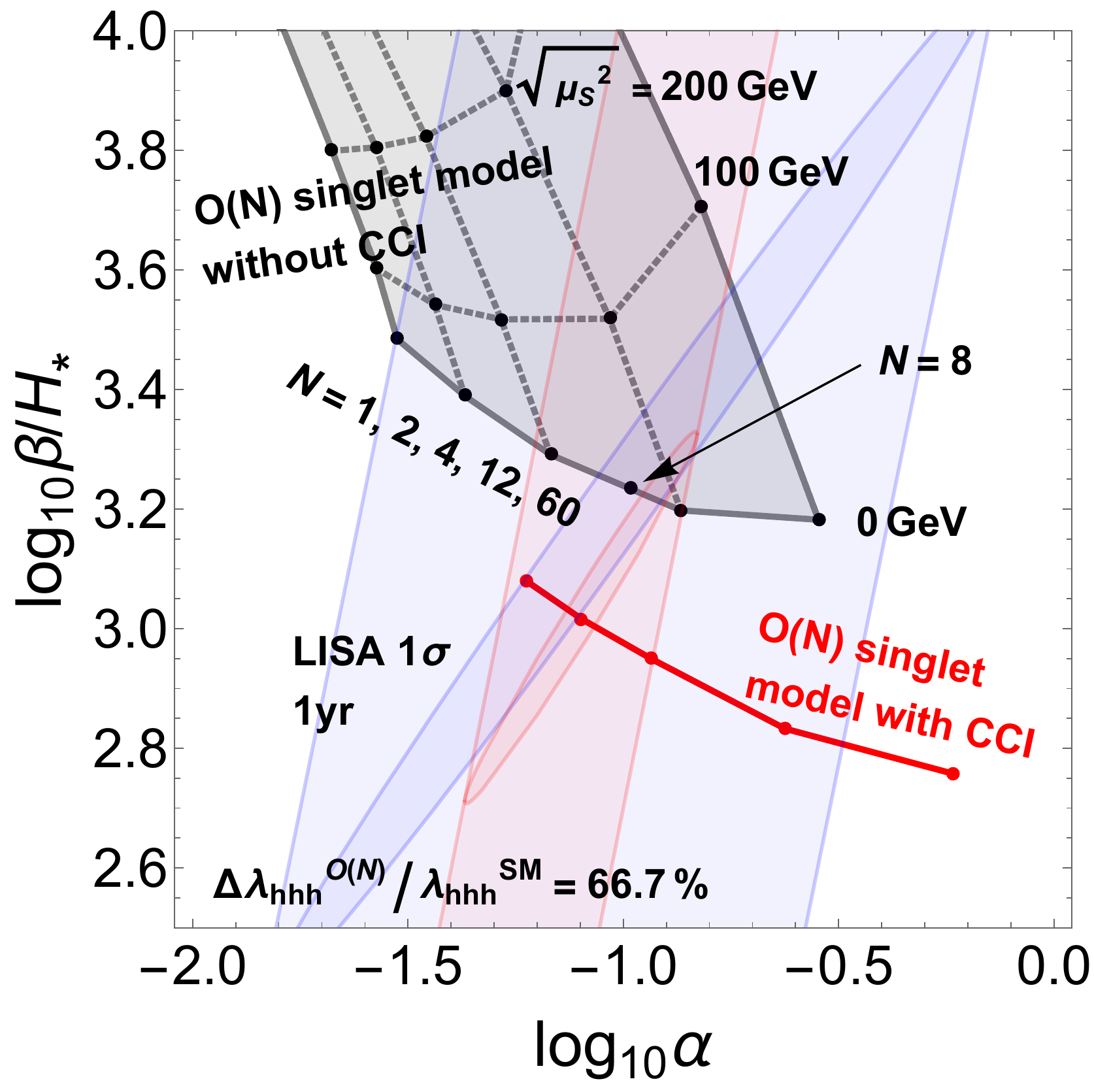} 
\includegraphics[width=0.32\columnwidth]{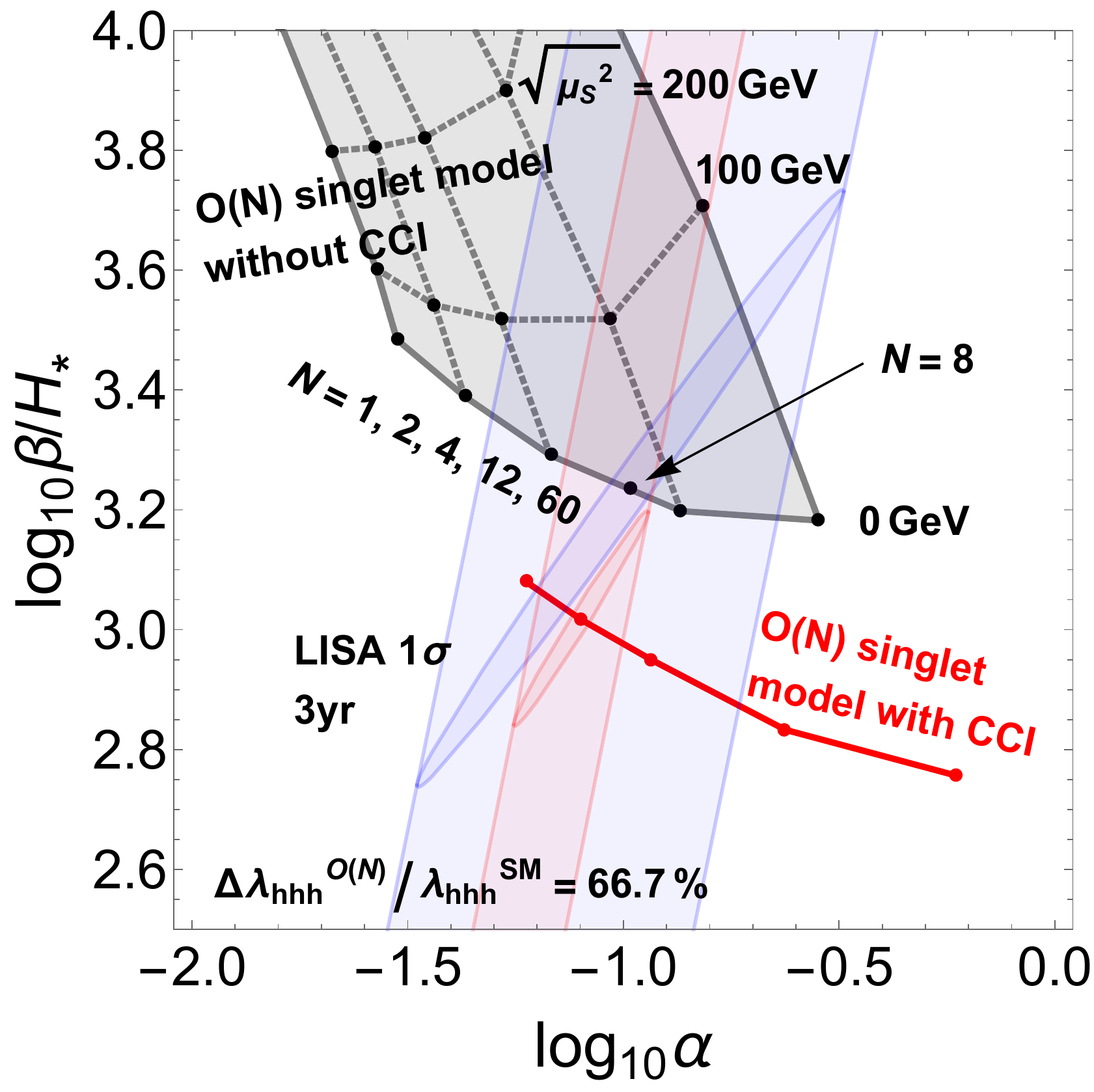} 
\includegraphics[width=0.32\columnwidth]{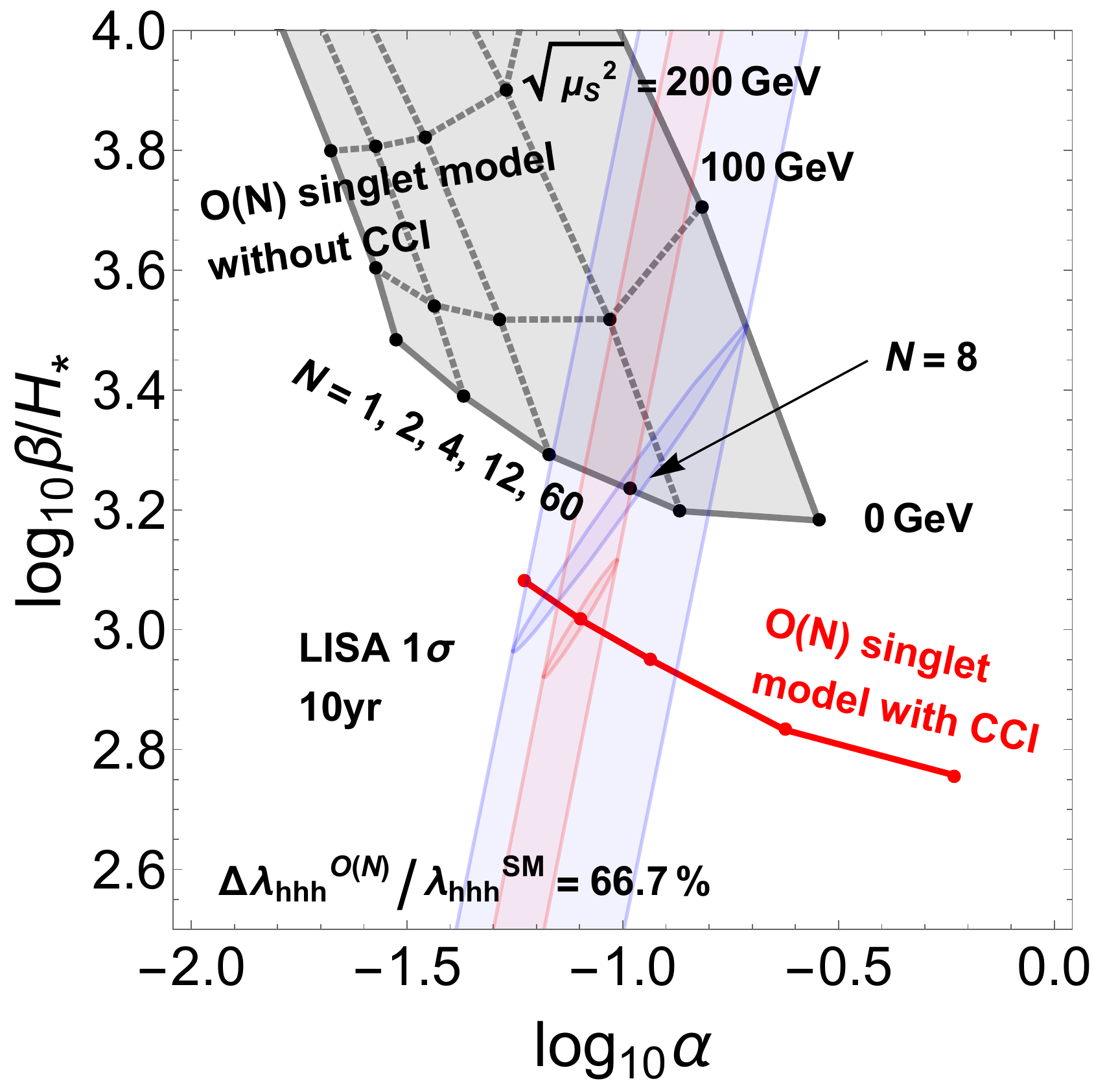} 
\caption{\small
LISA $1 \sigma$ contours for the $O(N)$ singlet models with and without CCI in $\alpha$-$\beta/H_*$ plane.
The same as the left panel of Fig.~\ref{fig:S51Synergy} 
($N = 2$ and $N = 8$ with and without CCI, respectively) 
except that the result of 3-parameter analysis is also shown.
}
\label{fig:A31_N=8}
\end{center}
\end{figure}

\begin{figure}
\begin{center}
\includegraphics[width=0.32\columnwidth]{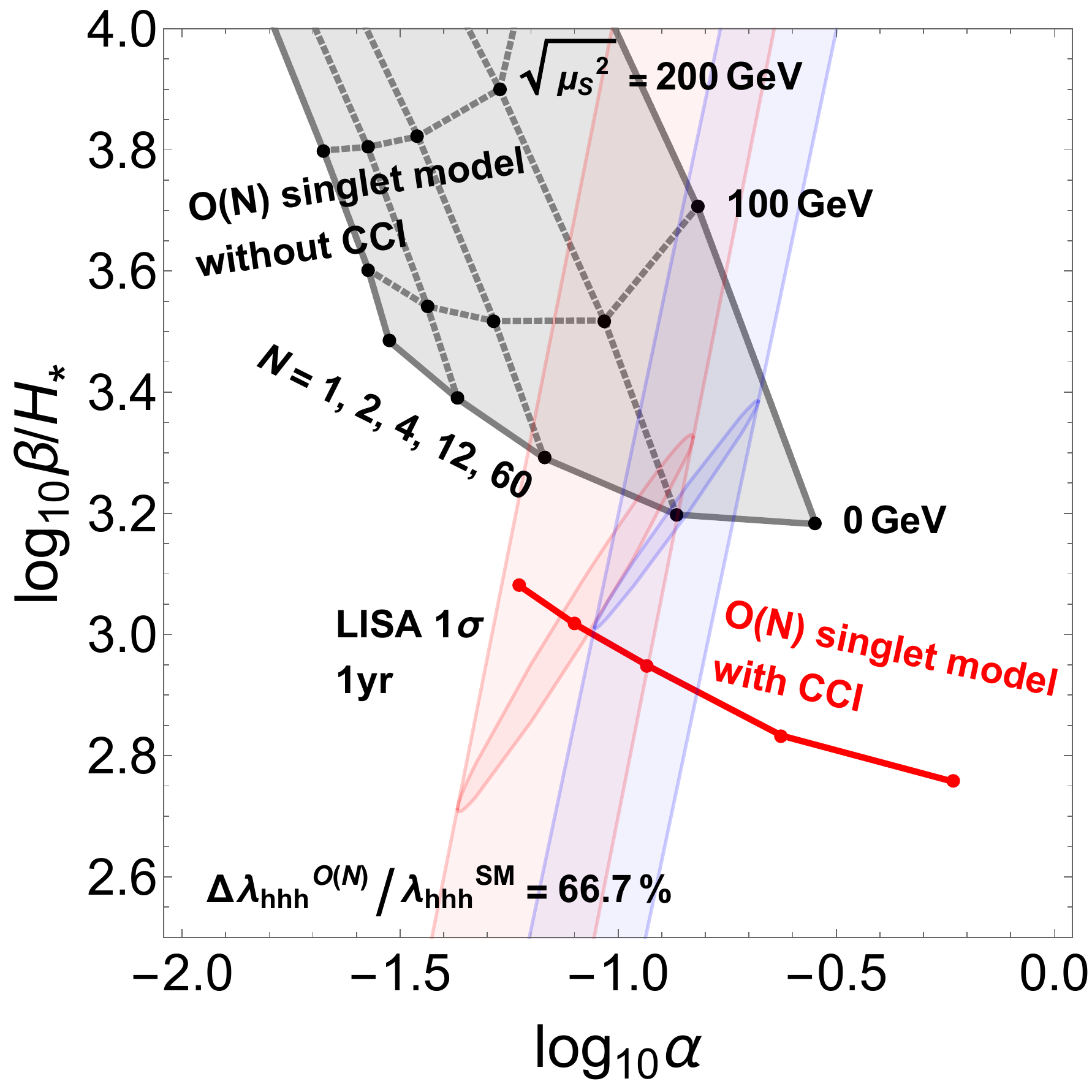} 
\includegraphics[width=0.32\columnwidth]{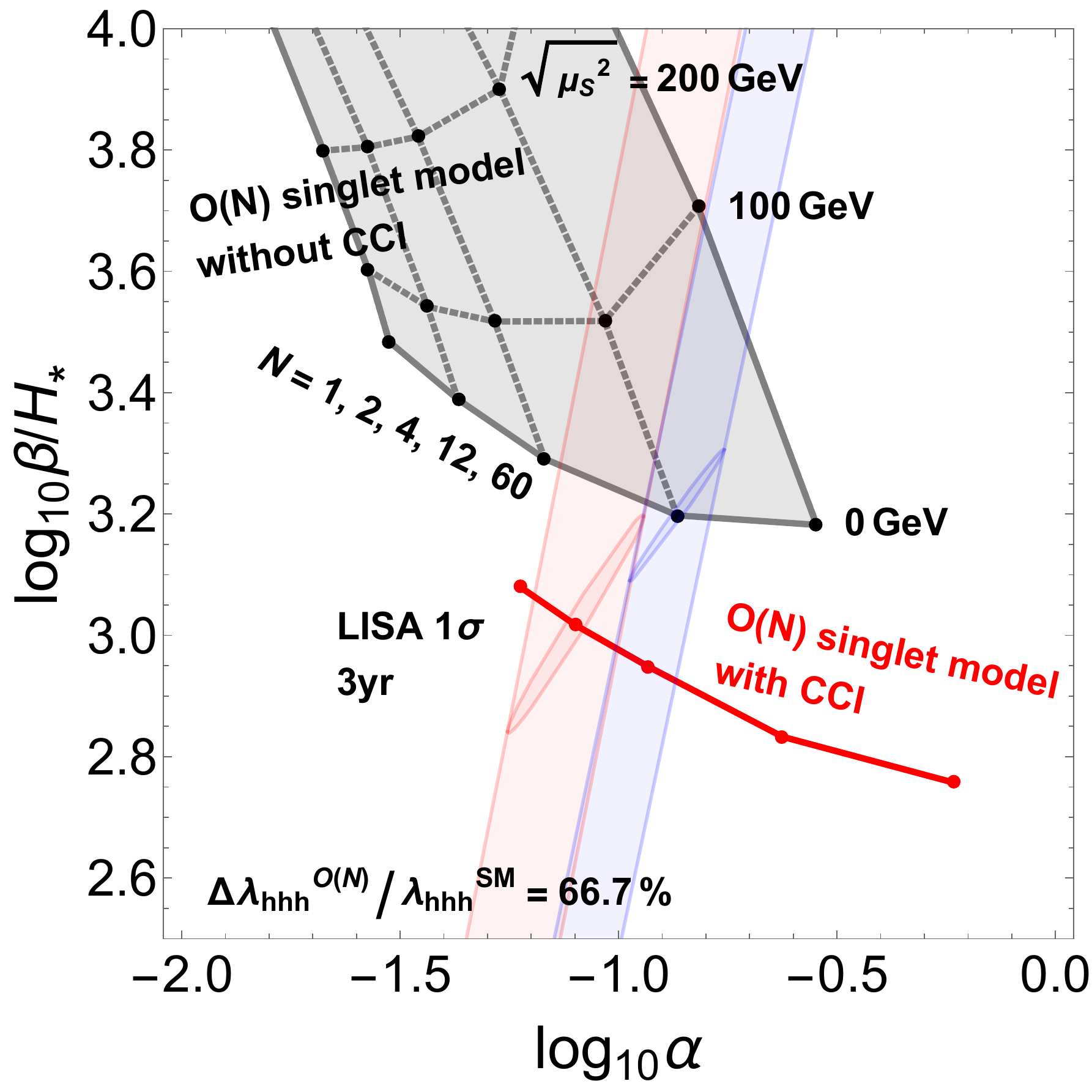} 
\includegraphics[width=0.32\columnwidth]{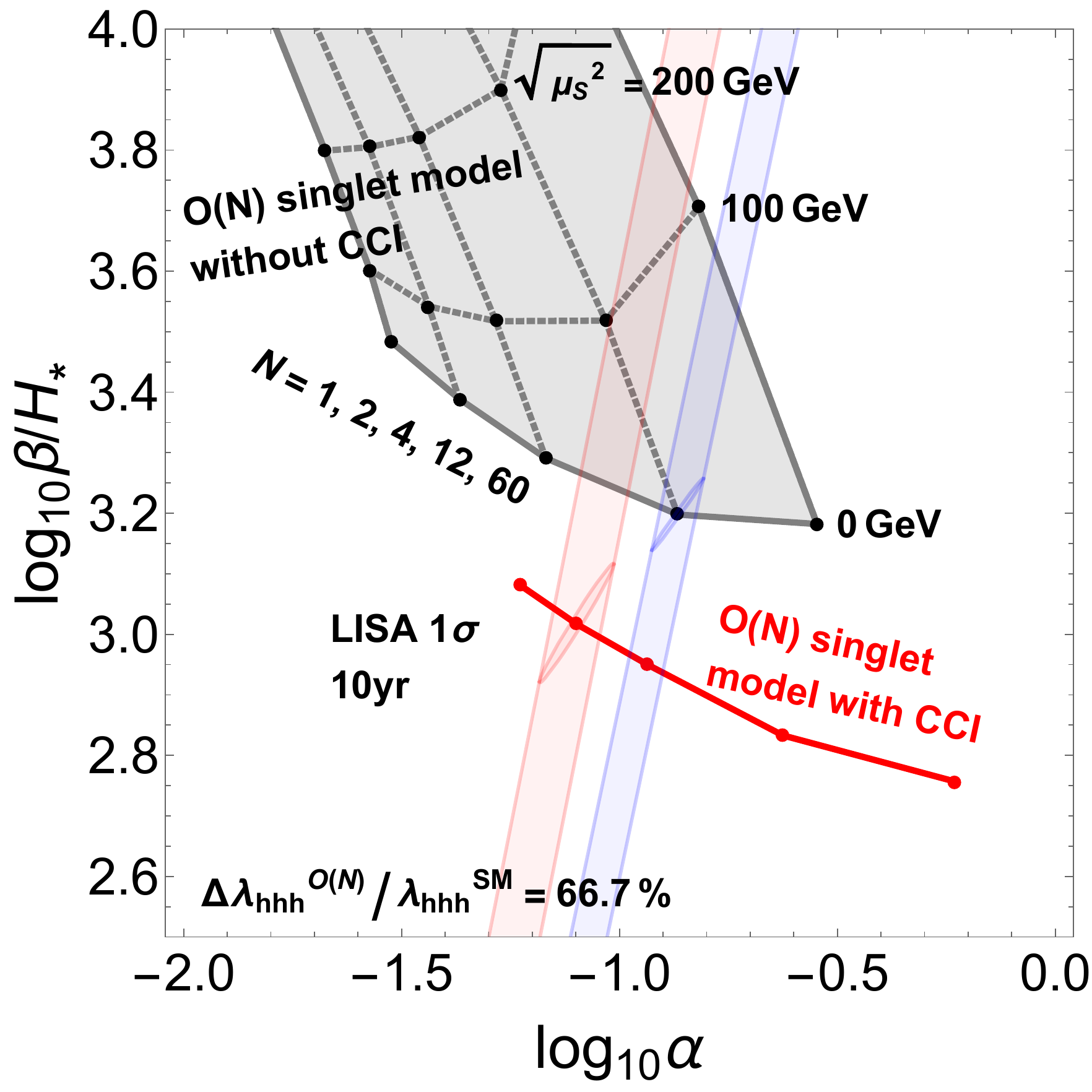} 
\caption{\small
LISA $1 \sigma$ contours for the $O(N)$ singlet models with and without CCI in $\alpha$-$\beta/H_*$ plane.
The same as the right panel of Fig.~\ref{fig:S51Synergy} 
($N = 2$ and $N = 12$ with and without CCI, respectively) 
except that the result of 3-parameter analysis is also shown.
}
\label{fig:A31_N=12}
\end{center}
\end{figure}

In Figs.~\ref{fig:A31_N=8} and \ref{fig:A31_N=12} we show the results of a Fisher analysis 
after marginalizing over the temperature just after the transition $T_*$.
The bands stretch in the $\beta/H_*$ direction,
but we still have the possibility to distinguish the models,
as seen in the right panel of Fig.~\ref{fig:A31_N=12}.

%%%%%%%%%%%%%%%
%%%%%%%%%%%%%%%

%%%%%%%%%%%%%%%
%%%%%%%%%%%%%%%

%%%%%%%%%%%%%%%%%%%%%%%%%%%%%%%%%%%%%%%%%%%%%%%%%%%%%%%
\subsubsection{Classically conformal $B-L$ model}
%%%%%%%%%%%%%%%%%%%%%%%%%%%%%%%%%%%%%%%%%%%%%%%%%%%%%%%

\begin{figure}
\begin{center}
\includegraphics[width=\columnwidth]{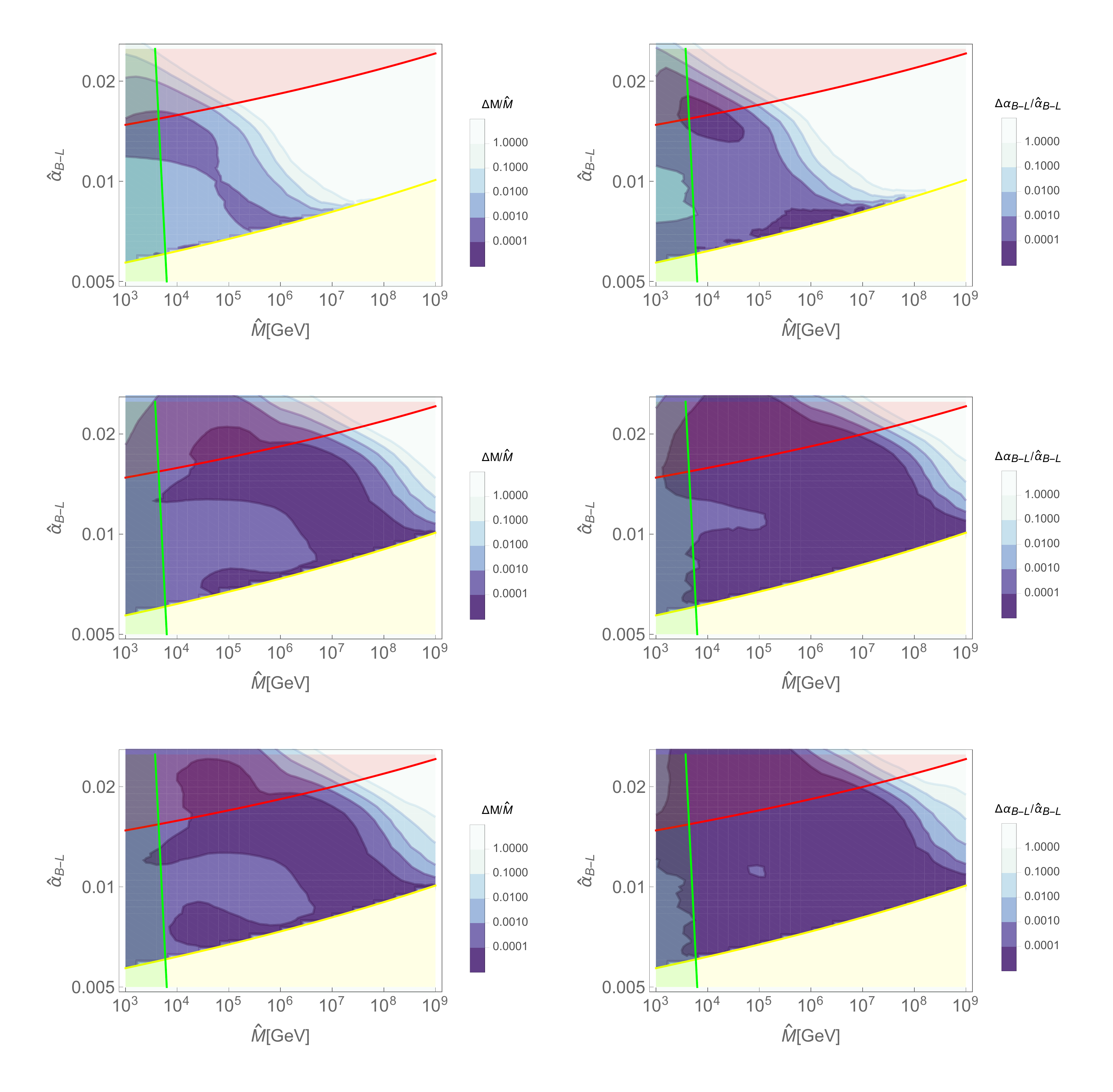} 
\caption{\small
1 $\sigma$ fractional error for $\Delta M / \hat{M}$ (left) and $\Delta \alpha_{B-L} / \hat{\alpha}_{B-L}$ (right) 
for the fiducial values $\hat{M}$ and $\hat{\alpha}_{B-L}$.
Each row corresponds to LISA (top), DECIGO (middle) and BBO (bottom).
Regions shaded in red, green and yellow are the same as Figs.~\ref{fig:S53TstarTR}--\ref{fig:S53alphabeta}.
The foreground is taken to be $S_{\rm WD} + S_{\rm NSBH}$.
}
\label{fig:A33DelMDelalpha}
\end{center}
\end{figure}

In Fig.~\ref{fig:A33DelMDelalpha} we show the result of a Fisher analysis
corresponding to the result in Sec.~\ref{subsec:Clacon} after including $S_{NSBH}$. 
It is seen that the result changes only slightly compared to Fig.~\ref{fig:S53DelMDelalpha}.
This is because the amount of GWs produced in this model is so large that they dominate the foreground $S_{NSBH}$.

%%%%%%%%%%%%%%%
%%%%%%%%%%%%%%%

\clearpage

%%%%%%%%%%%%%%%%%%%%%%%%%%%%%%%%%%%%%%%%%%%%%%%%%%
\small
\bibliography{ref}
%%%%%%%%%%%%%%%%%%%%%%%%%%%%%%%%%%%%%%%%%%%%%%%%%%

%%%%%%%%%%%%%%%%%%%%%%%%%%%%%%%%%%%%%%%%%%%%%%%%%%
\end{document}